\documentclass[12pt]{article}
\usepackage[english]{babel}
\usepackage[latin1]{inputenc}
\usepackage[active]{srcltx}
\usepackage{amsfonts,amssymb,amsmath, epsfig}
\usepackage{graphicx,graphics,psfrag}
\usepackage[usenames]{color}
\usepackage{subfigure}
\usepackage{amsmath,amstext,amssymb,amsfonts, amscd}
\usepackage{tikz}
\usetikzlibrary{shapes,arrows,shadows,matrix}
\usetikzlibrary{shapes.gates.logic.US,trees,positioning}

\usepackage{hyperref}           

\def\Box{\leavevmode\vbox{\hrule
     \hbox{\vrule\kern4pt\vbox{\kern4pt}%
           \vrule}\hrule}}

\newcounter{appendix}
\def\appendix{\advance\c@appendix by 1
   \def\thesection{\Alph{section}}
   \ifnum\c@appendix=1 \setcounter{section}{-1} \fi
   \@startsection {section}{1}{\z@}{-3.5ex plus -1ex minus 
   -.2ex}{2.3ex plus .2ex}{\Large\bf}}

\def\paragraph#1{{\bf #1\ }}

\newtheorem{lemma}{Lemma}[section]

\newtheorem{proposition}[lemma]{Proposition}

\newtheorem{remark}{Remark}[section]

\title{Self-Organized Hydrodynamics with congestion and path formation in
crowds}
\author{Pierre Degond$^{(1,2)}$, Jiale Hua$^{(1,2,3)}$}
\date{}
\begin{document}

\maketitle

\begin{center}
1-Universit\'e de Toulouse; UPS, INSA, UT1, UTM ;\\ 
Institut de Math\'ematiques de Toulouse ; \\
F-31062 Toulouse, France. \\
2-CNRS; Institut de Math\'ematiques de Toulouse UMR 5219 ;\\ 
F-31062 Toulouse, France.\\
email: pierre.degond@math.univ-toulouse.fr\\
3-current address: Universit\'e de Grenoble and CNRS, \\
Laboratoire Jean Kuntzmann, BP 53, 38041 Grenoble Cedex, France \\
email: jiale.hua@imag.fr
\end{center}

\begin{abstract}
A continuum model for self-organized dynamics is numerically investigated. The model describes systems of particles subject to alignment interaction and short-range repulsion. It consists of a non-conservative hyperbolic system for the density and velocity orientation. Short-range repulsion is included through a singular pressure which becomes infinite at the jamming density. The singular limit of infinite pressure stiffness leads to phase transitions from compressible to incompressible dynamics. The paper proposes an Asymptotic-Preserving scheme which takes care of the singular pressure while preventing the breakdown of the CFL stability condition near congestion. It relies on a relaxation approximation of the system and an elliptic formulation of the pressure equation. Numerical simulations of impinging clusters show the efficiency of the scheme to treat congestions. A two-fluid variant of the model provides a model of path formation in crowds. 
\end{abstract}

\medskip
\noindent
{\bf Acknowledgments:} 
This work has been supported by the French 'Agence Nationale pour la Recherche (ANR)' in the frame of the contracts 'Panurge' (ANR-07-BLAN-0208-03), 'Pedigree' (ANR-08-SYSC-015-01) and 'MOTIMO' (ANR-11-MONU-009-01) .

\medskip
\noindent
{\bf Key words: } Self-propelled particles, orientation dynamics, self-organization, hydrodynamic limit, volume exclusion, congestion, jamming,  finite volumes, Asymptotic-Preserving scheme, herds, crowds, path formation

\medskip
\noindent
{\bf AMS Subject classification: } 35L60, 35Q82, 82C22, 82C70, 92D50, 65M08, 65Z05, 76N99, 76L05
\vskip 0.4cm

\setcounter{equation}{0}
\section{Introduction}
\label{sec_intro}

The modeling of self-organized dynamics is at the core of an intense scientific activity. Self-organized dynamics occurs in systems of active agents such as bacterial suspensions and sperm \cite{Koch_Subramanian_AnnRevFluidMech11}, insect swarms \cite{Buhl_etal_Science06}, fish schools \cite{Aoki_JapanFisheries82, Couzin_etal_JTB02}, bird flocks \cite{Ballerini_etal_PNAS08, Lukeman_etal_PNAS10}, mammal herds or pedestrian crowds \cite{Helbing_Molnar_PRE95, Kretz_etal_JStatMech06, Moussaid_etal_PlocCB12}. These systems exhibit large-scale coherent structures that spontaneously emerge from the interactions between the agents but are not directly encoded in the interaction rules. Many mathematical models have been proposed to account for the emergence of large-scale order \cite{Aoki_JapanFisheries82, Bertin_etal_JPA09, Buhl_etal_Science06, Cucker_Smale_JapanJMath07, Couzin_etal_JTB02, Dorsogna_etal_PRL06, Gregoire_Chate_PRL04,  Mogilner_LEK_JMB99, Toner_Tu_PRE98, Topaz_etal_BMB06}. One of the most paradigmatic models is the Vicsek particle system \cite{Vicsek_etal_PRL95} which consists of self-propelled particles interacting through alignment interaction. Each particle moves with a constant speed and updates its direction so as to align with its neighbors up to a certain noise level. The Vicsek model exhibits phase transitions from disordered to fully aligned states when the noise is decreased or when the density is increased. 

In practice, more refined flocking models must be used to reproduce flocking behavior observed in nature. Among these refined models, the three zone model \cite{Aoki_JapanFisheries82, Couzin_etal_JTB02} offers a good compromise between physical accuracy and simplicity. It considers three kinds of interactions: long-range attraction, medium-range alignment '\`a la Vicsek' and short-range repulsion. In particular, short-range repulsion is believed to play a key role in the observed morphogenetic features of self-organized dynamics. The flock can indeed be seen as the region where all the particles are as close as they can be, given the short range repulsion, i.e. as the region where the particle density reaches a maximum value. The existence of a maximal allowed value of the density resulting from volume exclusion is referred to as the congestion constraint. 

Macroscopic models of fluid type (or hydrodynamic models) are particularly well suited to the modeling of large-scale self-organized structures by contrast to particle models (also known as 'Individual-Based Models' or IBM) which focus on the inter-particle interaction scale. Indeed, the numerical complexity of IBM's increases with the number of individuals faster than linearly. They become computationally very intensive for large particle systems which are the most relevant ones for the observation of self-organization. Additionally, in some cases, self-organization can be directly encoded into the fluid model and facilitates the observation of the resulting morphogenetic features. In particular, this can be realized with the congestion constraint which results in a clear-cut phase transition between compressible and incompressible fluid regions, the latter being those where the congestion constraint is reached. 

The goal of the present work is to study a hydrodynamic model of self-organized dynamics with a congestion constraint. The starting point is the hydrodynamic model of \cite{Degond_Motsch_M3AS08}, which will be referred to as 'Self-Organized Hydrodynamics' or in short, SOH. This model is derived through a macroscopic limit of the Vicsek particle system \cite{Vicsek_etal_PRL95}. The SOH model resembles the isentropic compressible Euler equations with two major differences. First, the local mean velocity $\boldsymbol{\Omega}$ is constrained to have unit norm $|\boldsymbol{\Omega}|=1$ (specifically, $\boldsymbol{\Omega}$ is the mean velocity direction rather the mean velocity itself), and second, the pressure gradient term in the momentum balance equation is projected onto the plane normal to $\boldsymbol{\Omega}$, in order to preserve the norm constraint. This makes the system non-conservative and non-classical since the state variables belong to a curved manifold. 

The SOH model only takes into account the alignment interaction. In the present work, we take into account the short-range repulsion by considering a singular pressure when the density $\rho$ approaches the congestion density $\rho^*$. The SOH model with volume exclusion constraint has been theoretically investigated in \cite{Degond_etal_JSP10} where the limit of an increasingly steeper singular pressure has been studied. This limit process results in a model where clear-cut phase transitions between uncongested and congested  areas occur, the former corresponding to a compressible model and the latter, to an incompressible one. 
In \cite{Degond_etal_JSP10}, the dynamics of the boundary between these areas (or boundary of the flock) has been studied. 

The present work is devoted to the derivation and study of adequate numerical methods for this model. In \cite{Degond_etal_JCP11}, a numerical method for the classical isentropic compressible Euler equations with congestion constraint has been derived. The key point is the use of an 'Asymptotic-Preserving' (AP) scheme previously developed for the small Mach-number limit of compressible flows \cite{Cordier_etal_JCP12, Degond_Tang_CICP11, Tang_KRM12}. Here, we adapt this scheme to the constrained SOH model. This adaptation is not straightforward, because of the non-conservative character of the SOH model and of the unit norm constraint on the velocity. In \cite{Motsch_Navoret_MMS11}, it has been shown that a relaxation method is best suited to the numerical resolution of the unconstrained SOH model. The present paper shows that the relaxation method of \cite{Motsch_Navoret_MMS11} and the AP methodology of \cite{Degond_etal_JCP11} can be combined in order to resolve the constrained SOH model. 

As an application of the presented methodology, a model of path formation in crowds is presented. The spontaneous emergence of trails in pedestrian counterflows in corridors is a well documented phenomenon \cite{Kretz_etal_JStatMech06, Moussaid_etal_PlocCB12}. It has inspired several kinds of modeling approaches, both at the discrete \cite{Helbing_Molnar_PRE95, Moussaid_etal_PNAS11} and continuous \cite{Appert_etal_NHM11, Burger_etal_KRM11} levels. Our model intends to describe path formation in a crowd of people heading towards opposite directions. It requires a two-fluid extension of the model to account for the existence of two kinds of pedestrians having opposite target directions. Our model shows that the congestion constraint prevents the flow to cross through high density regions, forcing the emergence of paths of oppositely moving fluids. 

Throughout the paper, the key property of the developed scheme is the Asymptotic-Preserving (AP) property. It is defined as follows. Consider a continuous physical model ${\cal{M}}^\varepsilon$ which involves a perturbation parameter $\varepsilon$ (here, $\varepsilon$ is the parameter describing the stiffness of the pressure and ${\cal{M}}^\varepsilon$ represents the SOH with stiffened pressure) which can range from $\varepsilon = {\mathcal O}(1)$ to $\varepsilon \ll 1$ values. Let ${\cal{M}}^{0}$ be the limit of ${\cal{M}}^\varepsilon$ when $\varepsilon \to 0$ (here ${\cal{M}}^{0}$ is the constrained SOH model). Let now ${\cal{M}}^\varepsilon_\Delta$ be a numerical scheme which provides a consistent discretization of ${\cal{M}}^\varepsilon$ with discrete time and space steps $(\Delta t, \Delta x)=\Delta$. The scheme ${\cal{M}}^\varepsilon_\Delta$ is said to be \textit{Asymptotic-Preserving (AP)} if its stability condition is independent of $\varepsilon$ and if its limit ${\cal{M}}^0_\Delta$ as $\varepsilon \to 0$ provides a consistent discretization of the continuous limit model ${\cal{M}}^{0}$. The AP property is illustrated by the commutative diagram of fig. \ref{fig:ap}. The literature about AP schemes is recent, yet increasingly abundant and applied to various contexts (see e.g.  \cite{Carrillo_etal_JCP08, Jin_SISC99, Klar_SINUM99}). 

\begin{figure}[h!]
\begin{center}
\begin{tikzpicture}[description/.style={fill=white,inner sep=2pt}]
\matrix (m) [matrix of math nodes, row sep=3em,
column sep=4em, text height=3ex, text depth=0.5ex]
{ {\cal{M}}^\varepsilon &  {\cal{M}}^{0} \\
{\cal{M}}^\varepsilon_\Delta &   {\cal{M}}^0_\Delta \\ };

\path[->,font=\scriptsize]
(m-1-1) edge node[auto] {$\varepsilon \to 0 $} (m-1-2)
(m-2-1) edge node[auto] {$ \Delta \to 0 $} (m-1-1);

\path[<-,font=\scriptsize]
(m-1-2) edge node[auto] {$ \Delta \to 0$} (m-2-2)
(m-2-2) edge node[auto] {$ \varepsilon \to 0 $} (m-2-1);
\end{tikzpicture}
\caption{Asymptotic-Preserving (AP) property: the upper horizontal arrow translates the assumption that the continuous model ${\cal{M}}^\varepsilon$ tends to the limit model ${\cal{M}}^{0}$ when $\varepsilon \to 0$. The left vertical arrow expresses that ${\cal{M}}^\varepsilon_\Delta$ is a consistent discretization of ${\cal{M}}^\varepsilon$ when the discretization parameter $\Delta \to 0$. The lower horizontal arrow indicates that the scheme ${\cal{M}}^\varepsilon_\Delta$ has a limit ${\cal{M}}^0_\Delta$ when $\varepsilon \to 0$ for fixed $\Delta$. Finally, the right vertical arrow expresses the AP-property: it says that the limit scheme ${\cal{M}}^0_\Delta$ is a consistent discretization of the limit model ${\cal{M}}^{0}$ when $\Delta \to 0$.}
\label{fig:ap}
\end{center}
\end{figure}

Constrained hydrodynamic models obtained as limits of hydrodynamic models involving a singular non-linear pressure have first been proposed in \cite{Bouchut_etal_JNLS00} and studied in \cite{Berthelin_M3AS2002} for the compressible Euler equations. They have then been extended to traffic flow to describe the dynamics of traffic jams \cite{Berthelin_etal_ARMA08}. The SOH model is derived from a kinetic formulation of the Vicsek model \cite{Degond_Motsch_M3AS08}. This kinetic description has been shown to be a valid description of the particle system in \cite{Bolley_etal_AMl12}. Variants of the SOH model taking into account non-isotropic observation \cite{Frouvelle_M3AS12}, phase transitions \cite{Degond_etal_submitted1, Frouvelle_Liu_SIMA12}, viscosity \cite{Degond_etal_submitted2, Degond_Yang_M3AS10}, trajectory control by curvature \cite{Degond_Motsch_JSP11} and precession \cite{Degond_Liu_M3AS12} can be found. Alternate hydrodynamic models derived from the Vicsek system can be found in \cite{Czirok_Vicsek_PhysicaA00, Ratushnaya_PhysicaA07} as well as related models in \cite{Carrillo_etal_KRM09, Carrillo_etal_M3AS10}.

The paper is organized as follows. The model framework is given in section \ref{sec_background}. Then, the numerical
method is introduced in section \ref{sec_numerical_method}. Numerical results are given in section \ref{sec_numerical_results}. The application to path formation in crowds is reported in section \ref{sec_two_fluids}. A conclusion is drawn in section \ref{sec_conclu}. Finally, for the reader's convenience, the expression of the scheme in the two-dimensional framework is exposed in Appendix 1.

\setcounter{equation}{0}
\section{Model framework}
\label{sec_background}

The Self-Organized Hydrodynamics (SOH) model is written as follows:
\begin{align}
&\rho_t + \nabla_{\boldsymbol{x}} \cdot \boldsymbol{q} = 0,\qquad \boldsymbol{q}  = \rho \boldsymbol{\Omega}, \label{eq:1}\\
&\rho \big(\boldsymbol{\Omega}_t + c (\boldsymbol{\Omega} \cdot
\nabla_{\boldsymbol{x}}) \boldsymbol{\Omega} \big) +
\lambda \left( \text{Id} - \boldsymbol{\Omega} \otimes \boldsymbol{
    \Omega} \right) \nabla_{\boldsymbol{x}} \big(p^\varepsilon(\rho) \big) =
0,\label{eq:2}\\
&|\boldsymbol{\Omega}|=1,\label{eq:22}
\end{align}
where $\rho(\boldsymbol{x},t) \in {\mathbb R}_+$ denotes the mass density,  $\boldsymbol{\Omega} (\boldsymbol{x},t) \in {\mathbb R}^2$, the mean velocity and $\boldsymbol{q} (\boldsymbol{x},t) \in {\mathbb R}^2$ the momentum  density, depending on the spatial position $\boldsymbol{x} \in {\mathbb R}^2$ and the time $t > 0$. The constants are such that $\lambda>0$ and $c \in {\mathbb R}$. The term $p^\varepsilon(\rho)=\varepsilon p(\rho)$ is the pressure, with a function $p(\rho)$ specified below. We denote by $\otimes$ the tensor product of two vectors. The matrix $\left( \text{Id} - \boldsymbol{\Omega} \otimes \boldsymbol{\Omega} \right)$ is the orthogonal projection onto the plane orthogonal to $\boldsymbol{\Omega}$. We do not specify any boundary conditions for the time being and suppose that initial conditions $\rho_0$ and $\boldsymbol{\Omega}_0$ are provided such that $|\boldsymbol{\Omega}_0|=1$ and $\rho_0 >0$.

Eq. (\ref{eq:1}) is the mass conservation (or continuity) equation while (\ref{eq:2}) is the momentum balance equation. We note that the model is similar to the isentropic compressible Euler model of  gas dynamics but for two important differences. First, the velocity $\boldsymbol{\Omega}$ is constrained to have unit norm, i.e. $\boldsymbol{\Omega} \in {\mathbb S}^1$, where ${\mathbb S}^1$ is the one-dimensional sphere. As a consequence, $\boldsymbol{\Omega}_t +c(\boldsymbol{\Omega}\cdot
\nabla_{\boldsymbol{x}} \boldsymbol{\Omega})$, which is a derivative of $\boldsymbol{\Omega}$, is orthogonal to $\boldsymbol{\Omega}$. This is why the pressure gradient term $\lambda \nabla_{\boldsymbol{x}}(p^\varepsilon(\rho))$ is multiplied by the projection operator $\left( \text{Id} - \boldsymbol{\Omega} \otimes \boldsymbol{\Omega} \right)$. In this way, the constraint (\ref{eq:22}) is satisfied at all times provided it is satisfied at initial time. This constraint expresses that $\boldsymbol{\Omega}$ should be seen as the velocity direction, or polarization vector (see \cite{Baskaran_Marchetti_PRL08} for another example of an active fluid model written in terms of its polarization vector). The actual velocity is a constant times the velocity direction. After some time rescaling, this constant can be chosen equal to unity (see \cite{Degond_Motsch_M3AS08} for details). 

The second difference is the presence of the constant $c$ in front of the velocity transport term $(\boldsymbol{\Omega} \cdot \nabla_{\boldsymbol{x}}) \boldsymbol{\Omega}$. In the isentropic Euler equations, $c=1$, i.e. the coefficient of the velocity transport term is the same as that of the mass transport term $\nabla_{\boldsymbol{x}} \cdot (\rho \boldsymbol{\Omega})$. In the SOH model, these coefficients are different, and may even be of different sign \cite{Frouvelle_M3AS12}. The difference reflects that, in swarming systems, velocity is not simply transported along with the fluid velocity itself. If $c \leq 1$, velocity propagates upstream the fluid. This situation occurs in car traffic, where drivers update their velocity according to the state of the flow in front of them. One may also find $c>1$, which occurs when particles update their velocity according to the flow behind them \cite{Frouvelle_M3AS12}. This occurs e.g. in locust swarms because locusts want to prevent themselves from being bitten by their congeners behind them. These considerations illustrate the loss of Galilean invariance of the model when $c \not = 1$ and the consecutive anisotropy of wave propagation described by these models. A discussion of the lack of Galilean invariance in flocking models and its consequences on the propagation of sound waves can be found in \cite{Tu_etal_PRL98}. 

In the SOH model derived in \cite{Degond_Motsch_M3AS08}, the pressure is just a linear function of the density. However, short-range repulsion can be included by introducing a non-linear pressure which blows up at the congestion density $\rho^*$. This non-linear pressure relation has been derived from a microscopic dynamics involving finite-sized particles undergoing short-range repulsion in \cite{Degond_etal_JSP10}. The congestion density $\rho^*$ corresponds to the jammed state where particles are in contact to each other. Additionally, we assume that the intensity of the repulsion force is very small unless the density $\rho$ is very close to $\rho^*$. The parameter $\varepsilon$ characterizes the width of the density range where the pressure is finite:  $p^\varepsilon(\rho)$ is supposed almost zero except in small neighborhood of $\rho^*$ and becomes infinite at exactly $\rho = \rho^*$. This sudden 'turn-on' of the repulsion force results in a sharp phase transition between uncongested and congested regions as shown below. 

\begin{figure}[htbp]
  \centering
\subfigure[$p(\rho)$ and $p^\varepsilon(\rho)$]{\label{Fig:15}  \includegraphics[scale=0.3]{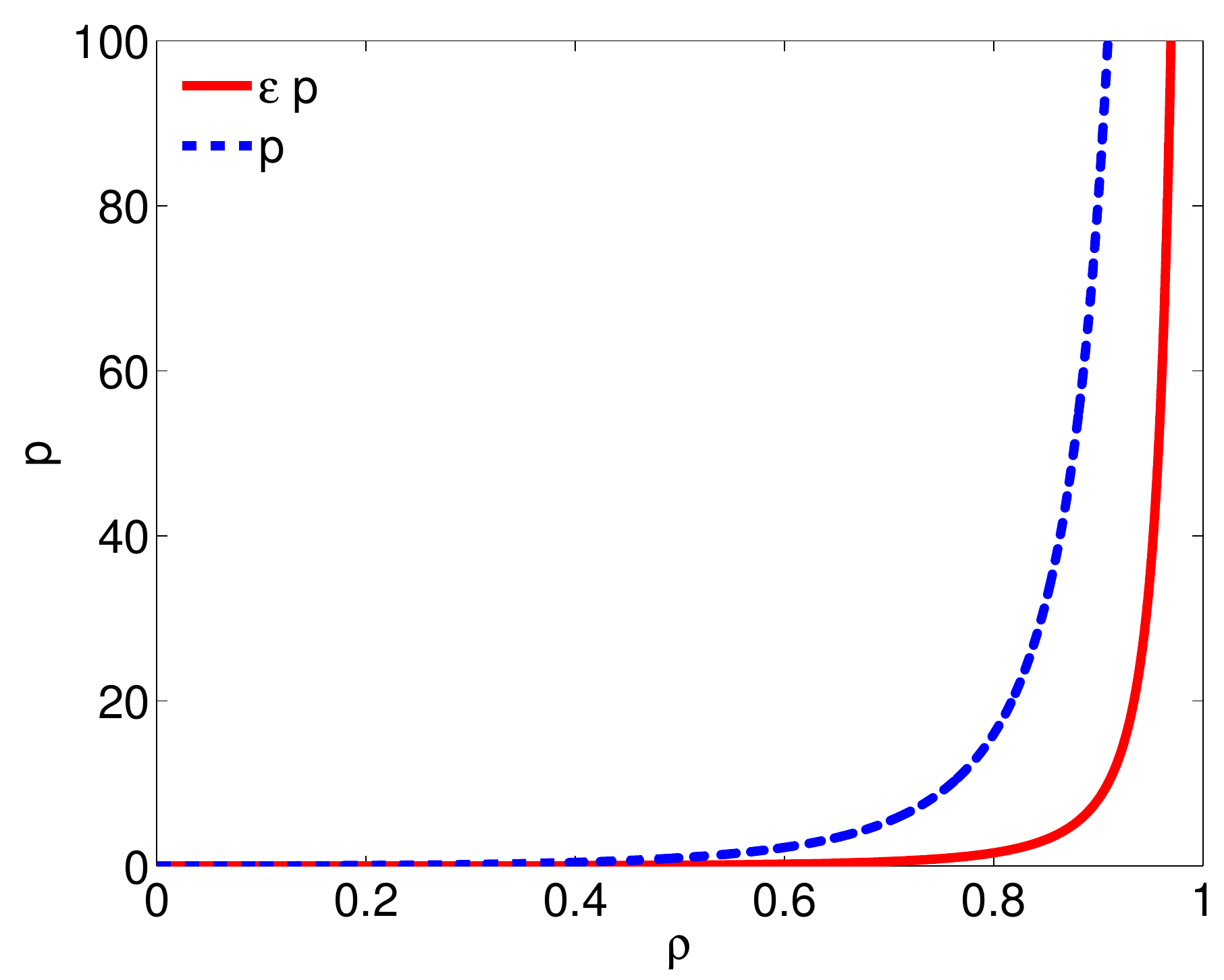}}
 \hspace{1cm}
\subfigure[$p_0^\varepsilon(\rho)$ and $p_1^\varepsilon(\rho)$]{\label{Fig:p0p1r} \includegraphics[scale=0.385]{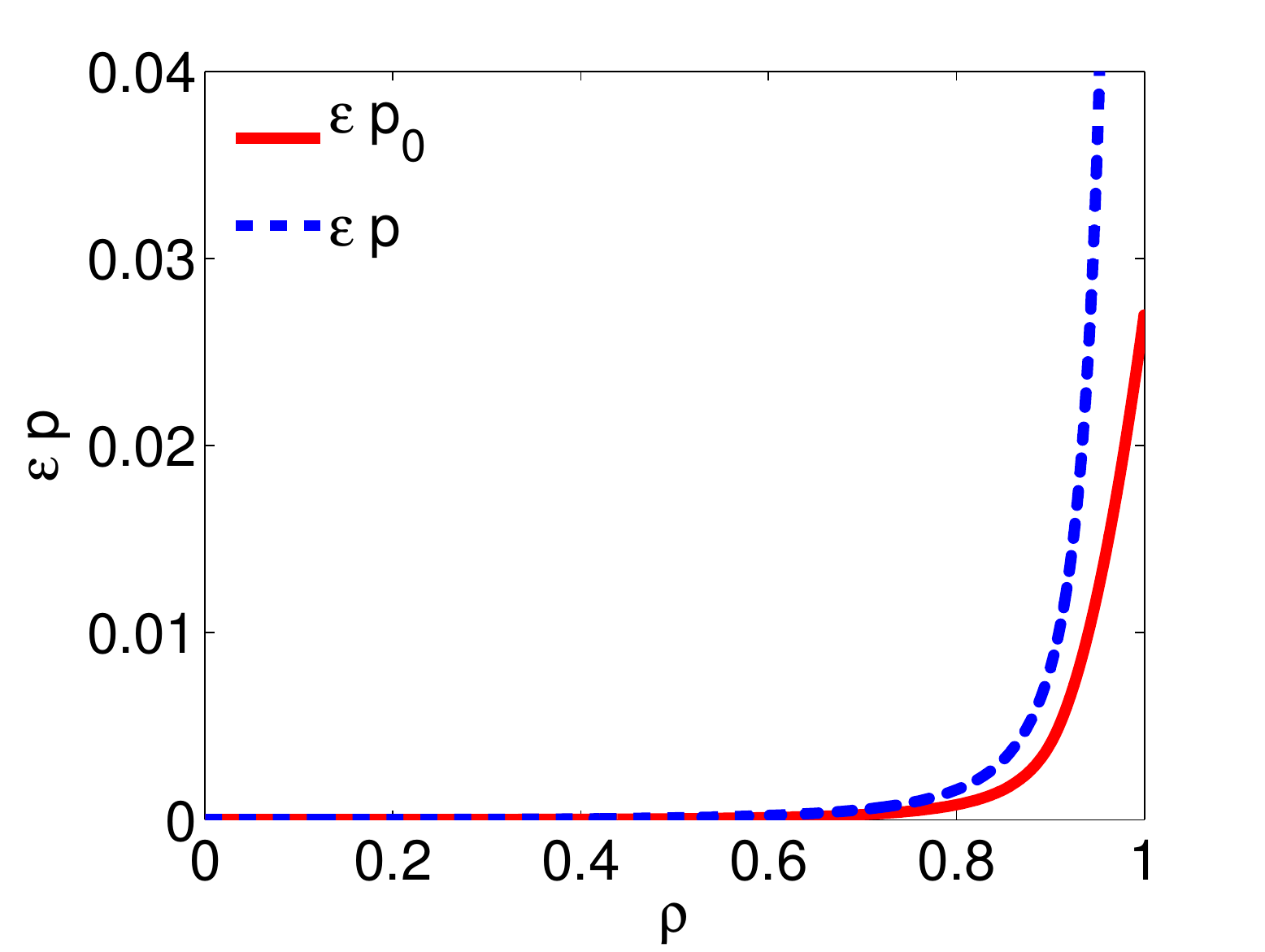}}
  \caption{Left: the pressure $p$ and the rescaled pressure $p^\varepsilon$ as functions of the density $\rho$. Right: the two components $p_0^\varepsilon(\rho)$ and $p_1^\varepsilon(\rho)$ as functions of the density $\rho$ (see section \ref{subsec_time_semi}). }
  \label{Fig:p0p1}
\end{figure}

These modeling hypotheses are translated in the following way: $p^\varepsilon(\rho)$ is defined by
\begin{gather}
\label{eq:20}
p^\varepsilon(\rho)=\varepsilon p(\rho),
\end{gather}
where $p(\rho)$ is such that $p(\rho) \to \infty$, when $\rho \to \rho^*$. To be more specific, we consider:
\begin{equation}
p(\rho) = \frac{1}{\left(\frac{1}{\rho} -
    \frac{1}{\rho^*}\right)^{\gamma}}, \quad \gamma>0,\label{eq:21}
\end{equation}
i.e. $p$ behaves like usual isentropic fluid pressure at low density: $p(\rho) \sim \rho^{\gamma}$ when $\rho \to 0$. From (\ref{eq:20}), (\ref{eq:21}), it follows that  $p^\varepsilon(\rho) = O(\varepsilon)$ for $\rho \ll \rho^*$ while  $p^\varepsilon(\rho) = O(1)$ for $\rho$ close to  $\rho^*$ (more precisely for $\rho = \rho^*-{\mathcal O}(\varepsilon^{1/\gamma})$). Fig.~\ref{Fig:15} displays typical functions $p$ and $p^\varepsilon$.

Solutions of (\ref{eq:1})-(\ref{eq:22}) depend on $\varepsilon$ and are denoted by $(\rho^\varepsilon, \boldsymbol{\Omega}^\varepsilon, \boldsymbol{q}^\varepsilon)$. Their formal limits as $\varepsilon\to 0$ are supposed to exist and are denoted by $(\rho, \boldsymbol{\Omega}, \boldsymbol{q})$:
$$ (\rho^\varepsilon, \boldsymbol{\Omega}^\varepsilon, \boldsymbol{q}^\varepsilon) \to 
(\rho, \boldsymbol{\Omega}, \boldsymbol{q}) \quad \mbox{ as } \quad \varepsilon \to 0.  $$
When $\varepsilon\to 0$, the singular pressure term $p^\varepsilon(\rho^\varepsilon)$ tends to different values
according to whether the limit satisfies $\rho < \rho^*$ (uncongested region) or $\rho = \rho^*$ (congested region). We denote by ${\mathcal C}(t)$ and ${\mathcal U}(t)$ the congested and uncongested regions respectively and by ${\mathcal I}(t)$ the interface between these two regions:
\begin{eqnarray*}
&&\hspace{-1cm}  
{\mathcal U}(t) = \{ x \in {\mathbb R}^2, \quad \rho(x,t) < \rho^* \}, \\
&&\hspace{-1cm}  
{\mathcal C}(t) = \{ x \in {\mathbb R}^2, \quad \rho(x,t) = \rho^* \}, \\ 
&&\hspace{-1cm}  
{\mathcal I}(t) = \partial {\mathcal U}(t) = \partial {\mathcal C}(t). 
\end{eqnarray*}

\paragraph{Uncongested region:} We consider a point $x \in {\mathcal U}(t)$, i.e. such that  $\rho(x,t) < \rho^*$. In this case, $p(\rho^\varepsilon) \to p( \rho) < \infty$ and consequently $p^\varepsilon(\rho^\varepsilon) = \varepsilon p(\rho^\varepsilon) \to 0$. Therefore, the formal limit is the presureless gas dynamics with
norm-one constraint on the velocity: in ${\mathcal U}(t)$, $(\rho,\boldsymbol{\Omega}, \boldsymbol{q})$ satisfies:  
\begin{align}
&\rho_t + \nabla_{\boldsymbol{x}}\cdot \boldsymbol{q} = 0,\qquad \boldsymbol{q}  = \rho \boldsymbol{\Omega},  \label{eq_mass_lim_uncon}\\
&\rho\left(\boldsymbol{\Omega}_t +c(\boldsymbol{\Omega}\cdot
\nabla_{\boldsymbol{x}})\boldsymbol{\Omega} \right)  =
0,   \label{eq_mom_lim_incon} \\
&|\boldsymbol{\Omega}|=1.   \label{eq_norm_lim_uncon}
\end{align}

\begin{figure}
\centering
\subfigure[Congestion region]{\label{Fig:lines_1}  
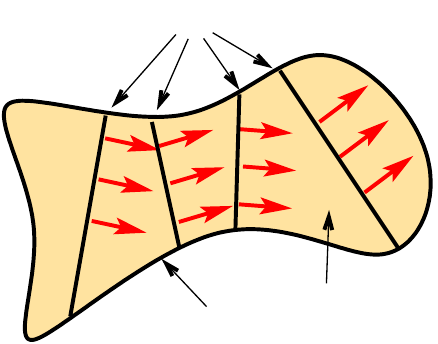}
\hspace{2cm}
\subfigure[Singularity of $\Omega$]{\label{Fig:lines_2} 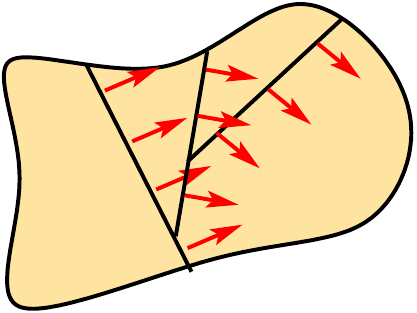}
\caption{The congestion region ${\mathcal C}(t)$ at a given time $t$ is enclosed by the curve ${\mathcal I}(t)$. Left: the vector field $\Omega(x,t)$ is constant along lines normal to itself. These lines form the family ${\mathcal N}(t)$. The quantity $\bar p(x,t)$ satisfies one-dimensional elliptic equations posed along these lines. It requires boundary conditions along the interface ${\mathcal I}(t)$. Right: a situation where the vector field $\Omega(x,t)$ has singularities}
\label{Fig:lines}
\end{figure}

\paragraph{Congested region:} We now consider a point $x \in {\mathcal C}(t)$, i.e. such that $\rho = \rho^*$. In this case, $p(\rho^\varepsilon) \to \infty$ and  $\lim_{\varepsilon \to 0} (\varepsilon p(\rho^\varepsilon))$ is undetermined. We assume that 
\begin{equation}
\lim_{\varepsilon\to 0}
\varepsilon p(\rho^{\varepsilon}) = \bar p \quad \mbox{ with } 0 \leq \bar p < \infty \quad \mbox{ almost everywhere.}
\label{eq_def_bar_p}
\end{equation}
Now, $\bar p$ becomes an additional unknown of the limit system:  in ${\mathcal C}(t)$, $(\rho,\boldsymbol{\Omega}, \boldsymbol{q}, \bar p)$ satisfies: 
\begin{align}
& \rho = \rho^*, \quad \quad \boldsymbol{q} = \rho^* \boldsymbol{\Omega}, \label{eq_mass_lim_con0} \\
& \nabla_{\boldsymbol{x}}\cdot \boldsymbol{\Omega} = 0,  \label{eq_mass_lim_con} \\
&\rho^* \left( \boldsymbol{\Omega}_t +c(\boldsymbol{\Omega}\cdot
\nabla_{\boldsymbol{x}})\boldsymbol{\Omega} \right) + \lambda
\left(\text{Id}-\boldsymbol{\Omega}\otimes\boldsymbol{
    \Omega}\right)\nabla_{\boldsymbol{x}}\bar{p} = 0,  \label{eq_mom_lim_con} \\
&|\boldsymbol{\Omega}|=1.  \label{eq_norm_lim_con}
\end{align}
The vector field $\boldsymbol{\Omega}$ is divergence-free and of unit-norm. Smooth vector fields like $\boldsymbol{\Omega}$ have a remarkable geometry (see Fig. \ref{Fig:lines_1} and \cite{Degond_etal_JSP10}): curves normal to $\boldsymbol{\Omega}$ are straight lines and $\boldsymbol{\Omega}$ is constant along these lines. We denote by ${\mathcal N}$ the family of straight lines normal to $\Omega$. The quantity $\bar p$ is now an unknown like the pressure in standard incompressible models. The governing equation for $\bar{p}$ is obtained by taking the divergence of (\ref{eq_mom_lim_con}) and using (\ref{eq_mass_lim_con}) to eliminate the time derivative. It is written for $x \in {\mathcal C}(t)$ : 
\begin{equation} 
- \lambda \nabla_{\boldsymbol{x}} \cdot \left( \left(\text{Id}-\boldsymbol{\Omega}\otimes\boldsymbol{
\Omega} \right) \nabla_{\boldsymbol{x}}\bar{p} \right) = c \rho^* \nabla_{\boldsymbol{x}} \left( (\boldsymbol{\Omega}\cdot
\nabla_{\boldsymbol{x}})\boldsymbol{\Omega} \right) . 
\label{eq_barp_elliptic}
\end{equation}
This leads to a family of one-dimensional elliptic equations for $\bar p$ posed along the lines of the family ${\mathcal N}$. Their inversion requires the knowledge of the boundary values of $\bar p$ on ${\mathcal I}(t)$. These boundary values, as well as the velocity of the interface ${\mathcal I}(t)$ have been determined in \cite{Degond_etal_JSP10}. They are linked to the jumps of $\rho$ and $\boldsymbol{\Omega}$ across ${\mathcal I}(t)$ and are found by solving a Riemann problem in the direction normal to ${\mathcal I}(t)$, supposing ${\mathcal I}(t)$ is smooth. However, if  ${\mathcal I}(t)$ is not smooth, which happens for instance at times when its topology changes, these quantities cannot be analytically known from \cite{Degond_etal_JSP10}. Since we target a general situation where topology changes can occur, we will not use the conditions of \cite{Degond_etal_JSP10} and do not need to recall their expression. 

The SOH model (\ref{eq:1})-(\ref{eq:22}) is hyperbolic, with characteristic speeds given by: 
\begin{equation} 
\xi_\pm = \cos \theta + \frac{1}{2} \left( (c-1) \cos \theta \pm \sqrt{ (c-1)^2 \cos^2 \theta + 4 \lambda \varepsilon p'(\rho^\varepsilon) \, \sin^2 \theta} \right) ,
\label{eq_char_speeds}
\end{equation}
where $\theta$ is the angle between $\boldsymbol{\Omega}$ and the propagation direction. If $\rho^\varepsilon \to \rho^*$ such that (\ref{eq_def_bar_p}) holds, then, using (\ref{eq:21}), we get 
\begin{equation} 
\varepsilon p'(\rho^\varepsilon) = {\mathcal O}(\varepsilon^{-1/\gamma}) \to \infty. \label{eq_lim_eps_p'}
\end{equation} 
Therefore, the characteristic speeds (\ref{eq_char_speeds}) tend to $\pm \infty$. This explains why the limit problem (\ref{eq_mass_lim_con})-(\ref{eq_mom_lim_con}) has elliptic characteristics, reflected in the elliptic equation (\ref{eq_barp_elliptic}) for $\bar p$. The second term of (\ref{eq_char_speeds}) is the sound speed 
$$ c_s = \frac{1}{2} \left( (c-1) \cos \theta \pm \sqrt{ (c-1)^2 \cos^2 \theta + 4 \lambda \varepsilon p'(\rho^\varepsilon) \, \sin^2 \theta} \right), $$ 
while the first term is the fluid velocity projected on the propagation direction $ u = \cos \theta$. We can define the Mach number as the ratio of these two velocities 
$${\mathcal M} = \frac{u}{c_s} = \frac{\cos \theta}{\frac{1}{2} \left( (c-1) \cos \theta \pm \sqrt{ (c-1)^2 \cos^2 \theta + 4 \lambda \varepsilon p'(\rho^\varepsilon) \, \sin^2 \theta} \right)}. $$
Due to (\ref{eq_lim_eps_p'}), we generically have ${\mathcal M} \to 0$ when $\varepsilon \to 0$. Thus, in the congested region, the $\varepsilon \to 0$ limit is a small Mach-number limit. We note that in the SOH model, the sound speed is anisotropic and depends on the angle between $\boldsymbol{\Omega}$ and the propagation direction. This has already been noticed in \cite{Tu_etal_PRL98} for other hydrodynamic models of flocks. 

The uncongested and congested models can be unified by writing the limit system in the following way: for $x \in {\mathbb R}^2$, 
  \begin{align}
&\rho_t + \nabla_{\boldsymbol{x}}\cdot \boldsymbol{q} = 0,\qquad \boldsymbol{q}  = \rho \boldsymbol{\Omega},  \label{eq_mass_lim}\\
&\rho\left(\boldsymbol{\Omega}_t +c(\boldsymbol{\Omega}\cdot
\nabla_{\boldsymbol{x}})\boldsymbol{\Omega}\right) +
\lambda\left(\text{Id}-\boldsymbol{\Omega}\otimes\boldsymbol{
    \Omega}\right)\nabla_{\boldsymbol{x}}\bar{p} = 0,  \label{eq_mom_lim} \\
&|\boldsymbol{\Omega}|=1, \label{eq_geom_lim}\\
&(\rho-\rho^*)\bar{p}=0. \label{eq_const_lim} 
\end{align}
However, this expression does not specify what happens when the constraint (\ref{eq_const_lim}) shifts from $\bar p = 0$ (in ${\mathcal U}(t)$) to $\rho = \rho^*$ (in ${\mathcal C}(t)$) at the interface ${\mathcal I}(t)$. Therefore, it must be complemented by interface conditions like those of \cite{Degond_etal_JSP10}.

\begin{remark}
The pressureless gas dynamics model is not strictly hyperbolic. This results in some unpleasant features such as the formation of singularities or the generation of weak instabilities \cite{Bouchut_Birkhauser94}. According to the physical context, it is possible that a non-zero pressure prevails in the uncongested region ${\mathcal U}$. This is easily modeled by considering a pressure $p^\varepsilon(\rho)$ of the form 
\begin{equation}
p^\varepsilon(\rho) = \varepsilon p(\rho) + p_B(\rho), 
\label{eq_background_pressure}
\end{equation}
where $p$ is given by (\ref{eq:21}) as previously and the background pressure $p_B(\rho)$ is independent of $\varepsilon$ and has finite limit $p_B(\rho^*)$ when $\rho \to \rho^*$. For instance, we can take 
\begin{equation}
p_B(\rho) = \kappa \rho^\gamma, 
\label{eq_background_pressure_2}
\end{equation}
with the parameter $\kappa$ controlling the strength of the background pressure. In this case, the  limit problem (\ref{eq_mass_lim_uncon})-(\ref{eq_norm_lim_uncon}) in the uncongested region ${\mathcal U}$ is changed to the following form : 
\begin{align}
&\rho_t + \nabla_{\boldsymbol{x}}\cdot \boldsymbol{q} = 0,\qquad \boldsymbol{q}  = \rho \boldsymbol{\Omega}, \nonumber\\
&\rho\left(\boldsymbol{\Omega}_t +c(\boldsymbol{\Omega}\cdot
\nabla_{\boldsymbol{x}})\boldsymbol{\Omega} \right)  + \lambda\left(\text{Id}-\boldsymbol{\Omega}\otimes\boldsymbol{
    \Omega}\right)\nabla_{\boldsymbol{x}} p_B(\rho) =
0,  \nonumber \\
&|\boldsymbol{\Omega}|=1,  \nonumber
\end{align}
which is nothing but the SOH model with pressure $p_B$.
In the congested region ${\mathcal C}$, there is no change to system (\ref{eq_mass_lim_con0})-(\ref{eq_norm_lim_con}) except that now the pressure $\bar p$ satisfies the inequality $p_B(\rho^*) \leq \bar p < \infty$ instead of (\ref{eq_def_bar_p}). 
\label{rem_background_pressure}
\end{remark}

The numerical resolution of problem (\ref{eq_mass_lim})-(\ref{eq_const_lim}) is quite challenging since the interface ${\mathcal I}(t)$ is a moving free boundary. It requires two different fluid solvers: a compressible one in the congested region ${\mathcal C}(t)$ and an incompressible one in the uncongested region ${\mathcal U}(t)$. The numerical discretization of the interface ${\mathcal I}(t)$ must be matched with the mesh used inside each of the domains ${\mathcal C}(t)$ and ${\mathcal U}(t)$. Since ${\mathcal I}(t)$ is a moving interface, some remeshing must be performed at each time step. Alternate methodologies such as level-sets may be envisionned. In all cases, these methods are computationally intensive and necessitate fine parameter tuning. Additionally, interface motion is induced by fluid motion, which itself is determined by the boundary conditions for $\bar p$ along the interface. However, in the case where ${\mathcal I}(t)$ changes topology, the formal asymptotic procedure of \cite{Degond_etal_JSP10} does not provide any information about these boundary conditions. In this circumstance, it is not known what interface motion takes place. Then, there is no other possibility than solving the original perturbation problem (\ref{eq:1})-(\ref{eq:22}), with a very small value of $\varepsilon$. This is the strategy developed in the present work. 

To solve the perturbation problem (\ref{eq:1})-(\ref{eq:22}) with an arbitrarily small value of $\varepsilon$, an Asymptotic-Preserving (AP) scheme is necessary. Indeed, the limit $\varepsilon \to 0$ corresponds to a small Mach-number limit. With a classical scheme (such as a shock-capturing scheme), the CFL stability condition leads to a time-step constraint $\Delta t \leq C_\varepsilon \Delta x$ with $C_\varepsilon \to 0$ as $\varepsilon \to 0$. Therefore, a classical scheme is impracticable for small values of $\varepsilon$. The AP property recalled in Figure \ref{fig:ap} guarantees that, for small values of $\varepsilon$, the scheme provides a consistent solution to the limit system (\ref{eq_mass_lim})-(\ref{eq_const_lim}). 

In this work, we adopt the AP-method of \cite{Cordier_etal_JCP12, Degond_Tang_CICP11, Tang_KRM12} which has already been used for the Euler system with congestion in \cite{Degond_etal_JCP11}. However, the SOH model is a non-conservative model. In general, non-conservative models may have multiple weak solutions of the shock-wave type. Additional information is necessary to single out the correct shocks: a suitable path in state space connecting the two states on each side of the shock must be defined \cite{LeFloch_IMA89}. The choice of the correct path depends on some microscopic information which is not included in the macroscopic model and which, in general, is not available. Here, we look for solutions of the SOH model which provide good approximations of the underlying microscopic Vicsek model \cite{Degond_Motsch_M3AS08}. Such solutions are considered as being the physically consistant solutions. In \cite{Motsch_Navoret_MMS11}, it is shown that these solutions are well-captured by a relaxation model. It consists of a gas-dynamics type conservative system for the density and momentum supplemented by a relaxation right-hand side which drives the velocity towards a vector of unit norm. In the present work, we rely on this relaxation model as the one which produces the physically consistant solution and propose a modification of the method of \cite{Motsch_Navoret_MMS11} which makes it AP in the small Mach-number limit. This modification is based on \cite{Degond_etal_JCP11, Degond_Tang_CICP11}. This numerical method is presented in the next section.

\setcounter{equation}{0}
\section{The numerical method}
\label{sec_numerical_method}

\subsection{Relaxation method and splitting}
\label{subsec_relax}

We first present the relaxation method of \cite{Motsch_Navoret_MMS11}. We consider the following relaxation system: 
\begin{gather}
\partial_{t}\rho^{\varepsilon,\beta} + \nabla_{\boldsymbol{x}}\cdot \boldsymbol{q}^{\varepsilon,\beta} = 0,\label{Eq:Euler_rho_relaxation}\\
\partial_{t}\boldsymbol{q}^{\varepsilon,\beta} + c\nabla_{\boldsymbol{x}}\cdot\left(\frac{\boldsymbol{q}^{\varepsilon,\beta}\otimes \boldsymbol{q}^{\varepsilon,\beta}}{\rho^{\varepsilon,\beta}}\right) + \lambda\nabla_{\boldsymbol{x}} \left(p^\varepsilon(\rho^{\varepsilon,\beta})\right) =  \frac{1}{\beta}(1-|\frac{\boldsymbol{q}^{\varepsilon,\beta}}{\rho^{\varepsilon,\beta}}|^2)\boldsymbol{q}^{\varepsilon,\beta},\label{Eq:Euler_q_relaxation}
\end{gather}
where the geometric constraint $|\boldsymbol{\Omega}|=1$ is removed. System
(\ref{Eq:Euler_rho_relaxation}), (\ref{Eq:Euler_q_relaxation}) is a conservative  model with a relaxation term. We have:

\begin{proposition} \cite{Motsch_Navoret_MMS11}
The relaxation model \eqref{Eq:Euler_rho_relaxation}-\eqref{Eq:Euler_q_relaxation} formally converges to the SOH model
\eqref{eq:1}-\eqref{eq:22} as $\beta \to 0$.
\end{proposition}

\noindent
The left-hand side of (\ref{Eq:Euler_rho_relaxation}), (\ref{Eq:Euler_q_relaxation}) coincides with the isentropic Euler system if and only if $c=1$. If $c \not = 1$, the characteristic speeds are given by 
\begin{gather}\label{eq:6}
  c\frac{q}{\rho},\quad c\frac{q}{\rho}\pm \sqrt{(c^2-c)\frac{q^2}{\rho^2}+
   \lambda \varepsilon p'(\rho)}.
\end{gather}
where $q$ is the component of the momentum along the propagation direction. These characteristic speeds are real and distinct if $c\geq 1$ (provided that $\rho>0$) and may be complex if $c<1$ for small values of $\varepsilon p'(\rho)$. Therefore, although the limit system (the SOH model) is always hyperbolic, the relaxation system (\ref{Eq:Euler_rho_relaxation}), (\ref{Eq:Euler_q_relaxation}) is strictly hyperbolic only for $c \geq 1$. We will see that, in spite of being non-hyperbolic, the relaxation system leads to a valid numerical solution of the SOH model in the relaxation limit when $c<1$. 

\begin{remark}
In \cite{Motsch_Navoret_MMS11}, a linear pressure was used and the relaxation system could stay hyperbolic even with $c<1$ for suitable choices of the parameter $\lambda$. Here, the non-linear pressure and the smallness of the parameter $\varepsilon$ require  $c\geq 1$ to guarantee hyperbolicity of the relaxation system for all values of $\rho >0$. 
\label{rem:hyperbolicity_MN}
\end{remark} 

The plan is therefore to discretize the relaxation system (\ref{Eq:Euler_rho_relaxation}), (\ref{Eq:Euler_q_relaxation}) with arbitrarily small values of $\beta$. To this aim, we use the
splitting method of \cite{Motsch_Navoret_MMS11}. Let $\Delta t$ be the time step and $(\rho^n, \boldsymbol{\Omega}^n, \boldsymbol{q}^n = \rho^n \boldsymbol{\Omega}^n) $ be an approximation of $(\rho, \boldsymbol{\Omega}, \boldsymbol{q}) $ at time $t^n = n \Delta t$. To pass from $(\rho^n, \boldsymbol{\Omega}^n, \boldsymbol{q}^n) $ to $(\rho^{n+1}, \boldsymbol{\Omega}^{n+1}, \boldsymbol{q}^{n+1}) $, we first solve the following
conservative equation with singular pressure $p^\varepsilon$: 
\begin{gather}
\partial_{t}\rho + \nabla_{\boldsymbol{x}}\cdot \boldsymbol{q} = 0,\label{Eq:Euler_rho_eps}\\
\partial_{t}\boldsymbol{q} + c \nabla_{\boldsymbol{x}}\cdot\left(\frac{\boldsymbol{q}\otimes \boldsymbol{q}}{\rho}\right) + \lambda \nabla_{\boldsymbol{x}} \left(p^\varepsilon(\rho)\right) = 0, \label{Eq:Euler_q_eps}
\end{gather}   
over the time-step $\Delta t$ with the AP-scheme developed in \cite{Degond_etal_JCP11} and initial conditions $(\rho^n, \boldsymbol{q}^n) $. This leads to intermediate values $(\rho^{n+1/2}, \boldsymbol{q}^{n+1/2}) $. Then, the relaxation system:
\begin{align}
&\rho_t  = 0, \label{eq_relax_rho}\\
&\boldsymbol{q}_t  = \frac{1}{\beta} \left( 1- \left|\frac{\boldsymbol{q}}{\rho} \right|^2 \right) \boldsymbol{q}, \label{eq_relax}
\end{align}
is solved over the time-step $\Delta t$ with initial condition  $(\rho^{n+1/2}, \boldsymbol{q}^{n+1/2})$ and yields $(\rho^{n+1}, \boldsymbol{q}^{n+1}) $. Finally, we set $\boldsymbol{\Omega}^{n+1} = \boldsymbol{q}^{n+1}/\rho^{n+1}$.

\subsection{Time semi-discrete scheme}
\label{subsec_time_semi}

We first focus on the time semi-discretization of the conservative step (\ref{Eq:Euler_rho_eps}), (\ref{Eq:Euler_q_eps}). Following \cite{Degond_etal_JCP11}, it is written as follows:
\begin{align}
&\frac{\rho^{n+1/2}-\rho^n}{\Delta t} + \nabla_{\boldsymbol{x}}\cdot \boldsymbol{q^{n+1/2}} = 0,\label{eq:4}\\
&\frac{\boldsymbol{q}^{n+1/2}-\boldsymbol{q}^{n}}{\Delta t} +c
\nabla_{\boldsymbol{x}}\cdot\left(\frac{\boldsymbol{q}^{n}\otimes
    \boldsymbol{q}^{n} }{\rho^{n}}\right) +  \lambda\nabla_{\boldsymbol{x}}
(p^\varepsilon_0(\rho^{n})) +\lambda\nabla_{\boldsymbol{x}} (
p_1^\varepsilon(\rho^{n+1/2})) = 0\label{eq:5}.
\end{align}
Here, we split the pressure $p^\varepsilon$ (see Fig. \ref{Fig:p0p1r}) into 
\begin{equation} 
p^\varepsilon = p_0^\varepsilon + p_1^\varepsilon, 
\label{eq_p_decomp}
\end{equation}
with 
\begin{gather}
       p_0^\varepsilon(\rho)=
  \begin{cases}
    \cfrac{1}{2} \varepsilon p(\rho),& \text{ if } \rho \leq \rho^*-\delta,\\[3ex]
    \begin{split}
     \varepsilon \frac{1}{2}\Big\{ &p( \rho^*-\delta)+p'( \rho^*-\delta)(\rho-
      \rho^*+\delta)\\
      &+\frac{1}{2}p''(\rho^*-\delta)(\rho-
      \rho^*+\delta)^2 \Big\}
    \end{split}
,& \text{ if } \rho>  \rho^*-\delta,
  \end{cases} \label{eq_p_decomp_1}
\end{gather}
and $\delta=\varepsilon^{\frac{1}{\gamma+2}}$.
We treat the $p_0^\varepsilon$ component of the pressure force explicitly and the $p_1^\varepsilon$ component implicitly. The implicit treatment of the pressure force in (\ref{eq:5}) as well as that of the mass flux in (\ref{eq:4}) provides the AP property in the small Mach-number limit \cite{Degond_Tang_CICP11}. However, if the full pressure force $p^\varepsilon$ is treated implicitly, some instabilities may appear \cite{Degond_Tang_CICP11}. Keeping the $p_0^\varepsilon$ part explicit removes these instabilities \cite{Degond_etal_JCP11}. The explicit part $p_0^\varepsilon$ is chosen to be $p^\varepsilon/2$ except and in a neighborhood $[\rho^*-\delta,\rho^*]$ of $\rho^*$ with $\delta$ adequately chosen. In this interval $p_0^\varepsilon$ is a quadratic function. Matching of $p_0^\varepsilon$ and of its derivatives up to second order is ensured at $\rho^* - \delta$. The choice $\delta = \varepsilon^{\frac{1}{\gamma+2}}$ guarantees that $p_0^{\varepsilon ''}(\rho^*-\delta)$ remains finite and consequently that  $p_0^\varepsilon$ itself remains finite over the whole range $[0,\rho^*]$ uniformly as $\varepsilon \to 0$. 

\begin{remark}
When a background pressure $p_B(\rho)$ prevails (see Remark \ref{rem_background_pressure}), the splitting (\ref{eq_p_decomp}) is replaced by (\ref{eq_background_pressure}) i.e. we let $p_0^\varepsilon(\rho) = p_B(\rho)$ and $p_1^\varepsilon(\rho) = \varepsilon p(\rho)$. 
\label{rem_background_pressure_splitting}
\end{remark}

Now, following \cite{Degond_etal_JCP11, Degond_Tang_CICP11}, we insert (\ref{eq:5}) into (\ref{eq:4}) to get an elliptic equation for $\rho^{n+1/2}$. 
\begin{gather}
  \begin{split}
    &\frac{\rho^{{n+1/2}} - \rho^{{n}}}{\Delta t}-\Delta t 
   \lambda \Delta_{\boldsymbol{x}}(p^\varepsilon_1(\rho^{{n+1/2}}))\\
    =&- \nabla_{\boldsymbol{x}}\cdot \boldsymbol{q}^{n} +\Delta t
    c\nabla_{\boldsymbol{x}}^2:\left(\frac{\boldsymbol{q}^{n}\otimes
        \boldsymbol{q}^n }{\rho^{n}}\right) + \Delta t\lambda \Delta_{\boldsymbol{x}}\left(p^\varepsilon_0(\rho^{{n}})\right).\label{eq:3}
  \end{split}
\end{gather}
Instead of solving (\ref{eq:3}) for $\rho^{n+1/2}$, we solve it for $p_1^{n+1/2}$ and then, deduce $\rho^{n+1/2}$ by inverting the relation $p_1=p^\varepsilon_1(\rho)$, which is possible thanks to the monotonicity of $p^\varepsilon_1$. This procedure is more stable and satisfies the congestion constraint $\rho^{n+1} \leq \rho^*$  automatically. Once $\rho^{n+1/2}$ is known, $\boldsymbol{q}^{n+1/2}$ is computed by solving (\ref{eq:5}).

We now consider the second (relaxation) step (\ref{eq_relax_rho}), (\ref{eq_relax}). We first note from (\ref{eq_relax_rho}) that the density is unchanged: $\rho^{n+1} = \rho^{n+1/2}$. Then,  $|\boldsymbol{\Omega}(t)|^2 = |\boldsymbol{q}(t)|^2 /  |\rho(t)|^2$ is a solution of:
\begin{gather*}
  \frac{1}{2}\partial_t |\boldsymbol{\Omega}|^2=\frac{1}{\beta}(1-|\boldsymbol{\Omega}|^2)|\boldsymbol{\Omega}|^2,
\end{gather*}
which can be solved analytically. We easily deduce that:
\begin{equation}
\boldsymbol{\Omega}^{n+1} = \left(\left| \boldsymbol{\Omega}^{n+1/2} \right|^2 + \left( 1- \left| \boldsymbol{\Omega}^{n+1/2} \right|^2 \right) \,  e^{-\frac{2}{\beta} \, \Delta t} \right)^{-\frac{1}{2}} \, \boldsymbol{\Omega}^{n+1/2}. 
\label{eq_relax_step}
\end{equation}
When $\beta\to 0$, \eqref{eq_relax_step} leads to a simple renormalization of the velocity field:
\begin{gather*}
\boldsymbol{\Omega}^{n+1}= \frac{\boldsymbol{\Omega}^{n+1/2}}{\left| \boldsymbol{\Omega}^{n+1/2} \right|} .
\end{gather*}

\subsection{Fully discrete scheme}
\label{subsec_fully_discrete}

The full space-time discretization is presented in the 1D case. For the sake of completeness, the 2D discretization is given in Appendix  1. Since the second step does not involve any spatial derivative, we focus on the first compressible step. For simplicity, we consider the domain $[0,1]$ and a uniform spatial mesh of step $\Delta x=\frac{1}{M}$, where $M$ is a positive
integer. We denote by $U_j^n=(\rho_j^{n},q_j^{n})^T$ the approximations of
$U=(\rho,q)^T$ at time $t^n=n\Delta t$ and positions $x_j=j\Delta x$, for $j = 0,1,\ldots,M$. We
discretize \eqref{eq:4}, \eqref{eq:5} with a semi-implicit version of the
local Lax-Friedrichs (or  Rusanov) method \cite{Leveque_Cambridge2002}  as follows:
\begin{gather}
   \begin{split}
     \frac{\rho^{n+1/2}_{j} - \rho^{n}_{j}}{\Delta t} + \frac{1}{\Delta
       x}\bigg[ & Q_{j+1/2}(U_{j}^{n},U_{j+1}^{n},U_{j}^{n+1/2},U_{j+1}^{n+1/2})
       \\
&-
       Q_{j-1/2}(U_{j-1}^{n},U_{j}^{n},U_{j-1}^{n+1/2},U_{j}^{n+1/2})\bigg]
     = 0,
   \end{split}
\label{Eq:rho_discrete}\\
\begin{split}
  \frac{q^{n+1/2}_{j} - q^{n}_{j}}{\Delta t} &+ \frac{1}{\Delta x}\left[F_{j+1/2}(U_{j}^{n},U_{j+1}^{n}) - F_{j-1/2}(U_{j-1}^{n},U_{j}^{n})\right]\\
  & + \frac{1}{2\Delta x}\left[\lambda
    p_1^\varepsilon(\rho_{j+1}^{n+1/2}) -\lambda p^\varepsilon_1(\rho_{j-1}^{n+1/2})\right] = 0.
\end{split}
\label{Eq:q_discrete}
\end{gather}
where the fluxes are given by:
\begin{align}
&Q_{j+1/2} = \frac{1}{2}\left[q_{j}^{n+1/2} + q_{j+1}^{n+1/2}\right] - \frac{1}{2}C_{j+1/2}^{n}(\rho_{j+1}^{n}  - \rho_{j}^{n}),\label{eq:23}\\
&F_{j+1/2}^n =
\frac{1}{2}\left[c\frac{(q_{j}^{n})^{2}}{\rho_{j}^{n}} +
  c\frac{(q_{j+1}^{n})^{2}}{\rho_{j+1}^{n}}+
  \lambda p^\varepsilon_0(\rho_{j+1}^n)+ \lambda p^\varepsilon_0(\rho_{j}^n)\right]- \frac{1}{2}C_{j+1/2}^{n}(q_{j+1}^{n}  - q_{j}^{n}).\label{eq:24}
\end{align}
The numerical fluxes consist of the sum of centered fluxes and decentering terms introducing diffusion. The latter are evaluated explicitly while some of the former are evaluated implicitly. More precisely, the central part of the mass flux in \eqref{eq:23} and of the $p_1^\varepsilon$ pressure contribution to the momentum flux in (\ref{Eq:q_discrete}) are solved implicitly. By contrast, the central part of the velocity transport flux and of the $p_0^\varepsilon$ pressure flux in the momentum flux (\ref{eq:24}) are kept explicit. This level of implicitness is sufficient to guarantee the AP character of the scheme \cite{Degond_Tang_CICP11} in the small Mach-number limit. The quantity $C_{j+1/2}^{n}$ is the local diffusion
coefficient and is given by:
\begin{eqnarray}
& & \hspace{-1cm} 
C_{j+1/2}^{n} = \max \{ C_j^n, C_{j+1}^n \} , \quad C_j^n = \left|\frac{q_{j}^{n}}{\rho_{j}^{n}}\right|+\sqrt{ (c^2 - c) \left| \frac{q_{j}^{n}}{\rho_{j}^{n}} \right|^2 + \lambda (p^\varepsilon_0)'(\rho_{j}^{n})} . 
\label{eq:C}
\end{eqnarray}
It is defined as the local maximal characteristic
speed related to the explicit pressure $p_0^\varepsilon$. Therefore, it remains
bounded as $\varepsilon$ goes to zero, thanks to the fact that $p_0^\varepsilon$ itself remains uniformly bounded when $\varepsilon \to 0$. By contrast, if a fully explicit scheme was used, $C_j^n$ in (\ref{eq:C}) would involve $p^\varepsilon_0 + p^\varepsilon_1$ and would not remain bounded as $\varepsilon \to 0$. The quantity $C_{j+1/2}^{n}$ provides the numerical viscosity which is needed to ensure the stability of the scheme. 

Based on this  discretization, we can apply the same strategy as
described in section \ref{subsec_time_semi} to get a discrete elliptic equation for $\rho$. We substitute \eqref{Eq:q_discrete} into \eqref{Eq:rho_discrete} and obtain:
\begin{gather*}
  \begin{split}
    \frac{\rho^{n+1/2}_{j} - \rho^{n}_{j}}{\Delta t} &+ \frac{q_{j+1}^{n} - q_{j-1}^{n}}{2\Delta x}- \frac{\Delta t}{4\Delta x^{2}}\lambda\left[p^\varepsilon_1(\rho_{j+2}^{n+1/2}) -2 p^\varepsilon_1(\rho_{j}^{n+1/2}) + p^\varepsilon_1(\rho_{j-2}^{n+1/2})\right]\\
    &- \frac{1}{2\Delta x}\left[C_{j+1/2}(\rho_{j+1}^{n} - \rho_{j}^{n}) - C_{j-1/2}(\rho_{j}^{n} - \rho_{j-1}^{n})\right]\\
    &- \frac{\Delta t}{2\Delta x^2}\left[F_{j+3/2}^{n} - F_{j+1/2}^{n}
      - F_{j-1/2}^{n} + F_{j-3/2}^{n}\right] = 0.
  \end{split}
\end{gather*}
This equation is consistent with the time semi-discrete case \eqref{eq:3}. Then we get a
nonlinear equation for $p_1$:
\begin{gather*}
   \begin{split}
    \rho((p_1)^{n+1/2}_{j})&  - \frac{\Delta t^2}{8\Delta x^{2}}\lambda\left[ (p^\varepsilon_1)_{j+2}^{n+1/2} -2  (p^\varepsilon_1)_{j}^{n+1/2} +  (p^\varepsilon_1)_{j-2}^{n+1/2}\right]\\
   = \ \rho^{n}_{j}& -\frac{\Delta t}{4\Delta x}(q_{j+1}^{n} - q_{j-1}^{n})+\frac{\Delta t}{4\Delta x}\left[C_{j+1/2}(\rho_{j+1}^{n} - \rho_{j}^{n}) - C_{j-1/2}(\rho_{j}^{n} - \rho_{j-1}^{n})\right]\\
    &+ \frac{\Delta t^2}{4\Delta x^2}\left[F_{j+3/2}^{n} - F_{j+1/2}^{n}
      - F_{j-1/2}^{n} + F_{j-3/2}^{n}\right].
  \end{split}
\end{gather*}
We use Newton iterations to solve this
nonlinear equation and  get $p_1^{n+1/2}$. The density $\rho^{n+1/2}$ is
then obtained by inverting the nonlinear function $p_1=p_1^\varepsilon(\rho)$ with
another series of Newton iterations. Once $\rho^{n+1/2}$
is solved, $q^{n+1/2}$ can be obtained by (\ref{Eq:q_discrete}), which yields: 
$$
q^{n+1/2}_{j} = \Phi(U_{j-1}^{n},U_{j}^{n},U_{j+1}^{n})  -\frac{\Delta t}{4\Delta x}\lambda\left[p^\varepsilon_1(\rho_{j+1}^{n+1/2}) - p^\varepsilon_1(\rho_{j-1}^{n+1/2})\right],
$$
with
$$
\Phi(U_{j-1}^{n},U_{j}^{n},U_{j+1}^{n}) = q^{n}_{j} - \frac{\Delta t}{2\Delta x}\left[F_{j+1/2}^n - F_{j-1/2}^n\right].
$$

\begin{remark}
(i) in the case where a background pressure prevails, we replace the splitting (\ref{eq_p_decomp}) by $p_0^\varepsilon(\rho) = p_B(\rho)$ and $p_1^\varepsilon(\rho) = \varepsilon p(\rho)$ (see Remark \ref{rem_background_pressure_splitting}). 

\noindent (ii) In the case where $c<1$, the previous scheme can be used as well. The only place where hyperbolicity is used is in the definition of the diffusion coefficient (\ref{eq:C}). In the case $c<1$ and when the quantity inside one of the square roots is negative, we replace it by its absolute value, i.e. the second formula of (\ref{eq:C}) is replaced by  \begin{eqnarray*}
& & \hspace{-1cm} 
\tilde C_j^n = \left|\frac{q_{j}^{n}}{\rho_{j}^{n}}\right|+ \left| (c^2 - c) \left| \frac{q_{j}^{n}}{\rho_{j}^{n}} \right|^2 + \lambda (p^\varepsilon_0)'(\rho_{j}^{n}) \right|^{1/2} . 
\end{eqnarray*}
\label{rem:background_nonhyperbolic}
\end{remark}

\setcounter{equation}{0}
\section{Numerical results}
\label{sec_numerical_results}

\subsection{Riemann problem test}
\label{subsec_Riemann}

In the one-dimensional case, $\rho$ and $\boldsymbol{\Omega}$ only depend on a one-dimensional variable $x$ and on time $t$. The vector field $\boldsymbol{\Omega}$ can be described by its angle $\theta$ with respect to the $x$-direction, i.e. $\boldsymbol{\Omega}=(\cos \theta, \sin \theta)$. Then, the SOH system can be written as follows:
\begin{align}
  &\rho_t +\cos\theta \, \rho_x-\rho \, \sin\theta \, \, \theta_x = 0, \label{eq:SOH1D_rho}\\
&\theta_t -\lambda \frac{{p^\varepsilon}'(\rho)}{\rho} \, \sin\theta \, \rho_x + c \, \cos\theta \, \, \theta_x=0. \label{eq:SOH1D_theta}
\end{align}
This system is hyperbolic with the characteristics speeds given by (\ref{eq_char_speeds}). However, this system is non-conservative, and shock waves are not uniquely defined \cite{LeFloch_IMA89}. A simple conservative form can be found \cite{Degond_etal_JSP10} and is written as follows:
\begin{align}
   &\rho_t +(\rho\cos\theta)_x = 0,\label{eq:15}\\
&\partial_t f_1(\theta) -\partial_x \left( cf_2(\theta)-\lambda g(\rho) \right)=0,\label{eq:16}
\end{align}
where 
$$f_1(\theta)=\mbox{ln} \left| \tan \left( \frac{\theta}{2} \right) \right|, \quad 
f_2(\theta)=\mbox{ln} |\sin\theta|, \quad g'(\rho)=\frac{{p^\varepsilon}'(\rho)}{\rho}.$$ 
From this conservative formulation, the solutions of the Riemann problem can be analytically derived. Indeed, by eliminating the shock speed in the Rankine-Hugoniot conditions associated to the conservative form (\ref{eq:15}), (\ref{eq:16}), the equation of the shock curves can be determined \cite{Leveque_Cambridge2002}. These equations relate the right state $(\rho_r, \theta_r)$ to the left state $(\rho_\ell,\theta_\ell)$ of the shocks. It is easy to see that for a given left state, there are two such curves and that they are given by the following equation:
\begin{gather}
  (\rho_r-\rho_\ell)(cf_2(\theta_r)-cf_2(\theta_\ell)-\lambda g( \rho_r)+\lambda g( \rho_\ell))=(\rho\cos(\theta_r)-\rho_\ell\cos(\theta_\ell))(f_1(\theta_r)-f_1(\theta_\ell)).
\label{eq_shock_curve}
\end{gather}
This equation can be solved locally. 

However, conservative forms of the SOH model (\ref{eq:SOH1D_rho}), (\ref{eq:SOH1D_theta}) are non unique. Other conservative forms than (\ref{eq:15}), (\ref{eq:16}) can possibly be found. Therefore, the analytical solutions of the Riemann problem derived from (\ref{eq_shock_curve}) may not be correct. 

The SOH model provides an approximation of the Vicsek particle model, as shown in \cite{Degond_Motsch_M3AS08}. Therefore, physically valid solutions of (\ref{eq:SOH1D_rho}), (\ref{eq:SOH1D_theta})  are those which are close to the underlying Vicsek particle model. In \cite{Motsch_Navoret_MMS11}, it is shown that the relaxation model provides such a consistant approximation. By contrast, some other numerical methods (such as standard shock-capturing methods) select physically 'wrong' solutions, which are different from those of the Vicsek particle model. Therefore, the numerical method developed in the present paper, which relies on the relaxation model of \cite{Motsch_Navoret_MMS11}, provides a consistant approximation of the physically correct solution. In order to determine whether the analytic solutions of the Riemann problem based on (\ref{eq_shock_curve}) are the physically correct ones, we can compare them to a reference solution obtained by the present numerical method. 

We test a shock of the first family (i.e. associated to the smallest characteristic speed) under the condition
$c=\lambda=1$. Since the conservative form requires that
$\sin \theta \neq 0$, the test example can be defined as follows:
\begin{gather*}
(\rho_\ell,\theta_\ell)=(0.8,0.14), \quad (\rho_r,\theta_r)=  (0.9969,1.4502),  
\end{gather*}
with the shock speed deduced from the Rankine-Hugoniot relation and given by
\begin{gather}
\sigma= -3.4136.\label{eq:18}
\end{gather}

\begin{figure}[htbp]
  \centering
\subfigure[density $\rho$]{\label{fig:1:a} \includegraphics[scale=0.3]{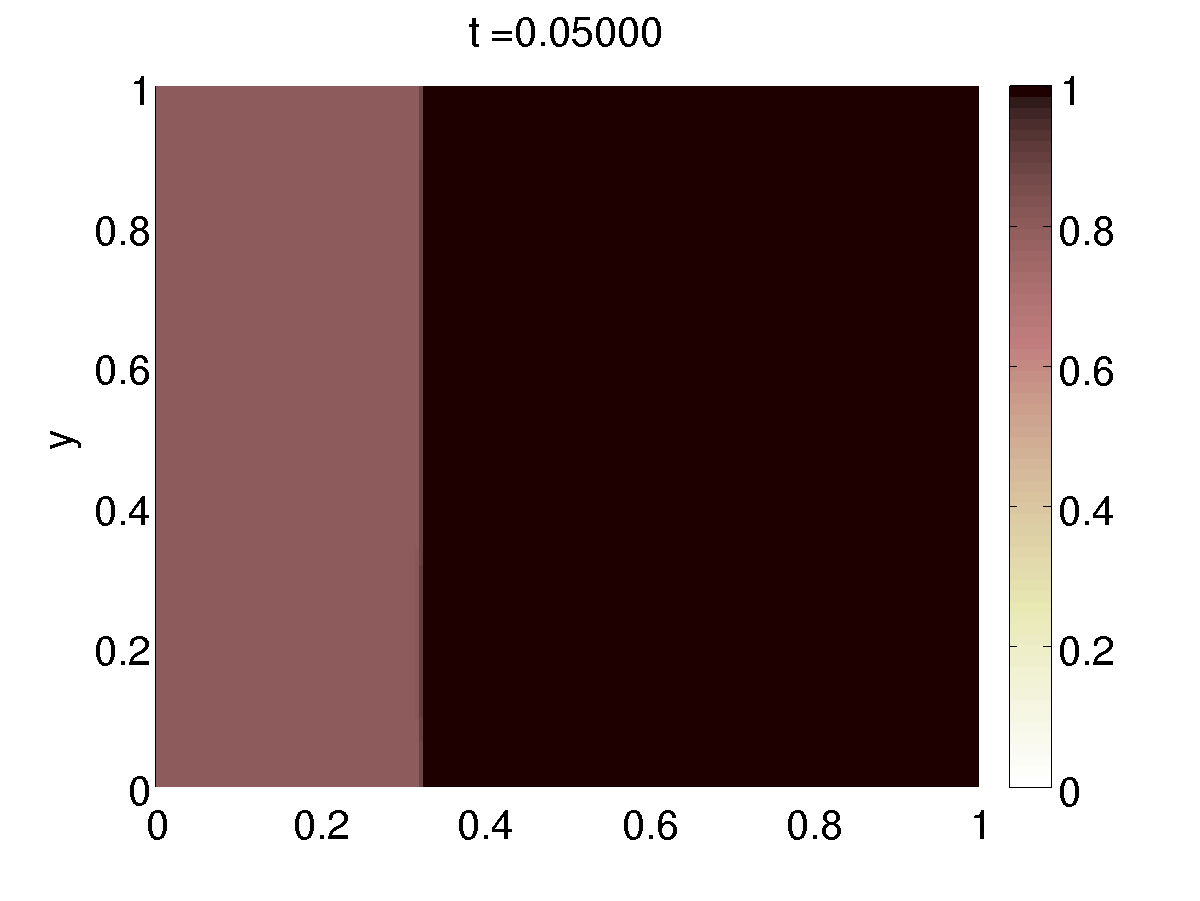}}
\subfigure[velocity $\boldsymbol{\Omega}$]{\label{fig:1:b} \includegraphics[scale=0.3]{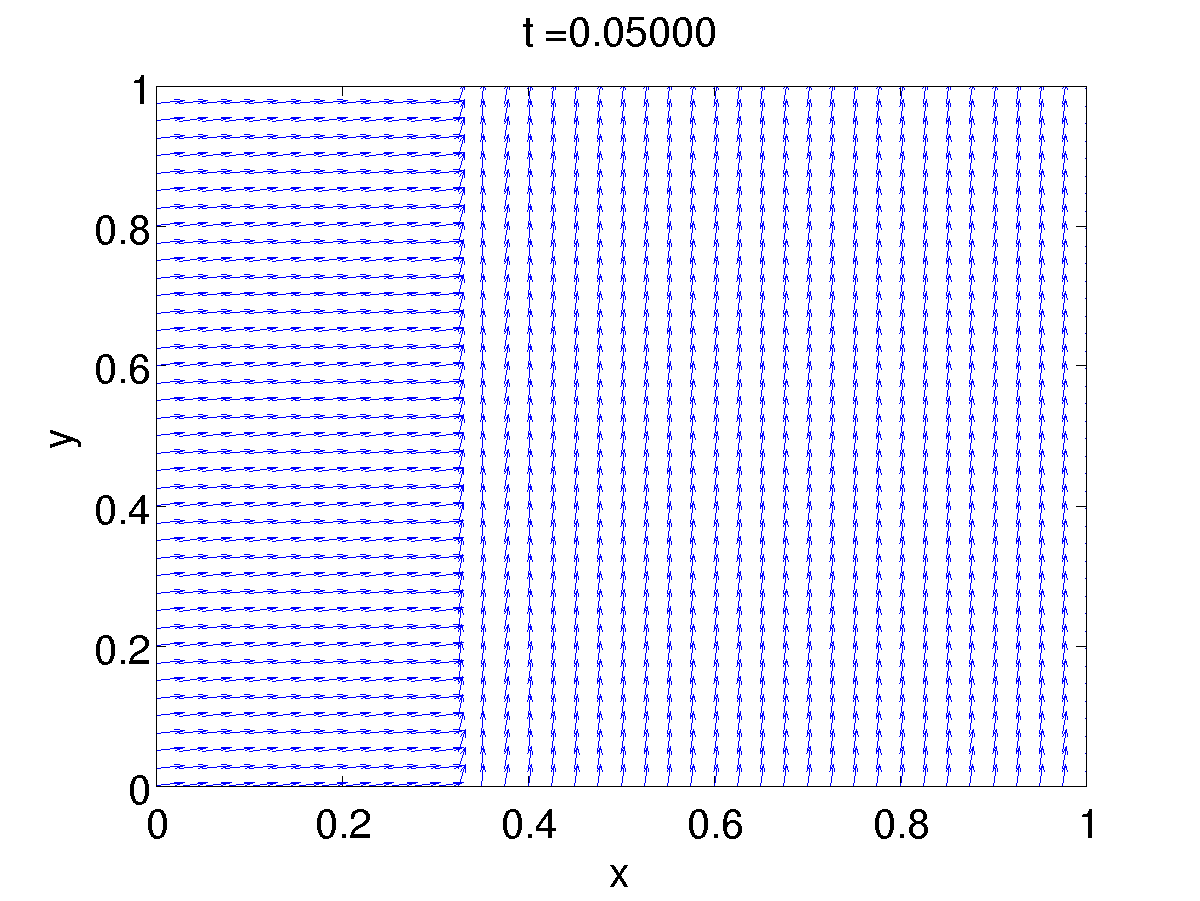}}
   \caption{Shock-wave test problem. Density $\rho$ (Fig. \ref{fig:1:a}) and
     vector field  $\boldsymbol{\Omega}$ (Fig. \ref{fig:1:b}) as functions of $x$ and $y$ at time $t = 0.05$. The density is color-coded, with color map indicated to the right of Fig. \ref{fig:1:a}. The field $\boldsymbol{\Omega}$ is indicated by blue arrows.}
  \label{fig:1}
\end{figure}

\begin{figure}[htbp]
  \centering
\subfigure[density
$\rho$  at $t=0$]{\label{fig:2:a} \includegraphics[scale=0.22]{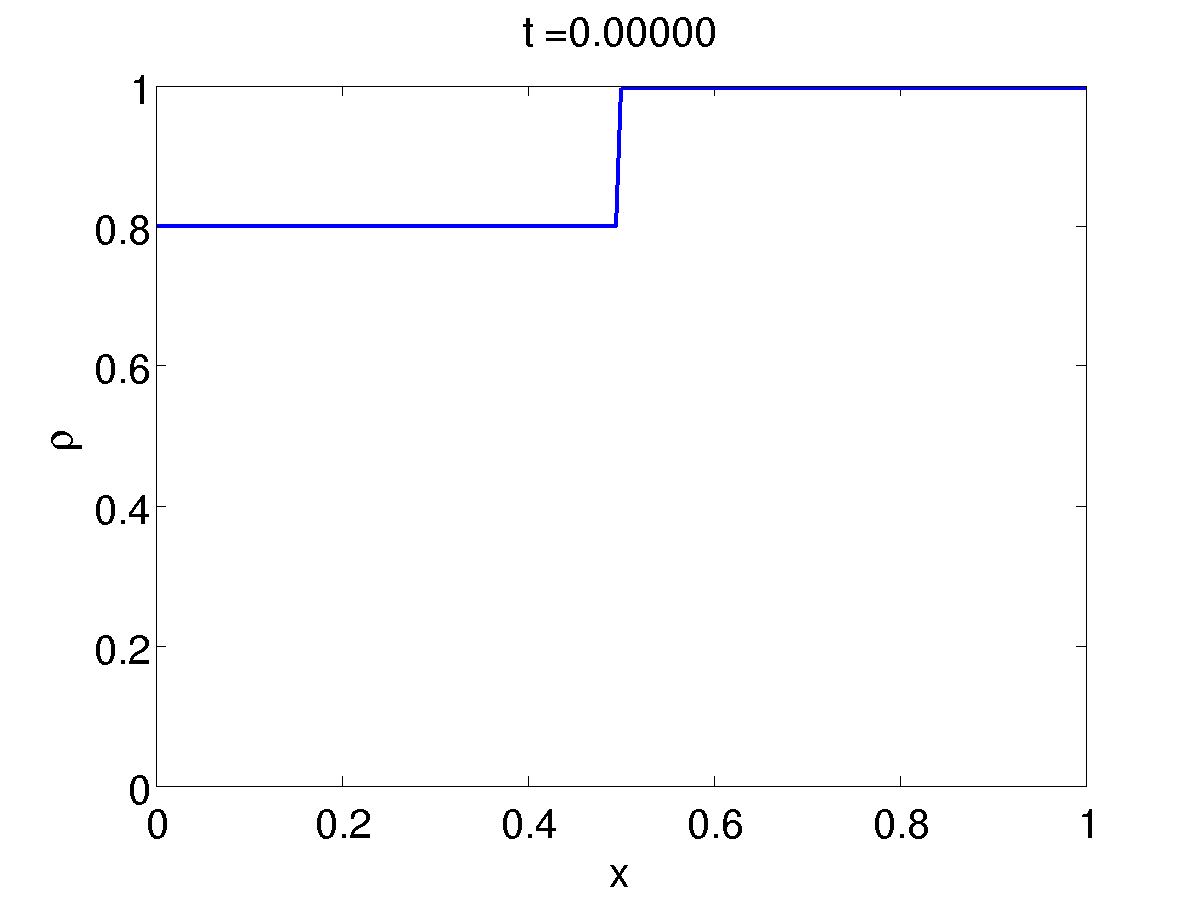}}
\hspace{1cm}
\subfigure[density
$\rho$ at $t=0.14$]{\label{fig:2:c} \includegraphics[scale=0.22]{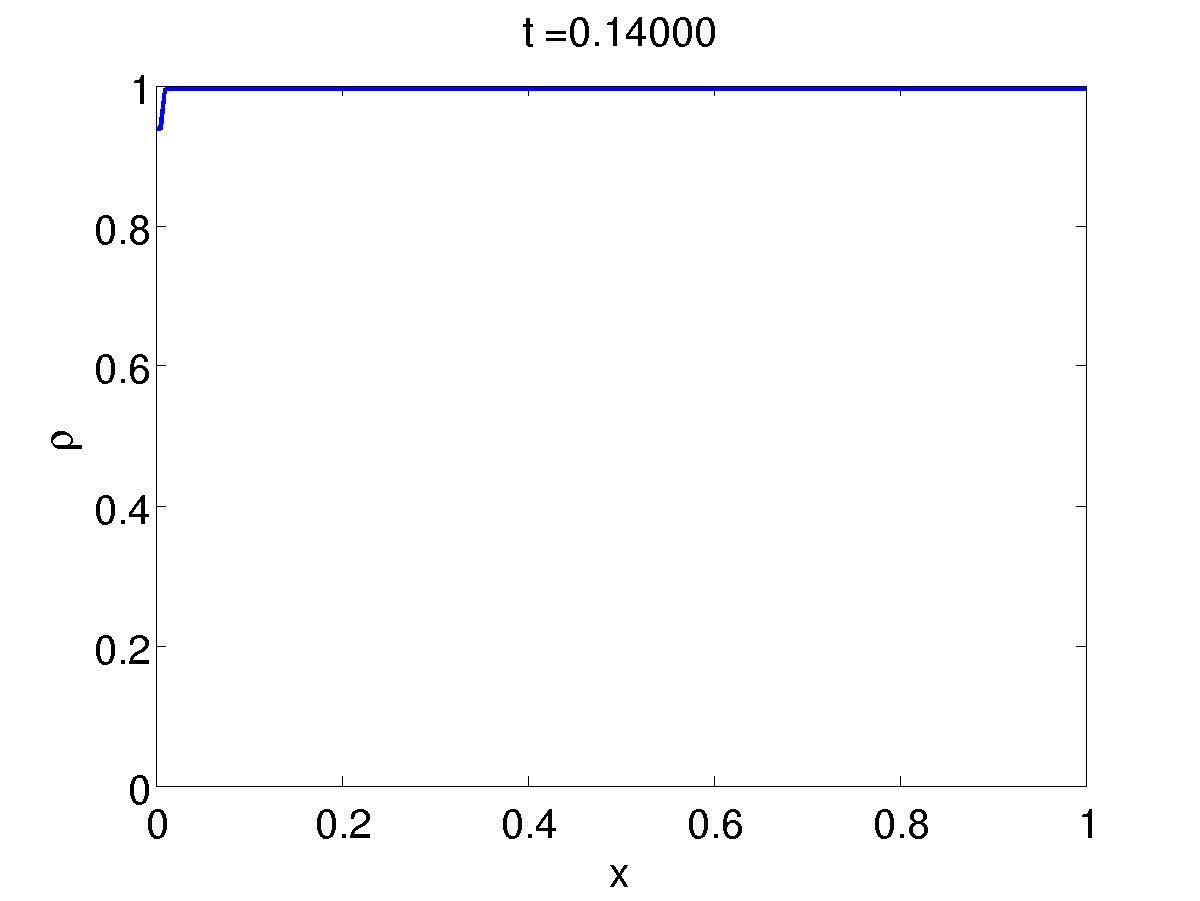}}\\
\subfigure[momentum $\boldsymbol{q}$ at $t=0$]{\label{fig:2:b} \includegraphics[scale=0.22]{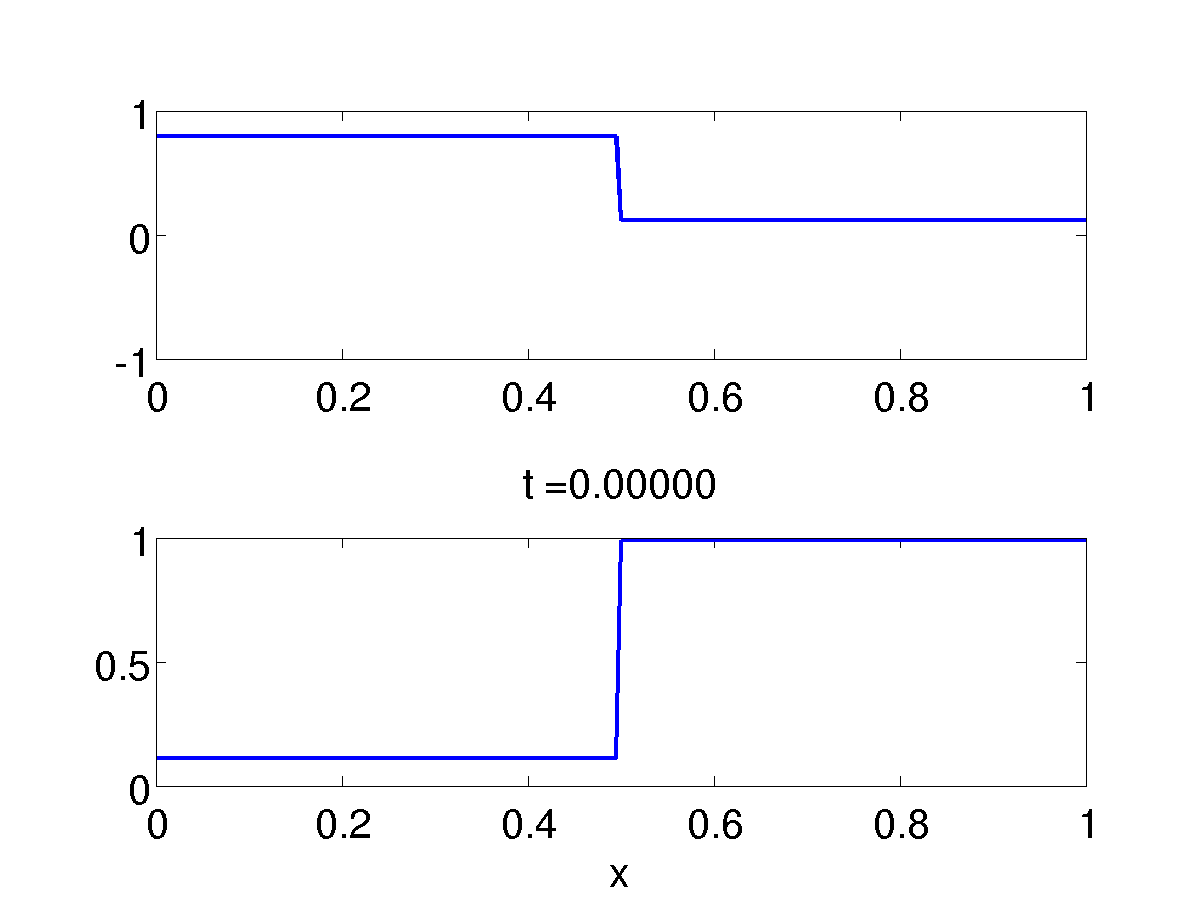}}
\hspace{1cm}
\subfigure[momentum $\boldsymbol{q}$
at $t=0.14$]{\label{fig:2:d} \includegraphics[scale=0.22]{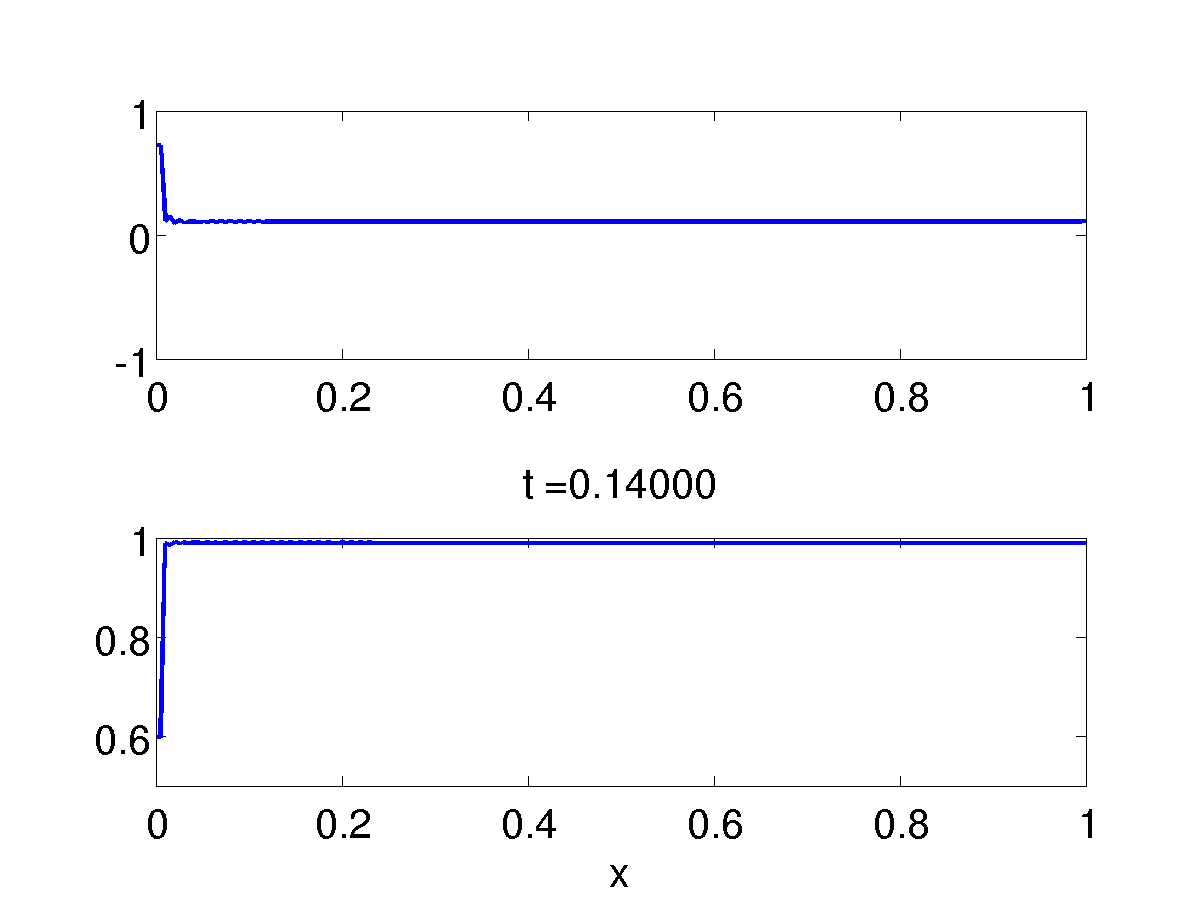}}
   \caption{Shock-wave test problem. Cross-section along the axis $y=0.5$ of the density $\rho$ (figs \ref{fig:2:a} and \ref{fig:2:c}), $x$-component of the momentum $\boldsymbol{q}$ (figs \ref{fig:2:b} and \ref{fig:2:d}, top) and $y$-component of momentum $\boldsymbol{q}$ (figs \ref{fig:2:b} and \ref{fig:2:d}, bottom) at time $t=0$ (left) and $t=0.14$ (right) as functions of $x$.}
  \label{fig:2}
\end{figure}

The numerical simulation also yields a shock wave, as can be seen on Fig.
\ref{fig:1} obtained with parameters $\varepsilon=10^{-4}, \beta=10^{-7}$,  mesh sizes $\Delta x=0.005$, $\Delta t=0.0005$ at time $t = 0.05$. In these simulations, one-dimensional simulations have been performed using a two-dimensional code by imposing periodic boundary conditions along the top and bottom horizontal boundaries. The initial condition has been set independent of the $y$-coordinate. Fig. \ref{fig:1:a} shows the density map in the two-dimensional domain, with a color coding (brown for $\rho = 0.8$ and black for $\rho =0.9969$). Fig. \ref{fig:1:b} represents the vector field $\boldsymbol{\Omega}$ with blue arrows. Arrows are almost horizontal for $\theta = 0.14$ and almost vertical for $\theta = 1.4502$. Fig. \ref{fig:2} displays cross-sections of the various components of the  solution along the axis $y=0.5$ as functions of $x$ at times $t=0$ (left figures) and $t=0.5$ (right figures). Figs. \ref{fig:2:a} and \ref{fig:2:c} show the density $\rho$ and figs. \ref{fig:2:b} and \ref{fig:2:d} show the $x$ and $y$-components of the momentum $\boldsymbol{q}$ (top and bottom pictures respectively). By recording the position of the shock at time $t=0.5$, it is possible to determine the physical shock speed. By inspection of Fig. \ref{fig:2}, it is found to be $\hat{\sigma}\approx - 3.5714$. This is different from the prediction \eqref{eq:18} and the difference is significant, given the fine mesh size. Indeed, if the shock speed was given by \eqref{eq:18}, the shock would reach the boundary at time $t=0.1465$. This is different from what we observe in Fig. \ref{fig:2} by more than 10 time steps. 

These numerical evidences show that the conservative form (\ref{eq:15})-(\ref{eq:16}) does not provide the true physical solution, while the relaxation method, as shown in \cite{Motsch_Navoret_MMS11}, does.

\subsection{2D simulations of collisions of herds}
\label{subsec_simu_collision}

\subsubsection{Initial conditions}
\label{subsubsec_2D_ini}

\begin{figure}[htbp]
  \centering
\subfigure[density $\rho$]{\label{fig:3:a} \includegraphics[scale=0.38]{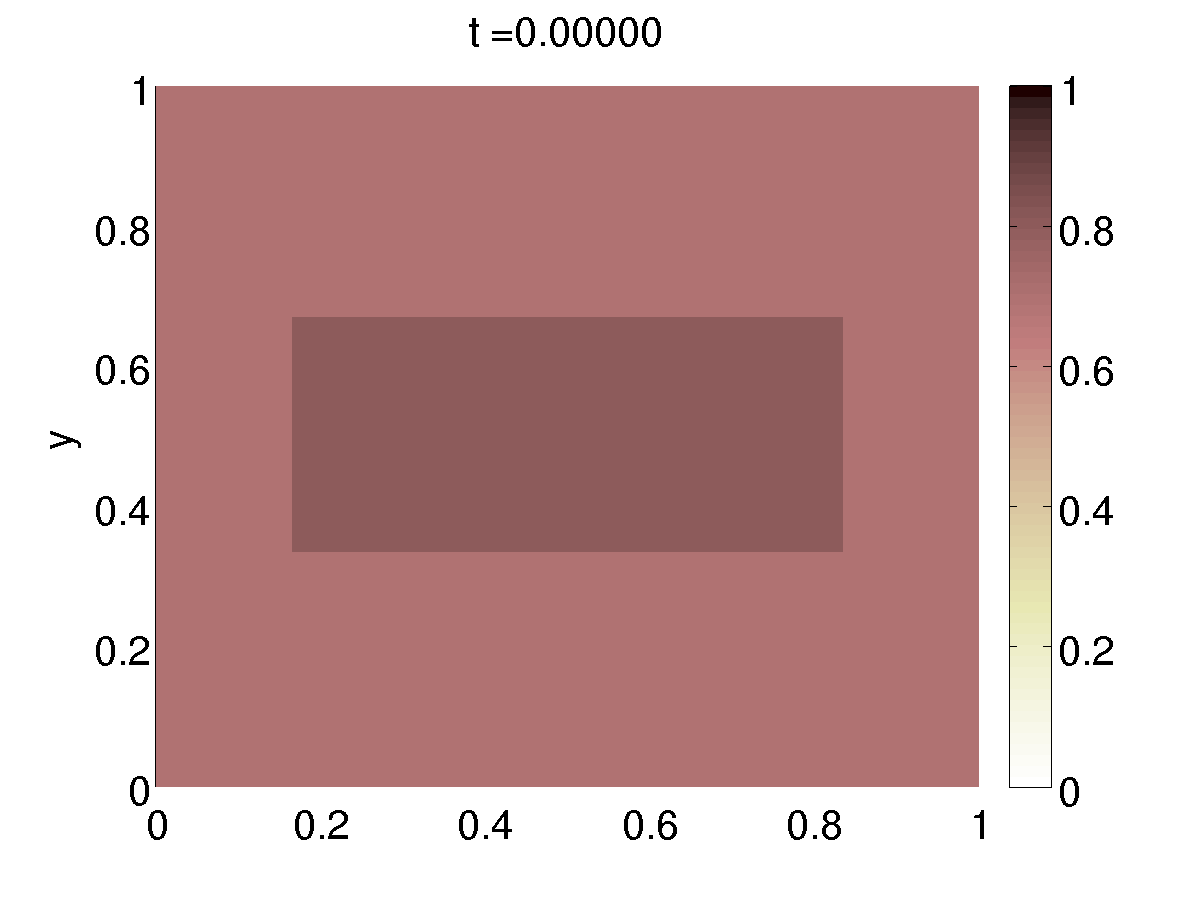}}
\subfigure[$\boldsymbol{\Omega}$]{\label{fig:3:b} \includegraphics[scale=0.38]{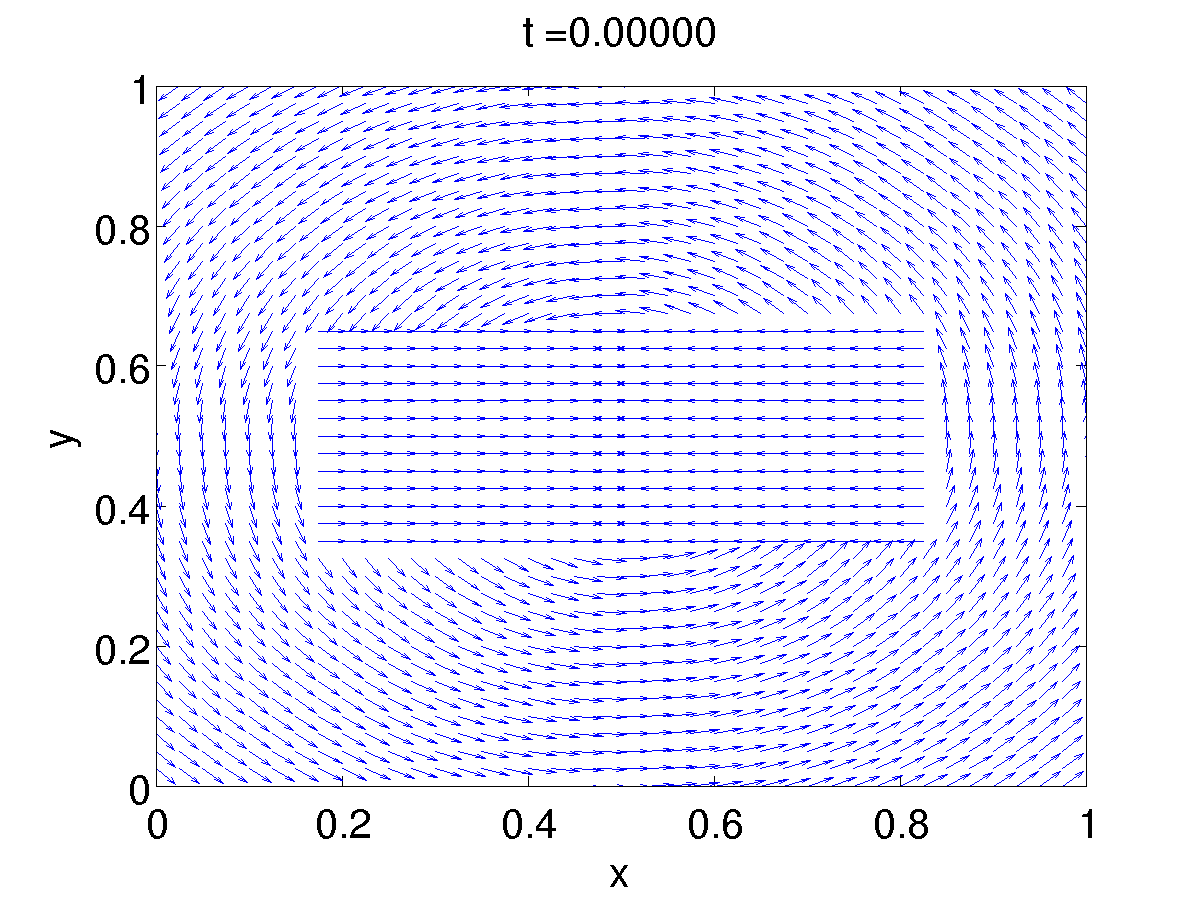}}
   \caption{Initial data for $\rho$ (Fig. \ref{fig:3:a}) and $\boldsymbol{\Omega}$ (Fig. \ref{fig:3:b}), as functions of $x$ and $y$. The density is color-coded, with color map indicated to the right of Fig. \ref{fig:3:a}. The field $\boldsymbol{\Omega}$ is indicated by blue arrows.}
  \label{fig:3}
\end{figure}

In this section, we show simulation results in a truly 2D setting. Our main focus is the dynamics of the interaction of two clustered regions, which cannot
be studied theoretically with the methods of \cite{Degond_etal_JSP10}, as explained at the end of section \ref{sec_background}. The initial datum describes
two clusters with densities close to the congestion density moving towards each other, inside a lower density fluid. Since the speed of the
flow must be of norm one everywhere, the lower density fluid (background flow) is supposed to adopt a swirling motion around the center of the domain $D=[0,1]\times[0,1]$. This leads to the following initial condition: 
\begin{gather*}
   \rho (\boldsymbol{x},0) = 0.8\times \boldsymbol{1}_{A\cup
     B}+0.7\times\boldsymbol{1}_{D\backslash (A\cup B)},\\
 \boldsymbol{\Omega}(\boldsymbol{x},0)=
  \boldsymbol{1}_A\begin{pmatrix}
    1\\ 0
  \end{pmatrix}
+\boldsymbol{1}_B\begin{pmatrix}
    -1\\ 0
  \end{pmatrix}+\boldsymbol{1}_{D\backslash (A\cup B)}\boldsymbol{v}(\boldsymbol{x}),\\
A=\left[\frac{1}{6},\frac{1}{2}\right]\times\left[\frac{1}{3},\frac{2}{3}\right],
\quad
B=\left[\frac{1}{2},\frac{10}{12}\right]\times\left[\frac{1}{3},\frac{2}{3}\right],\\
\boldsymbol{v}(\boldsymbol{x})= \frac{1}{\sqrt{(x-0.5)^2+(y-0.5)^2}}\begin{pmatrix}
    -(y-0.5)\\ x-0.5
  \end{pmatrix},
\end{gather*}
where $\boldsymbol{1}_A$ denotes the indicator function of the set $A$. 
The definition of $\boldsymbol{v}$ corresponds to a background flow
rotating around the center of the domain $\boldsymbol{P}=(0.5,0.5)^t$
counterclockwise. The initial data can be seen in Figure \ref{fig:3}. Fig \ref{fig:3:a} shows the density $\rho$ in the two-dimensional domain, in color coding (larger densities are shown in darker brown color, and lower densities, in lighter brown color). The dark inner rectangle shows the large density clusters, surrounded by lower density regions in lighter brown. Fig. \ref{fig:3:b} provides a view of the vector field $\boldsymbol{\Omega}$ represented by blue arrows, in the two-dimensional domain. Inside the large density rectangle, the arrows are horizontal, and point towards the right in the left-hand half of the rectangle and to the left in the right-hand half of the rectangle. This shows that the inner rectangle is formed by two high density clusters moving one towards each other. The arrows outside the inner rectangle indicate that the flow is circulating counterclockwise around the inner rectangle. 

The numerical parameters are chosen as follows: 
\begin{equation}
\varepsilon=10^{-4}, \quad \beta=10^{-7}, \quad \lambda=1, \quad \Delta x=0.005, \quad \Delta t=0.0005. 
\label{eq:parameters}
\end{equation}
The presence of a relatively large background density is necessary otherwise instabilities develop and generate negative pressures \cite{Degond_etal_JCP11}. Developments are currently undertaken to prevent these instabilities and ensure the positivity of the pressure in all cases. We first investigate the case where $c=1$, without background pressure (in this case, we use the decomposition (\ref{eq_p_decomp}) of the pressure). Then, we add a background pressure, and use the decomposition (\ref{eq_background_pressure}) instead. Then, we keep a background pressure, and investigate successively the cases $c=2$ and $c=0.5$. In the latter case, the relaxation system is not hyperbolic, and we use the method described in Remark \ref{rem:background_nonhyperbolic} (ii) to overcome the appearance of complex valued characteristic speeds.

\subsubsection{Case $c=1$, without background pressure} 
\label{subsubsec_c=1_nobp}

\begin{figure}[htbp]
  \centering
\subfigure[density $\rho$]{\label{fig:4:a} \includegraphics[scale=0.38]{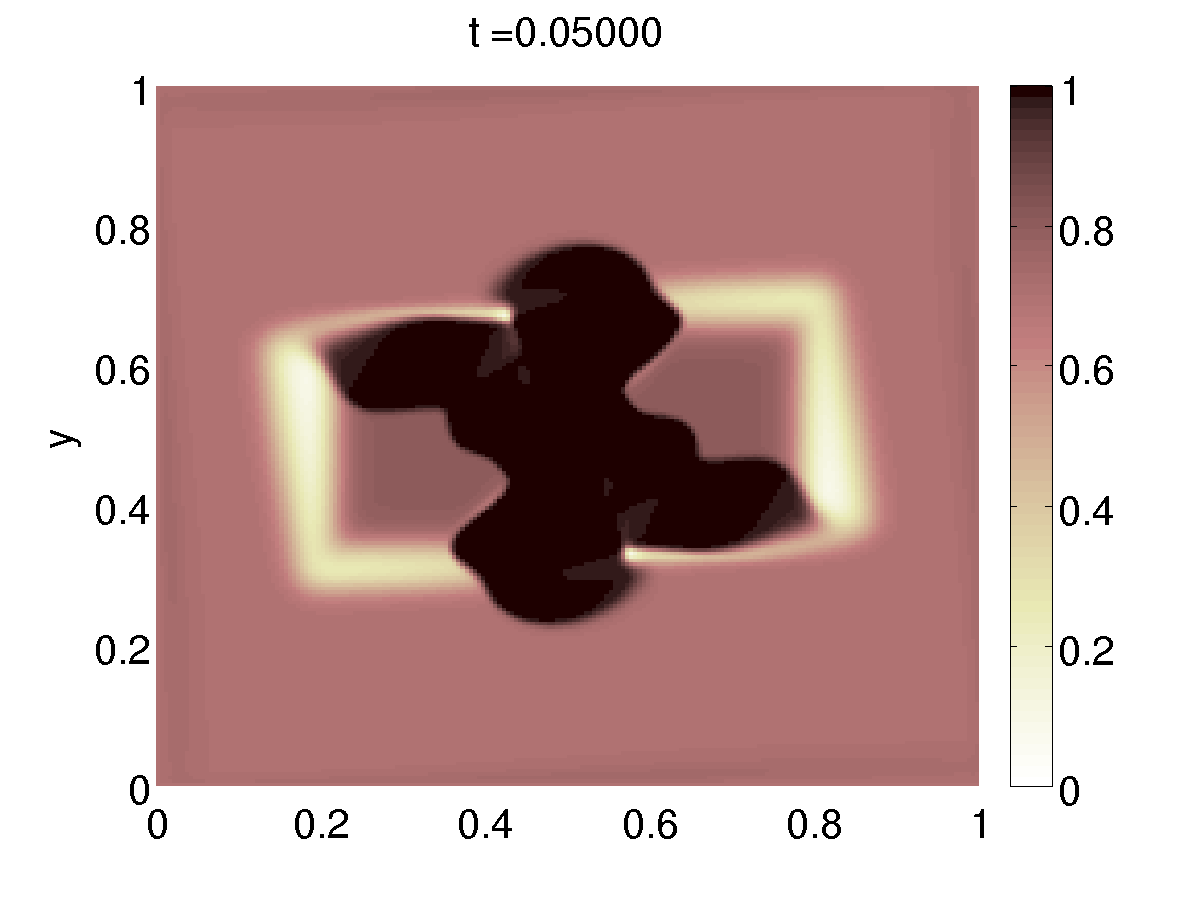}}
\subfigure[$\boldsymbol{\Omega}$]{\label{fig:4:b} \includegraphics[scale=0.38]{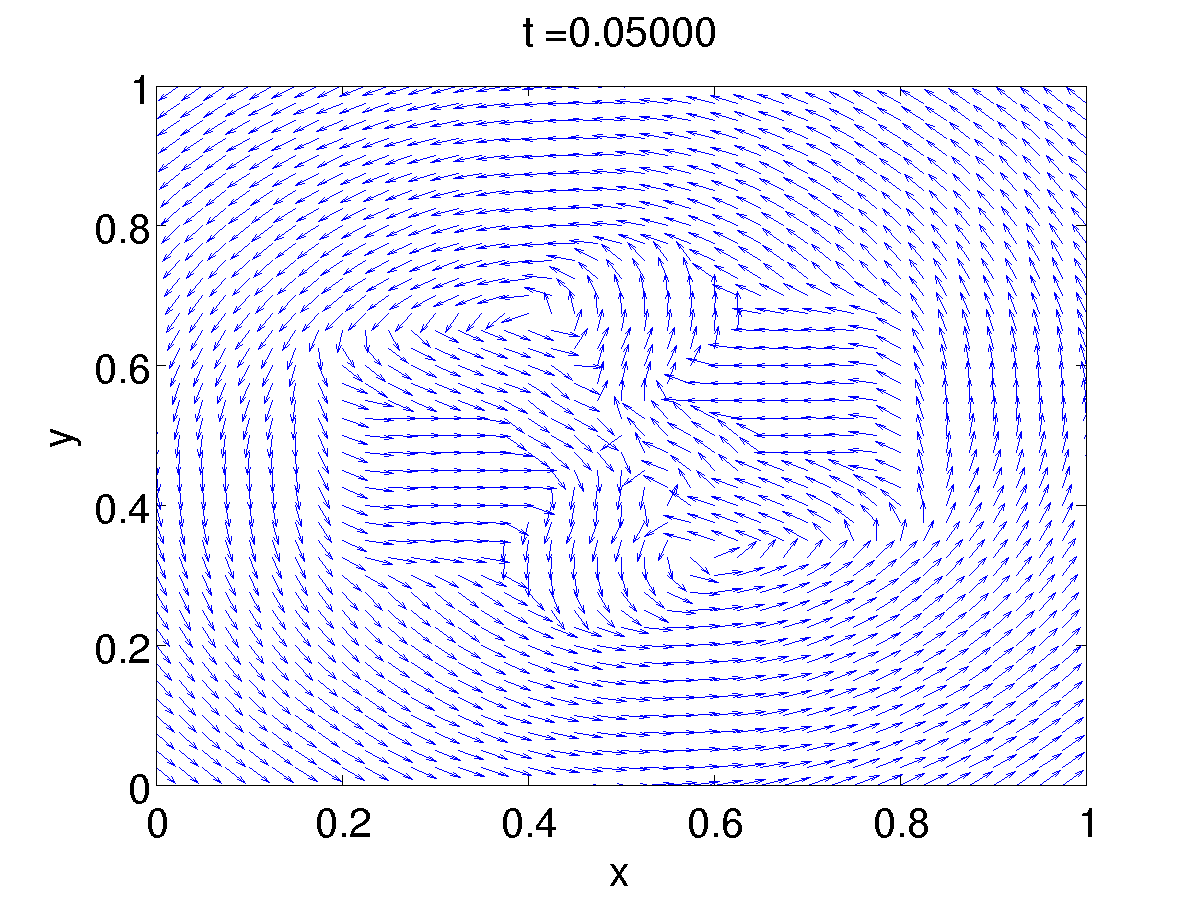}}
   \caption{Cluster collision. No background pressure and $c=1$. Density $\rho$ (Fig. \ref{fig:4:a}) and velocity $\boldsymbol{\Omega}$ (Fig. \ref{fig:4:b}) at $t=0.05$, as functions of $x$ and $y$. The density is color-coded, with color map indicated to the right of Fig. \ref{fig:4:a}. The field $\boldsymbol{\Omega}$ is indicated by blue arrows.}
  \label{fig:4}
\end{figure}

Figure \ref{fig:4} displays the density $\rho$ and the vector field $\boldsymbol{\Omega}$ at time $t=0.05$, using the same representation as in figure \ref{fig:3}. As we can see on Fig \ref{fig:4:a}, the two clusters initially located in the inner rectangle are moving one towards each other and generate a congestion region (displayed in black color) between them. The congestion constraint generates a deflected wave which propagates in the vertical direction and compress the background flow, also bringing it close to the congestion density. This compression occurs in the central parts of the top and bottom horizontal boundaries of the high density rectangle, as indicated by the black and dark brown colors in this areas. Finally, due to its rotation about the high density inner rectangle, the background flow impinges against the high density clusters along the top left and bottom right horizontal boundaries of the high density inner rectangle. This also produces some congestion there, as indicated by the black and dark brown colors in these areas. The two moving clusters leave a  vacuum region behind them, which is shown by the light beige color along the outer vertical edges of the high density rectangle. Similarly, due to the rotation of the background flow, a vacuum region is created along the bottom left and top right horizontal boundaries of the high density inner rectangle shown in light beige color in these areas. 

Figure \ref{fig:4:b} highlights the geometrical features of the vector field $\boldsymbol{\Omega}$. Except in the congested regions, the vector field is not affected. Therefore, the swirling pattern of the background flow away from the inner rectangle is clearly visible. Similarly, $\boldsymbol{\Omega}$ is not changed in the lower left part of the left cluster and upper right part of the right cluster: it is directed in the horizontal direction and points to the right in the left cluster and to the left in the right cluster. However, $\boldsymbol{\Omega}$ is strongly affected in the congested region. The flow abruptly changes direction towards a globally vertically directed flow in the central part of the clustered region. This flow is directed downwards in the left cluster and upwards in the right cluster. This is consistent with the direction of the swirling flow, which seems to force the circulation inside the congested region to adopt a similar counterclockwise rotation as for the background flow. As a result, an S-shaped shock line appears inside each cluster, separating the regions of unperturbed horizontal motion to those of vertical motion. The upwards and downwards vertical motions are separated by a slip line located in the middle between the two clusters. This slip line is almost vertical (slightly tilted in the south-east to north-west direction). This slip line ends up with two counterclockwise rotating vortices. These vortices are generated by the vertical flows inside the clusters that short-circuit the swirling motion. They are located where the vertical slip line rejoins the swirling motion at the upper left and lower right horizontal boundaries of the high density clusters. The vertical circulation which takes place in the inner part of the clusters penetrate the background flow on the sides of the vortices and generates a shock line with a hemi-circular shape. This hemi-circular shape reproduces the shape of the transition between congested and uncongested densities that can be oberved in Fig. \ref{fig:4:a}, near the central parts of the top and bottom horizontal boundaries of the high density inner rectangle. 

From these observations, we deduce that $\boldsymbol{\Omega}$ is not smooth. The constraint $|\boldsymbol{\Omega}|=1$ contributes to generating complex patterns. Their geometric features are outlined in section \ref{sec_background} and in Fig. \ref{Fig:lines_1}. Indeed, as a consequence of the fact that $\boldsymbol{\Omega}$ is constant along straight lines normal to itself, its integral lines are parallel curves to each other. This feature is clearly seen on Fig. \ref{fig:4:b}. However, at singularities, the parallel curves are interrupted (see Fig. \ref{Fig:lines_2}). Here, due to the collision
of the two clusters, vortex singularities, discontinuities and slip lines form. The resulting topology takes a form that is reminicent of domain walls in micromagnetism. Due to the time dynamics, these domain walls evolve in the course of time and can change topology.

\begin{figure}[htbp]
  \centering
\subfigure[density $\rho$]{\label{fig:5:a} \includegraphics[scale=0.38]{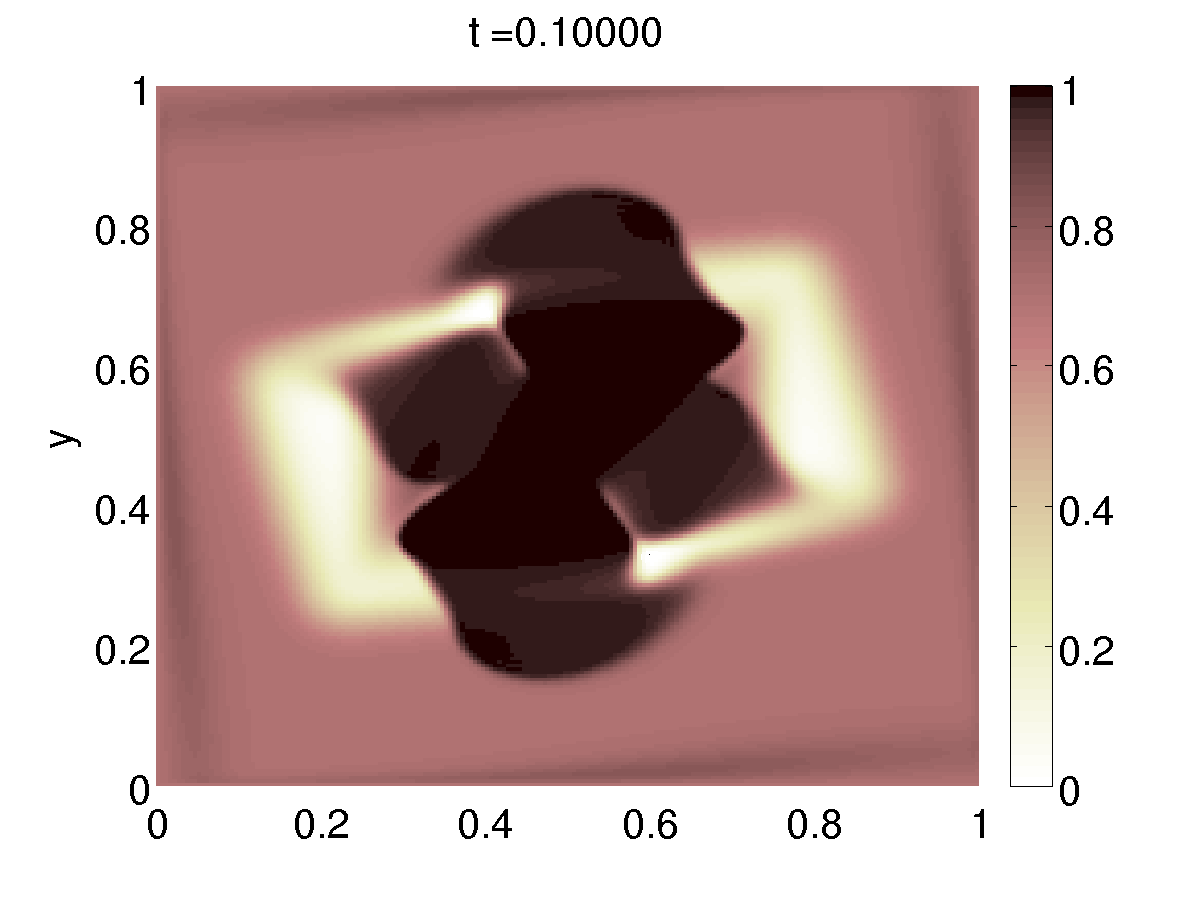}}
\subfigure[$\boldsymbol{\Omega}$]{\label{fig:5:b} \includegraphics[scale=0.38]{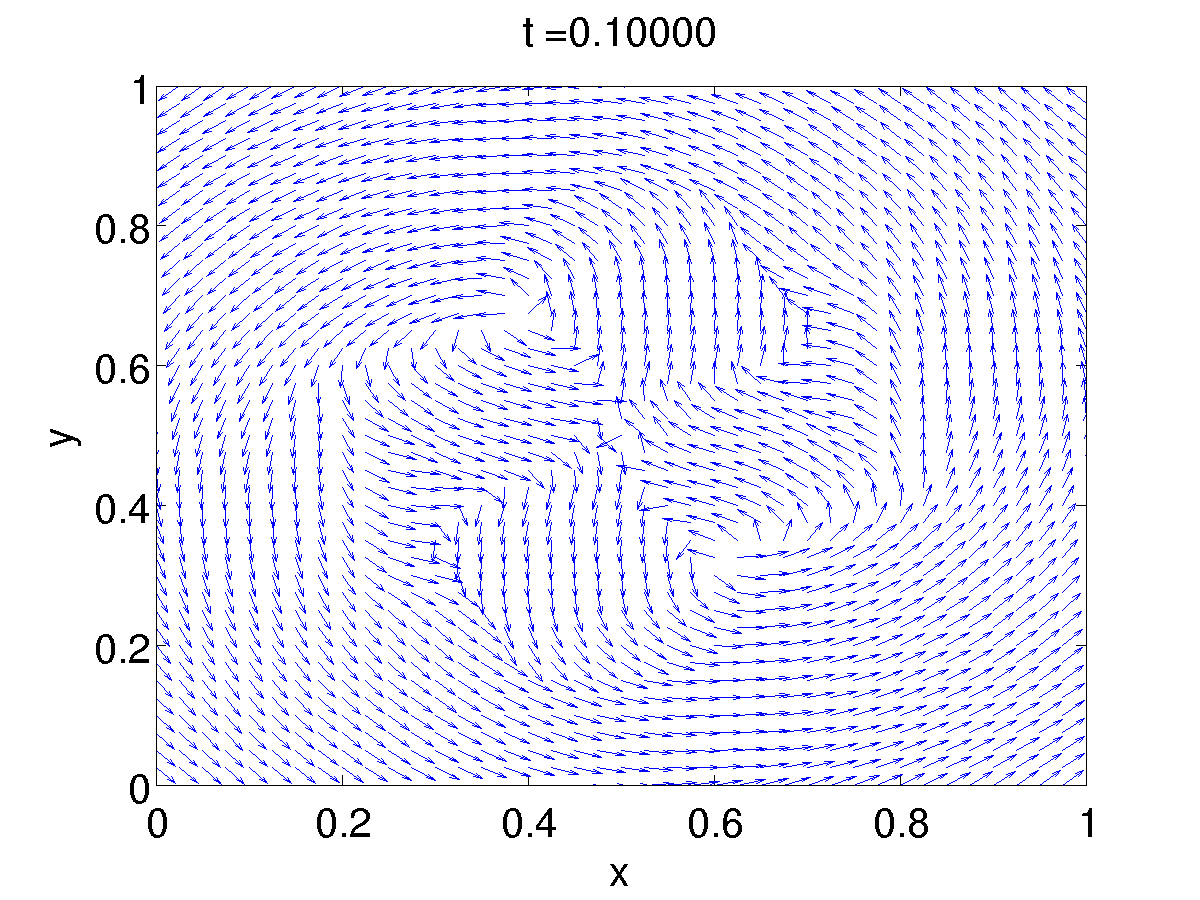}}
   \caption{Cluster collision. No background pressure and $c=1$. Density $\rho$ (Fig. \ref{fig:5:a}) and velocity $\boldsymbol{\Omega}$ (Fig. \ref{fig:5:b}) at $t=0.1$, as functions of $x$ and $y$. The density is color-coded, with color map indicated to the right of Fig. \ref{fig:5:a}. The field $\boldsymbol{\Omega}$ is indicated by blue arrows.}
  \label{fig:5}
\end{figure}

As time goes on, the two herds interact and form a larger
congestion region that spreads towards the low density zones. This can be seen in Figure \ref{fig:5}, which shows the solution at time $t=0.1$ (see density $\rho$ in Fig. \ref{fig:5:a} and vector field $\boldsymbol{\Omega}$ in Fig. \ref{fig:5:b}, with the same representation as in the previous figures). In Fig. \ref{fig:5:a}, we note that, due to wave dispersion, the densities inside the congested regions of the background flow and of the cluster tails have decreased below the congestion density (the color in Fig \ref{fig:5:a} is brown and not black in these regions). Congestion becomes strictly restricted to the compression region between the two clusters. The vacuum region following the tail of the clusters has also widened, because the clusters have moved further, leaving a region devoid of particles behind them. The general flow structure discussed above for Fig. \ref{fig:4:b} still remains on Fig. \ref{fig:5:b}. However, the slip line at the interface between the two clusters in the middle of the high density region has been transformed into a saddle singularity. Simultaneously, the horizontal flow inside the clusters seem to be more strongly affected. A splay field structure appears near the center of the clusters close to the saddle singularity. There, the horizontal flow splits into upwards and downwards components. The roughly straight top and bottom horizontal boundaries of the congestion region, which can be seen in Fig. \ref{fig:5:a} as an abrupt transition from black to brown corresponds to a contact discontinuity. Indeed, in Fig. \ref{fig:5:b}, along the same line, there is no discontinuity of the field $\boldsymbol{\Omega}$. Therefore, across this line, the density suffers an abrupt discontinuity while the velocity is continuous. Finally, the entire structure of the high density inner rectangle has rotated by an angle of about $10$ degrees in the counterclockwise direction, i.e. in the same rotation direction as the background swirling flow.

\subsubsection{Case $c=1$, with background pressure}
\label{subsubsec_c=1_bp}

We now investigate how the presence of a background pressure modifies the density patterns and the flow structure. We change the pressure relation to (\ref{eq_background_pressure}) and we take $p_B$ as (\ref{eq_background_pressure_2}) with $\kappa = 1$ (see also remark \ref{rem_background_pressure}). The other parameters are given by (\ref{eq:parameters}). The simulation is more robust since the background pressure prevents the appearance of negative densities.

\begin{figure}[htbp]
  \centering
\subfigure[density $\rho$]{\label{fig:7:a} \includegraphics[scale=0.38]{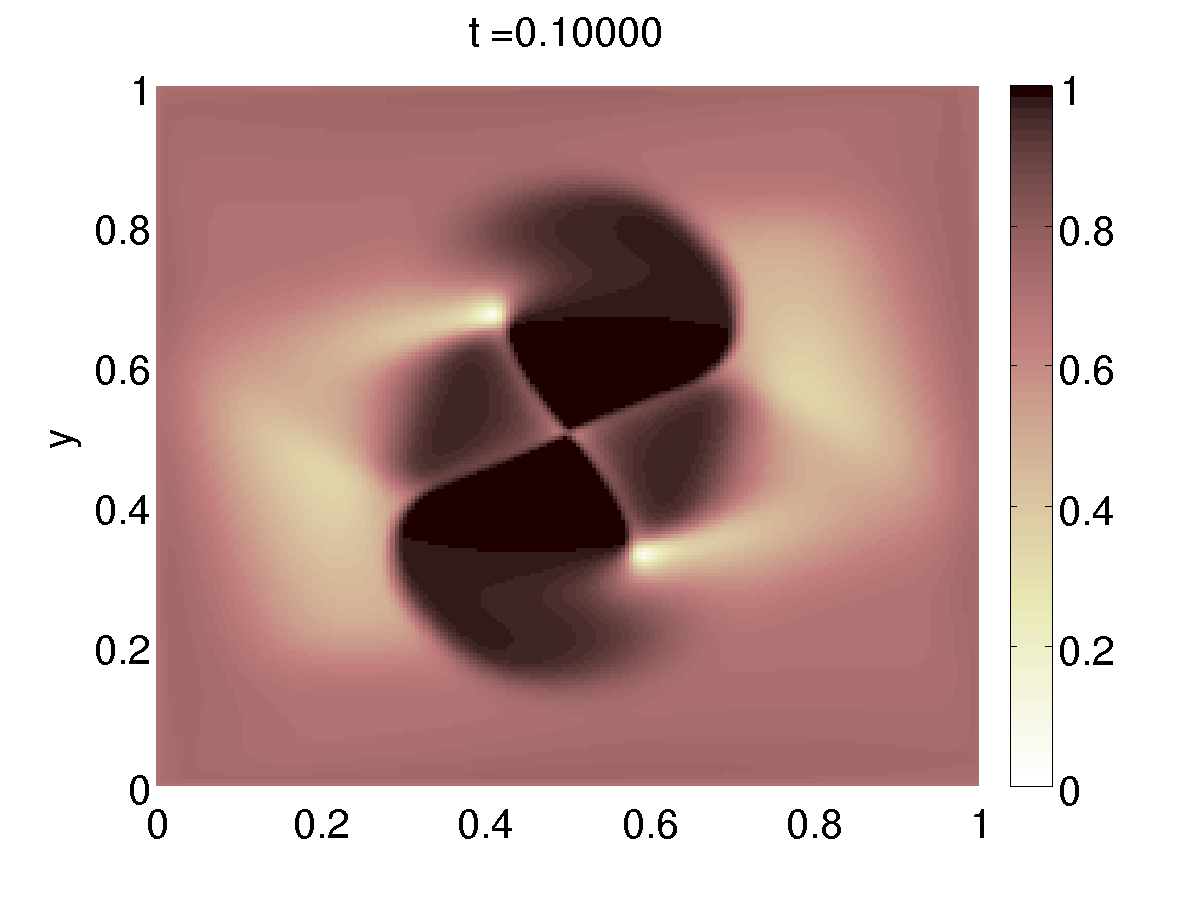}}
\subfigure[$\boldsymbol{\Omega}$]{\label{fig:7:b} \includegraphics[scale=0.38]{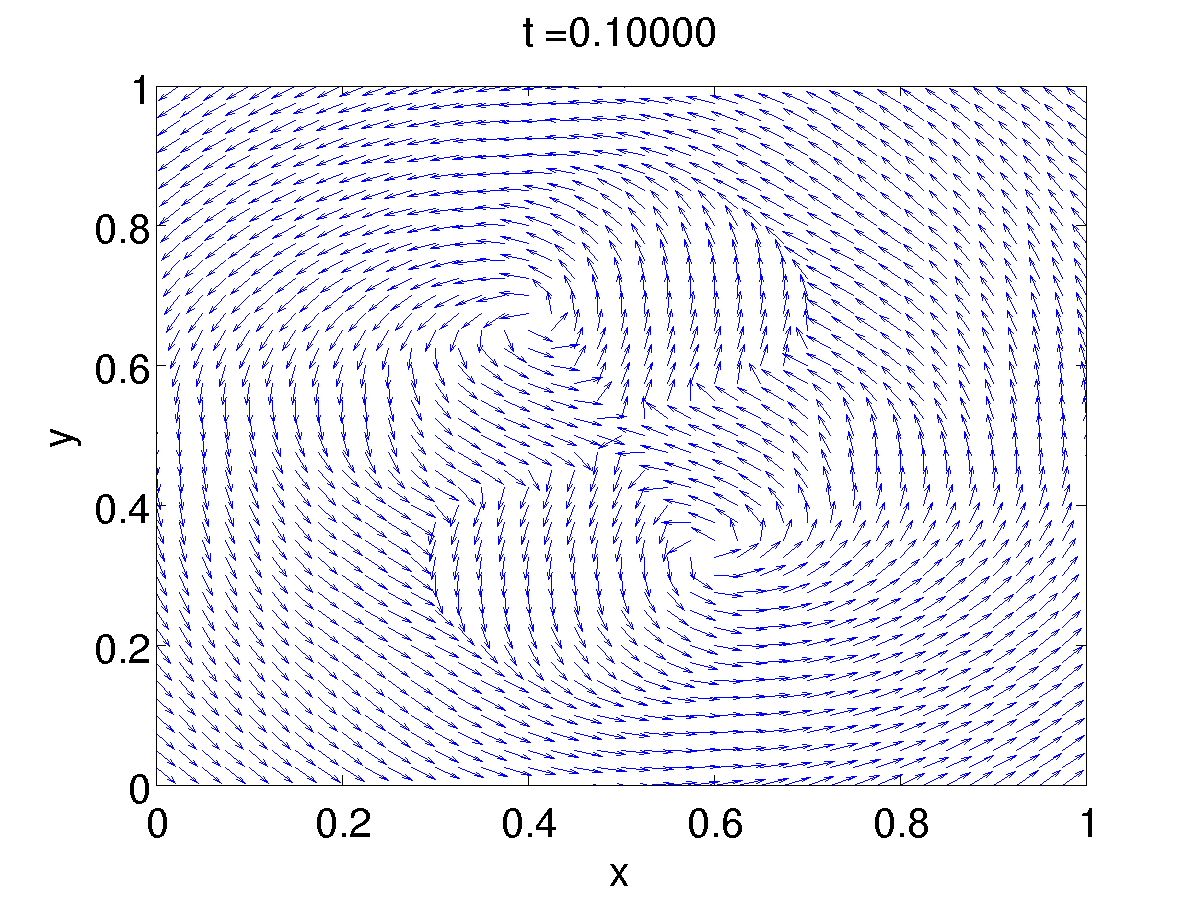}}
   \caption{Cluster collision with background pressure and $c=1$. Density $\rho$ (Fig. \ref{fig:7:a}) and velocity $\boldsymbol{\Omega}$ (Fig. \ref{fig:7:b}) at $t=0.1$, as functions of $x$ and $y$. The density is color-coded, with color map indicated to the right of Fig. \ref{fig:7:a}. The field $\boldsymbol{\Omega}$ is indicated by blue arrows.}
  \label{fig:7}
\end{figure}

The results at time $t=0.1$ are shown in Fig. \ref{fig:7} (see density $\rho$ in Fig. \ref{fig:7:a} and vector field $\boldsymbol{\Omega}$ in Fig. \ref{fig:7:b}, with the same representation as in the previous figures). This figure must be compared with Fig. \ref{fig:5}, where the results without background density are displayed, the other parameters being the same. We can see that the effect of the background pressure is to smear out the vacuum regions and to propagate the compression wave faster. Indeed, on Fig. \ref{fig:7:a}, we remark that the vacuum regions have partially filled up due to the development of rarefaction waves at the outer vertical boundaries of the high density rectangle. These rarefaction waves bring the fluid from the backgound low density regions into the vacuum regions, as can be seen on Fig. \ref{fig:7:b}. Indeed, in these vacuum regions, the direction of the vector field $\boldsymbol{\Omega}$ is oblique and points from the background flow towards the vacuum region. By contrast, the direction of $\boldsymbol{\Omega}$ for the corresponding pressureless case in Fig. \ref{fig:5:b} is vertical. Like in the pressureless case (Fig. \ref{fig:5:a}), a vertical flow is created by the collision of the clusters. It arises from the deflection of the cluster velocities into the vertical direction by the congestion constraint. This vertical flow impinges on the background flow and creates a hemi-circular wave-front (see Fig. \ref{fig:7:a}: middle part of the top and bottom horizontal boundaries of the high-density rectangle). However, by contrast to the pressureless case, the wave front propagates further and has a more regular circular shape. This is obviously due to acoustic wave propagation in the low-density background flow, which conveys information about velocity and modifies the flow direction downstream. Indeed, in Fig. \ref{fig:7:b}, it can be seen that the direction of $\boldsymbol{\Omega}$ is abruptly changed along the shock curve that bounds the region of compression. The acoustic wave conveys information about the vertical velocity of the cluster into the low density background. These changes of flow direction contribute to pre-compress the gas, far downstream the congestion region. Similar to the pressureless case, we notice that the nearly straight top and bottom horizontal boundaries of the congestion region (see transition from black to brown on Fig. \ref{fig:7:a}) are contact discontinuities. Indeed, across this line, there appears no discontinuity of $\boldsymbol{\Omega}$ on Fig. \ref{fig:7:b}. The density in the cluster tails (Fig. \ref{fig:7:a}) is smaller than in the pressureless case (Fig. \ref{fig:5:a}). This is also due to the development of rarefaction waves that contribute to fill in what was the vacuum region in the pressureless case. We notice that the two counterclockwise rotating vortices and the saddle singularities in the middle of the cluster interaction region remain (compare Figs. \ref{fig:7:b} and  \ref{fig:5:b}). Therefore, the general structures of the flow are similar in the background or no-background pressure cases, but the details of these structures are quantitatively affected in a perceivable way by the presence or not of the background pressure.

\subsubsection{Case $c\not=1$, with background pressure}
\label{subsubsec_cnot=1_bp}

\medskip
\paragraph{Influence of $c$: theoretical analysis.} Before investigating the influence of the parameter $c$ on the simulation results, we perform some analysis of the effect of this parameter on the propagation of waves. We first digress on the physical significance of the characteristic speeds (\ref{eq_char_speeds}).

The characteristic speeds (\ref{eq_char_speeds}) are the propagation speeds of the waves generated by a small perturbation $(\delta \rho, \delta \boldsymbol{\Omega})$ of a uniform state $(\rho, \boldsymbol{\Omega})$, in the linear approximation. Since these waves are independent of the spatial coordinate normal to the propagation direction, it is legitimate to consider the one-dimensional system (\ref{eq:SOH1D_rho}), (\ref{eq:SOH1D_theta}). Introducing the variable $u= \cos \theta \in [-1,1]$, this system can be written in terms of the unknowns $(\rho,u)$ in matrix form as follows: 
\begin{equation}
\left( \begin{array}{c} \rho \\ u \end{array} \right)_t + A(\rho,u) \left( \begin{array}{c} \rho \\ u \end{array} \right)_x = 0, 
\label{eq:SOH_matrix_form}
\end{equation} 
with 
\begin{equation}
A(\rho,u) = \left( \begin{array}{cc} u &  \rho \\ - \bar \lambda (1-u^2) & c u \end{array} \right), 
\label{eq:SOH_matrix}
\end{equation} 
where $x$ is the coordinate along the propagation direction and $\bar \lambda = \lambda \frac{{p^\varepsilon}'(\rho)}{\rho}$. The quantity $u$ is the projection of the vector $\boldsymbol{\Omega}$ along the propagation direction. The characteristic speeds $\xi_\pm$ are the two eigenvalues of $A(\rho, u)$, i.e. the roots of the polynomial $P(\xi) = \det (A(\rho,u) - \xi \, \mbox{Id})$. Their expression, deduced from (\ref{eq_char_speeds}), is recalled here for the sake of convenience: 
\begin{equation} 
\xi_\pm = \frac{1}{2} \left( (1+c) u \pm \sqrt{ \Delta } \right) , \qquad \Delta = (1-c)^2 u^2 + 4 \bar \lambda  \, (1-u^2) . 
\label{eq_char_speeds_2}
\end{equation}

Now, the perturbation waves $(\delta \rho, \delta u)$ are not arbitrary, but are colinear to the associated eigenvectors. In other words, a simple wave propagating with velocity $\xi_\pm$ is such that 
$$ \left( \begin{array}{c} \delta \rho \\ \delta u \end{array} \right) \, \, \parallel \, \,  \Xi_\pm, $$ 
where $\Xi_\pm$ is the eigenvector of $A$ associated to $\xi_\pm$. This condition leads to the following relations between $\delta \rho$ and $\delta u$: 
\begin{equation}
\frac{1}{2} \big( (1-c) u \mp \sqrt{ \Delta } \big) \delta \rho + \rho \delta u = 0. 
\label{eq:eigenvectors}
\end{equation}
In (\ref{eq:eigenvectors}), the minus sign is associated to $\xi_+$ and vice-versa. 

The special case $u=1$ is particularly simple. It corresponds to an unperturbed state such that the vector $\boldsymbol{\Omega}$ is parallel to the propagation direction. In this case, we have $\xi_\pm = \frac{1}{2} \big( (1+c) u \pm |c-1| u \big)$. To simplify further, we must discuss the position of $c$ with respect to $1$. 

\begin{itemize}
\item[-] If $c>1$, we have: 
$$ \xi_- = u, \qquad \xi_+ = cu, $$
and the associated waves are such that 
\begin{equation}  
\delta u_- = 0, \qquad  \rho \, \delta u_+ = (c-1) \, \delta \rho_+, 
\label{eq:eigenvectors_c>1}
\end{equation}
where the indices '$+$' and '$-$' refer to the wave associated to $\xi_+$ and $\xi_-$. 

\item[-] If $c<1$, we have:
$$ \xi_- = cu, \qquad \xi_+ = u, $$
and the associated waves are such that 
\begin{equation}  
\rho \, \delta u_- = - (1-c) \, \delta \rho_-, \qquad \delta u_+ = 0  . 
\label{eq:eigenvectors_c<1}
\end{equation}

\end{itemize}

We notice that, in both cases, one characteristic speed coincides with the velocity $u$ and that the associated wave is a contact discontinuity: the velocity is continuous on both sides of the waves and the density varies arbitrarily. 

The other characteristic speed corresponds to the velocity $cu$. Therefore, it is the velocity associated to the transport of $\boldsymbol{\Omega}$. The corresponding eigenvector shows that velocity perturbations associated to this wave affect the density. Suppose that initially, $\delta u$ is decreasing with respect to $x$. This corresponds to a situation where particles in the rear are faster than particles in the front. In usual fluids, this contributes to stiffening the velocity front, ultimately leading to a shock, like in the Burgers equation. Simultaneously, a density bump forms in the region where the velocity fronts stiffens. Let us examine how density perturbations behave in the presence of a decreasing velocity perturbation profile in the situation where $c \not =1$. In the case $c>1$, the second relation (\ref{eq:eigenvectors_c>1}) shows that $\delta \rho_+$ has the same sign as $\delta u_+$ and is therefore, decreasing. Therefore, the influence of $c>1$ is to lower the density increase due to the stiffening of the shock. In the case $c>1$, we thus expect a lowering of the amount of congestion generated by the dynamics. By contrast, in the case $c<1$, the first relation (\ref{eq:eigenvectors_c<1}) shows that $\delta \rho_-$ has the opposite sign to that of $\delta u_-$ and is therefore, increasing. Therefore, the influence of $c<1$ is to raise the density increase due to the stiffening of the shock. In the case $c<1$, we thus expect an amplification of the amount of congestion generated by the dynamics.

\begin{figure}[htbp]
  \centering
\subfigure[density $\rho$]{\label{fig:10:a} \includegraphics[scale=0.38]{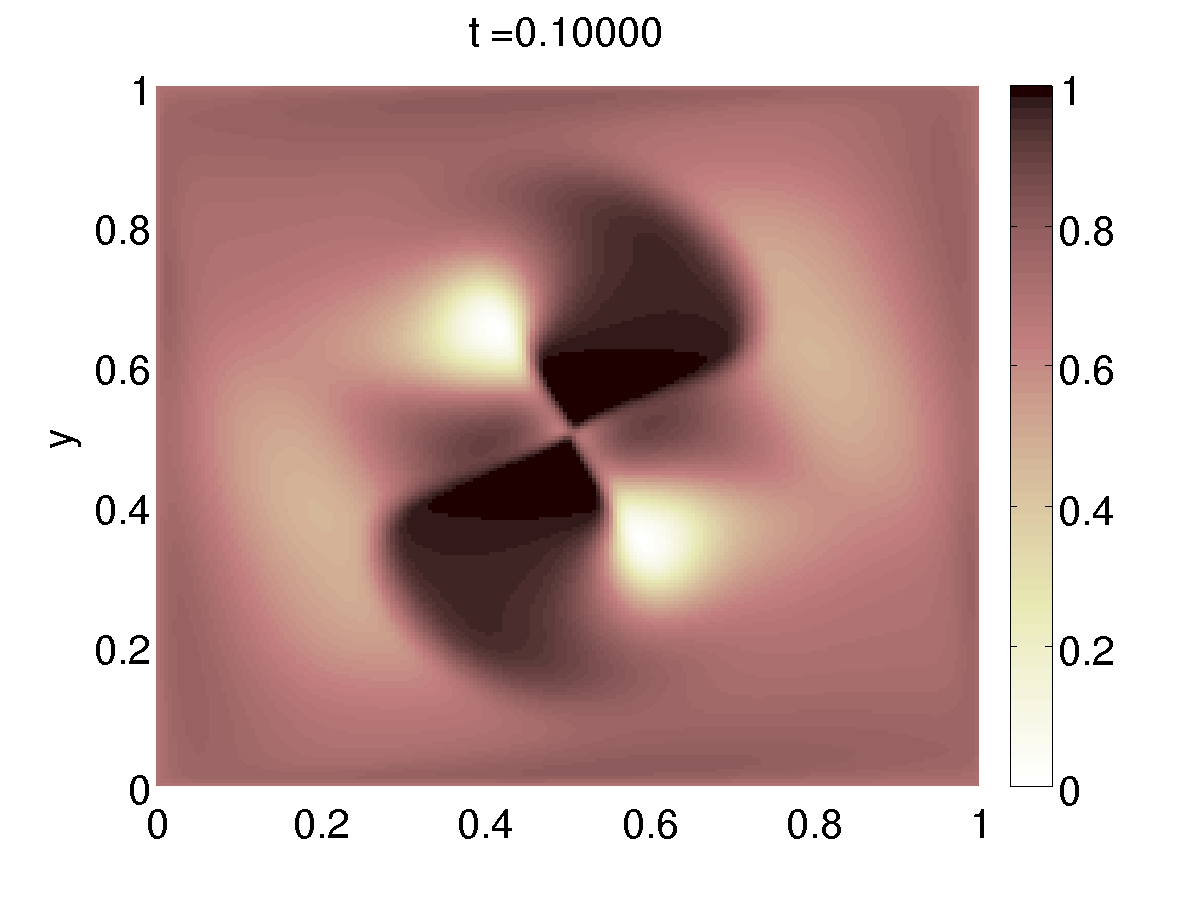}}
\subfigure[$\boldsymbol{\Omega}$]{\label{fig:10:b} \includegraphics[scale=0.38]{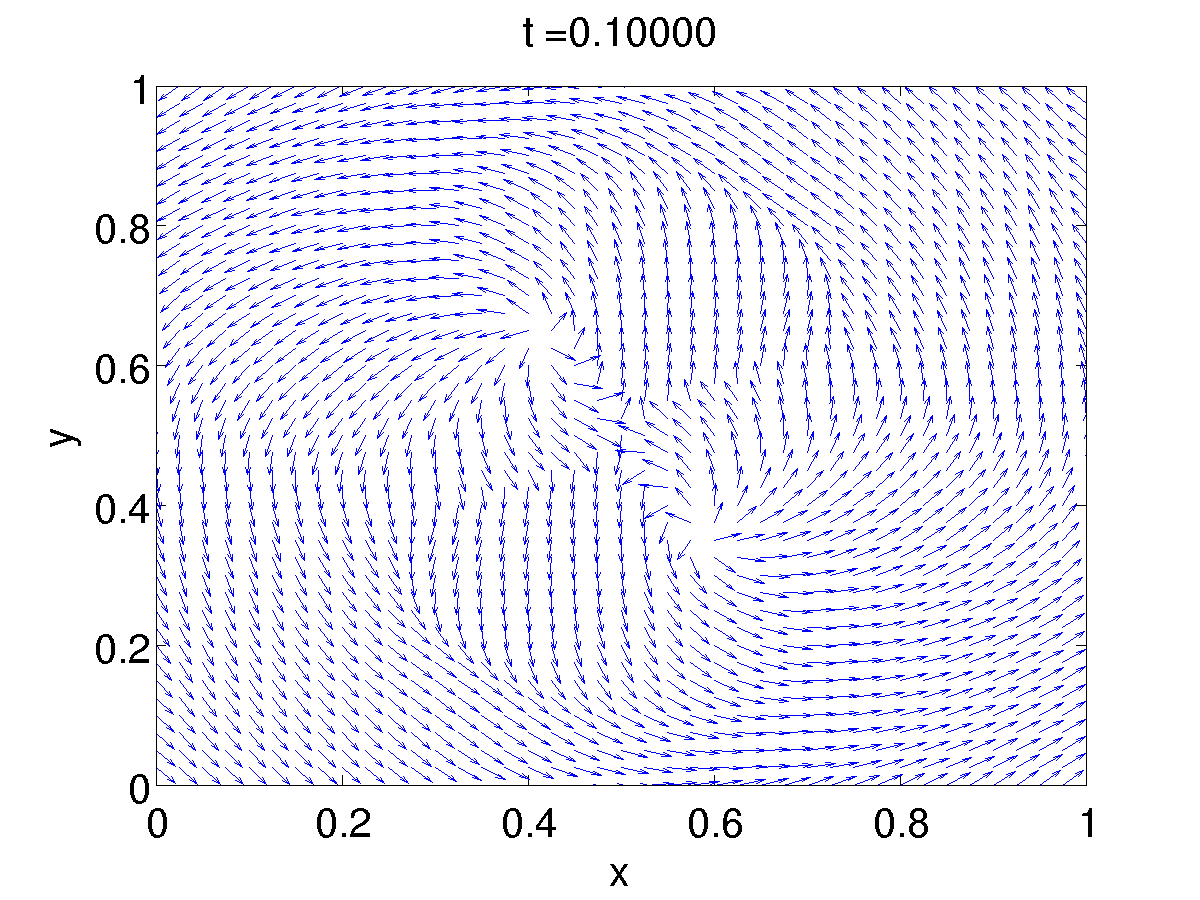}}
   \caption{Cluster collision with background pressure and $c=2$. Density $\rho$ (Fig. \ref{fig:10:a}) and velocity $\boldsymbol{\Omega}$ (Fig. \ref{fig:10:b}) at $t=0.1$, as functions of $x$ and $y$. The density is color-coded, with color map indicated to the right of Fig. \ref{fig:10:a}. The field $\boldsymbol{\Omega}$ is indicated by blue arrows.}
  \label{fig:10}
\end{figure}

\medskip
\noindent
\paragraph{Case $c>1$.} We now illustrate this analysis by numerical simulations. In Fig. \ref{fig:10}, we show simulation results obtained with $c=2$, the other parameters being given by (\ref{eq:parameters}). Again, we consider a background pressure given by (\ref{eq_background_pressure_2}) with $\kappa=1$. The density $\rho$ is shown in Fig. \ref{fig:10:a} and the vector field $\boldsymbol{\Omega}$, in Fig. \ref{fig:10:b}, with the same representation as in the previous figures. This figure must be compared with Fig. \ref{fig:7}, where the case $c=1$ with the same background pressure is displayed. 
As expected from the analysis of the role played by the constant $c$ above, the size of the congested area is smaller than in the case $c=1$ (compare the surface occupied by the black regions on Fig. \ref{fig:10:a} and Fig. \ref{fig:7:a}). The vertical flow created by the deflection of the clusters due to the congestion constraint impacts the background flow more deeply. The hemi-circular structure generated by this impact extends almost up to the top and bottom horizontal boundaries of the domain. But simultaneously, the area where the congestion density is reached is smaller. Therefore, most of the compression is due to the fast propagation of the velocity information, due to the large value $c=2$. This propagation pre-compress the fluid downstream the congestion and contributes to reducing the amount of congestion. Again, we observe that the horizontal boundary of the congested region is a contact discontinuity as no associated discontinuity can be observed on the chart of  $\boldsymbol{\Omega}$, in Fig. \ref{fig:10:b}. The two counterclockwise vortices have transformed into unstable nodes as seen from the flow patterns in Fig. \ref{fig:10:b}, and simultaneously, the area near these unstable nodes has emptied from particles. This area appears in light beige color in Fig. \ref{fig:10:a}, indicating that the density is very small there. The central saddle point has remained and the congested area adopts a remarkable X-pattern around this point. The vacuum regions along the vertical boundaries and bottom left and top right horizontal boundaries of the initially high density inner rectangle have been almost completely filled and are hardly noticed on Fig. \ref{fig:10:a}.

\begin{figure}[htbp]
  \centering
\subfigure[density $\rho$]{\label{fig:16:a} \includegraphics[scale=0.38]{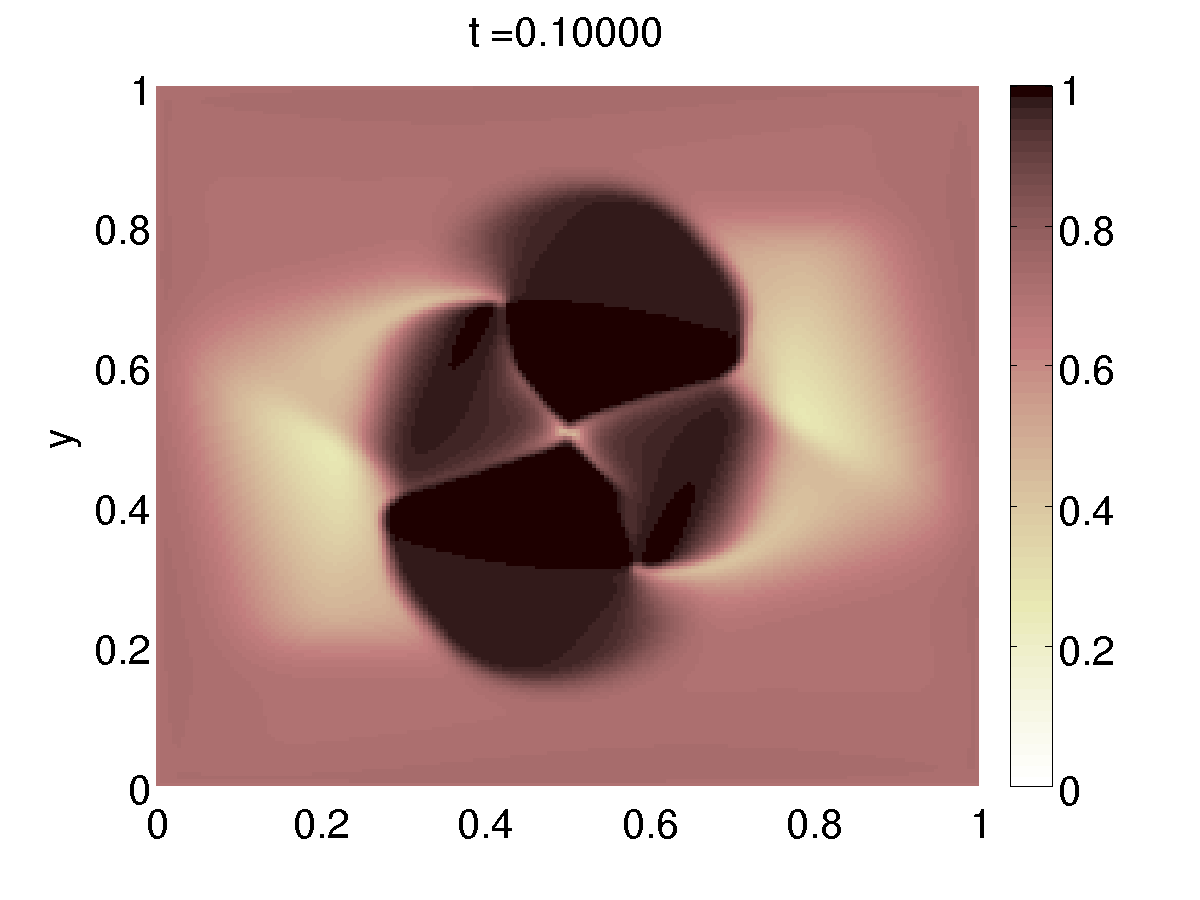}}
\subfigure[$\boldsymbol{\Omega}$]{\label{fig:16:b} \includegraphics[scale=0.38]{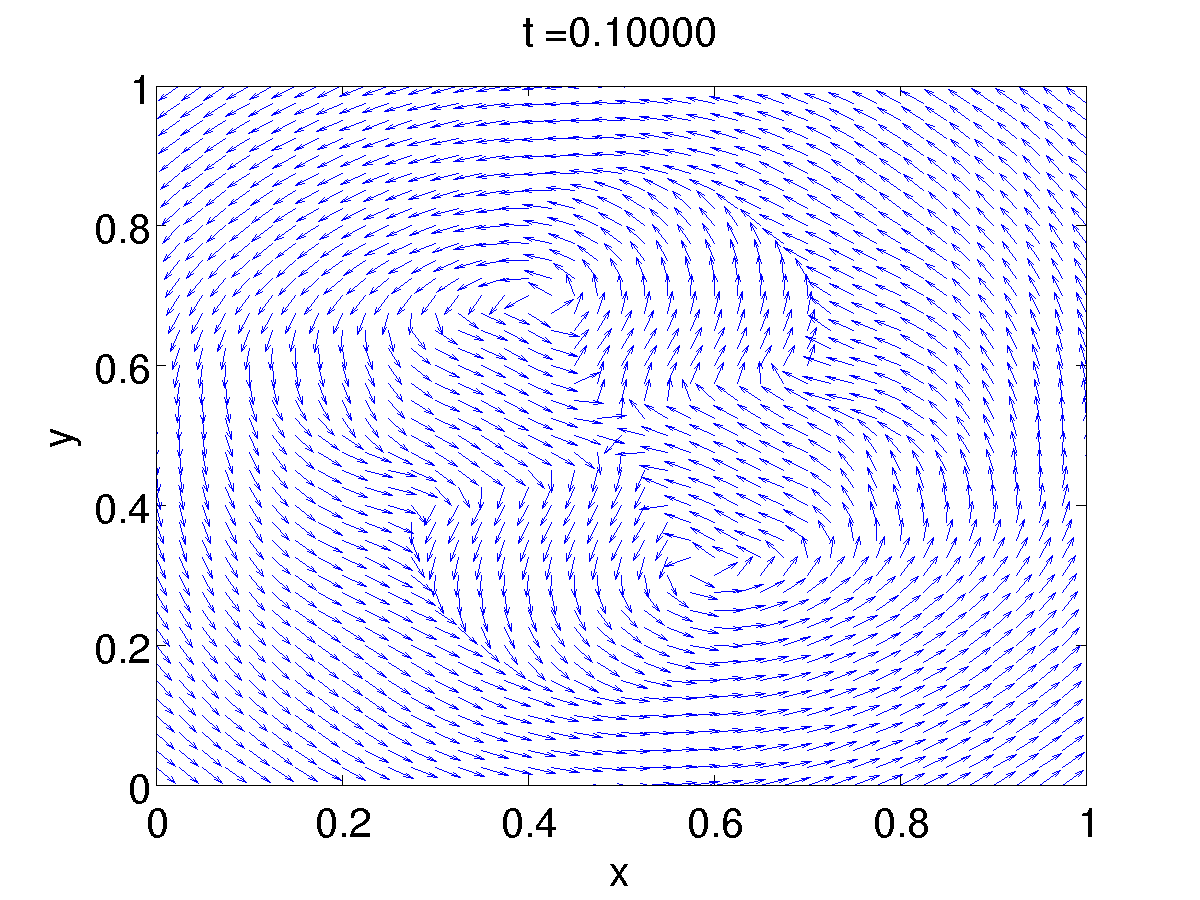}}
   \caption{Cluster collision with background pressure and $c=0.5$. Density $\rho$ (Fig. \ref{fig:16:a}) and velocity $\boldsymbol{\Omega}$ (Fig. \ref{fig:16:b}) at $t=0.1$, as functions of $x$ and $y$. The density is color-coded, with color map indicated to the right of Fig. \ref{fig:16:a}. The field $\boldsymbol{\Omega}$ is indicated by blue arrows. }
  \label{fig:16}
\end{figure}

\medskip
\noindent
\paragraph{Case $c<1$.} We now turn to the opposite case, namely the case $c=0.5$. The case $c<1$ is of broader significance than the case $c>1$. The case $c<1$ covers traffic systems and a larger class of biological swarming systems such as fish schools, mammal herds, \ldots \cite{Degond_Motsch_M3AS08, Frouvelle_M3AS12}. By contrast, the case $c>1$, which occurs with backwards vision \cite{Frouvelle_M3AS12}, would apply e.g. to locust swarms. As discussed in Sec. \ref{subsec_relax}, the relaxation approximation is not hyperbolic in general, although the SOH is. To bypass this problem, we adopt the strategy described in Remark \ref{rem:background_nonhyperbolic} (ii). In  Fig. \ref{fig:16}, we show the simulation results with $c=0.5$ in the presence of a background pressure, the other parameters being given by (\ref{eq:parameters}). The density $\rho$ is shown in Fig. \ref{fig:16:a} and the vector field $\boldsymbol{\Omega}$, in Fig. \ref{fig:16:b}, with the same representation as in the previous figures. This figure must be compared with Figs. \ref{fig:7} and \ref{fig:10}, where the cases $c=1$ and $c=2$ with the same background pressure are displayed. The qualitative features of the solution in Figs. \ref{fig:7}, \ref{fig:10} and \ref{fig:16} are fairly similar. This is a strong indication that the solution obtained with $c=0.5$ is correct and this validates the strategy consisting in using the relaxation system, in spite of its non-hyperbolic character. Unfortunately, there are no analytical solutions available for this model. So, we cannot provide a better validation than through this analogy for the time being. 

Again, the analysis of the role played by the constant $c$ above is confirmed by the inspection of Fig. \ref{fig:16:a}. Indeed, the size of the congested areas is larger than in the case $c=1$. For instance, in  Fig. \ref{fig:16:a}, we notice that there remains a congested area (black color) in the part of the cluster which has not yet been deflected (region just below (resp. above) the upper (resp. lower) vortex), while the same region in Fig. \ref{fig:7:a} is no more congested. The vertical flow created by the deflection of the clusters due to the congestion constraint penetrates the background flow less deeply. But, on the other hand, all the area included in the hemi-circular structure generated by this impact is very close to congestion (dark brown color on Fig. \ref{fig:16:a}), while only one side of this region is close to congestion in Fig. \ref{fig:7:a}. Propagation of velocity information is slow, due to the small value $c=0.5$. Therefore, the pre-compression of the fluid downstream the congestion is less efficient. The uncompressed downstream fluid does not oppose to the propagation of the congestion and therefore, the amount of congestion is larger. Again, the feature that the horizontal boundary of the congested region is a contact discontinuity can still be observed by comparing the density chart (Fig. Fig. \ref{fig:16:a}) and the chart of  $\boldsymbol{\Omega}$ (Fig. \ref{fig:16:b}). The two counterclockwise vortices are there (Fig. \ref{fig:10:b}), but the vacuum area close to these vortices is almost non-existing (see Fig. \ref{fig:16:a}). The central saddle point is also there. The vacuum regions along the vertical boundaries and bottom left and top right horizontal boundaries of the initially high density inner rectangle are fairly visible on Fig. \ref{fig:16:a}.

\subsubsection{Conclusion of the two-dimensional tests}
\label{subsubsec_2Dtests_conclu}

From these tests, we first conclude on the validity of the numerical
strategy developed here. This strategy, consisting of the combination
of the relaxation model of \cite{Motsch_Navoret_MMS11} and of the
AP-method of \cite{Degond_etal_JCP11} proves its effectiveness, even
in the cases where the relaxation model is not hyperbolic. The second
conclusion of these tests is that the density and flow patterns
generated by the SOH model can be very complex. The parameter $c$,
which tunes the speed of propagation of velocity information with
respect to the material flow propagation plays a key role in the type
of structures that are generated. The lower value of $c$, the stiffer
the dynamics is, with the generation of large congested areas and
stiff density and velocity gradients. The constraint of the norm one
velocity   generates various types of singularities such as vortices, nodes, slip lines, shock waves and saddle points. These structures are subject to topological changes as time evolves. It is also a remarkable feature that the method is able to reproduce these singularities and follow their evolution in time efficiently.

\setcounter{equation}{0}
\section{Application: a model of path formation in crowds}
\label{sec_two_fluids}

In this section, we show how the relaxation model with congestion \eqref{Eq:Euler_rho_relaxation}, \eqref{Eq:Euler_q_relaxation} can be used to simulate path formation in crowds. To this aim, we extend it to two-fluid flows to model different groups of pedestrians heading towards opposite directions. We denote by $(\rho_+,\boldsymbol{q}_+)$ (respectively $(\rho_-,\boldsymbol{q}_-)$) the density and momentum of pedestrians heading to the right (respectively to the left). The desired velocity of each group of pedestrians is denoted by $\boldsymbol{w}^{\pm}$. The system satisfied by $(\rho_+,\boldsymbol{q}_+, w_+, \rho_-,\boldsymbol{q}_-, w_-)$ is written as follows: 
\begin{align}
&\rho_{+,t} + \nabla_{\boldsymbol{x}}\cdot \boldsymbol{q}_+ = 0, \label{eq:peds_dens_+}\\
&\boldsymbol{q}_{+,t} +
\nabla_{\boldsymbol{x}}\cdot\left(\frac{\boldsymbol{q}_+\otimes
    \boldsymbol{q}_+ }{\rho_+}\right) +  \nabla_{\boldsymbol{x}}
(p^\varepsilon(\rho))  =
\frac{1}{\beta}(\rho_+ \boldsymbol{
  w}_+-\boldsymbol{q}_+), \label{eq:peds_vel_+} \\
&(\rho_+ \boldsymbol{ w}_+)_t+\nabla_{\boldsymbol{x}}\cdot\left({\boldsymbol{w}_+\otimes
    \boldsymbol{q}_+ }\right)=0, \label{eq:peds_desvel_+}\\
&\rho_{-,t} + \nabla_{\boldsymbol{x}}\cdot \boldsymbol{q}_- = 0, \label{eq:peds_dens_-}\\
&\boldsymbol{q}_{-,t} +
\nabla_{\boldsymbol{x}}\cdot\left(\frac{\boldsymbol{q}_-\otimes
    \boldsymbol{q}_- }{\rho_-}\right) +  \nabla_{\boldsymbol{x}}
(p^\varepsilon(\rho)) = \frac{1}{\beta}(\rho_-
\boldsymbol{w}_--\boldsymbol{q}_-), \label{eq:peds_vel_-}\\
&(\rho_- \boldsymbol{ w}_-)_t+\nabla_{\boldsymbol{x}}\cdot\left({\boldsymbol{w}_-\otimes
    \boldsymbol{q}_- }\right)=0, \label{eq:peds_desvel_-}\\
& \rho=\rho_++\rho_-. \label{eq:peds_dens}
\end{align}
Eqs. (\ref{eq:peds_dens_+}), (\ref{eq:peds_dens_-}) are the continuity equations and eqs. (\ref{eq:peds_vel_+}), (\ref{eq:peds_vel_-}), the momentum balance eqs. for each species of pedestrians. The left-hand side of these equations form two systems similar to the compressible isentropic gas dynamics equations. However, they are coupled by a single pressure $p^\varepsilon (\rho)$ inside the momentum balance equation, where $\rho$ given by (\ref{eq:peds_dens}) is the total density of pedestrians. The pressure has a stiff profile when the density reaches the congestion density $\rho^*$ given by (\ref{eq:20}), (\ref{eq:21}). The tendency for pedestrians to rejoin their target velocity is modeled by the relaxation terms at the right-hand sides of the momentum balance eqs. (\ref{eq:peds_vel_+}), (\ref{eq:peds_vel_-}), with relaxation rate $\beta^{-1}$. The desired velocity is a quantity attached to the pedestrians, and is therefore passively transported by the flow. This is expressed by eqs. (\ref{eq:peds_desvel_+}), (\ref{eq:peds_desvel_-}). Indeed, using the continuity eqs. (\ref{eq:peds_dens_+}), (\ref{eq:peds_dens_-}), these eqs. can be written in non-conservative forms as 
$$ \boldsymbol{ w}_{\pm,t} + ( \boldsymbol{u}_\pm \cdot \nabla_{\boldsymbol{x}} ) \boldsymbol{w}_\pm =0, \qquad \boldsymbol{u}_\pm = \frac{\boldsymbol{q}_\pm}{\rho_\pm}, $$
which express a passive transport of $\boldsymbol{ w}_\pm$ by the velocity $\boldsymbol{u}_\pm$. This model bears analogies with the the Aw-Rascle model of car traffic \cite{Aw_Rascle_SIAP00, Berthelin_etal_ARMA08} and its extension to pedestrian traffic \cite{Appert_etal_NHM11}.

\begin{figure}[htbp]
  \centering
\subfigure[Density $\rho_+$ at
$t=0$]{\label{fig:20:a}\includegraphics[scale=0.3]{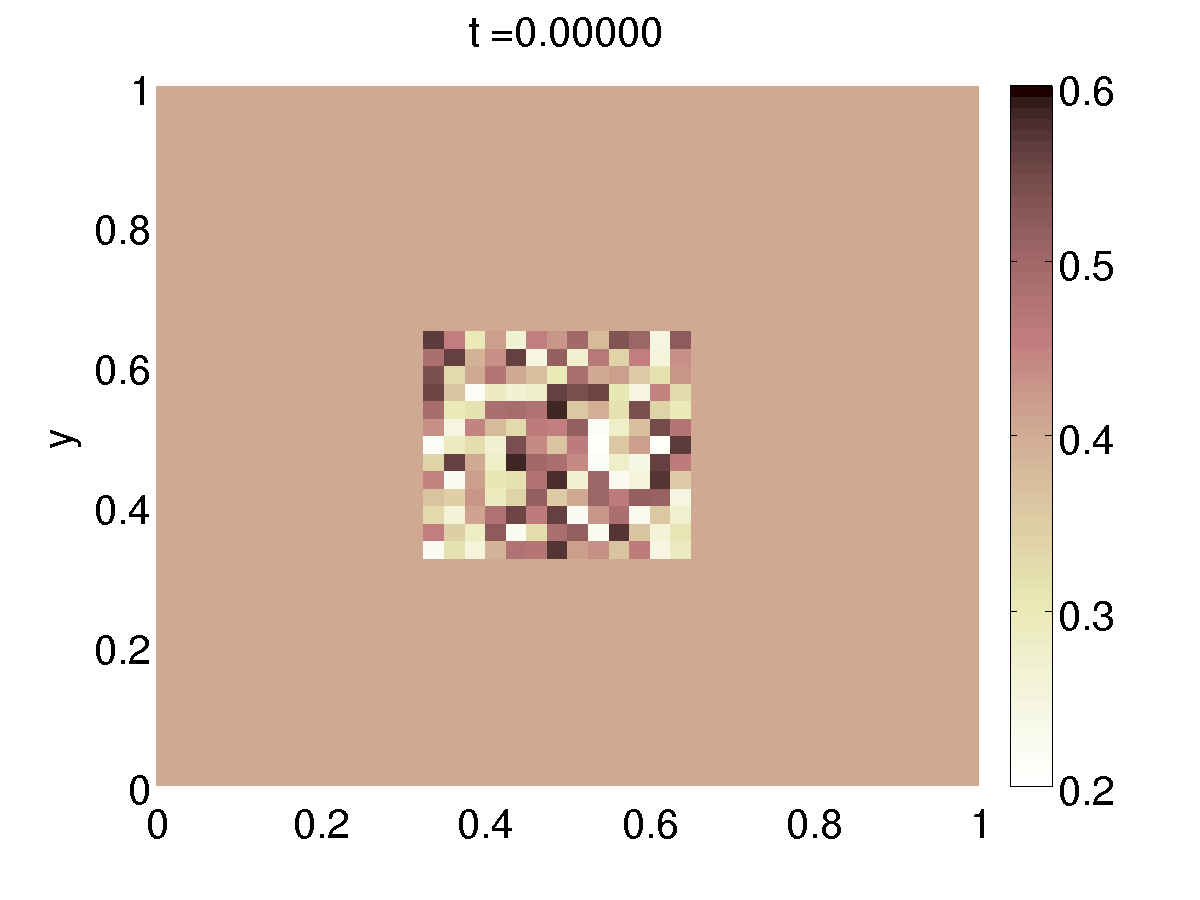}}
\subfigure[Density $\rho_-$ at
$t=0$]{\label{fig:20:b}\includegraphics[scale=0.3]{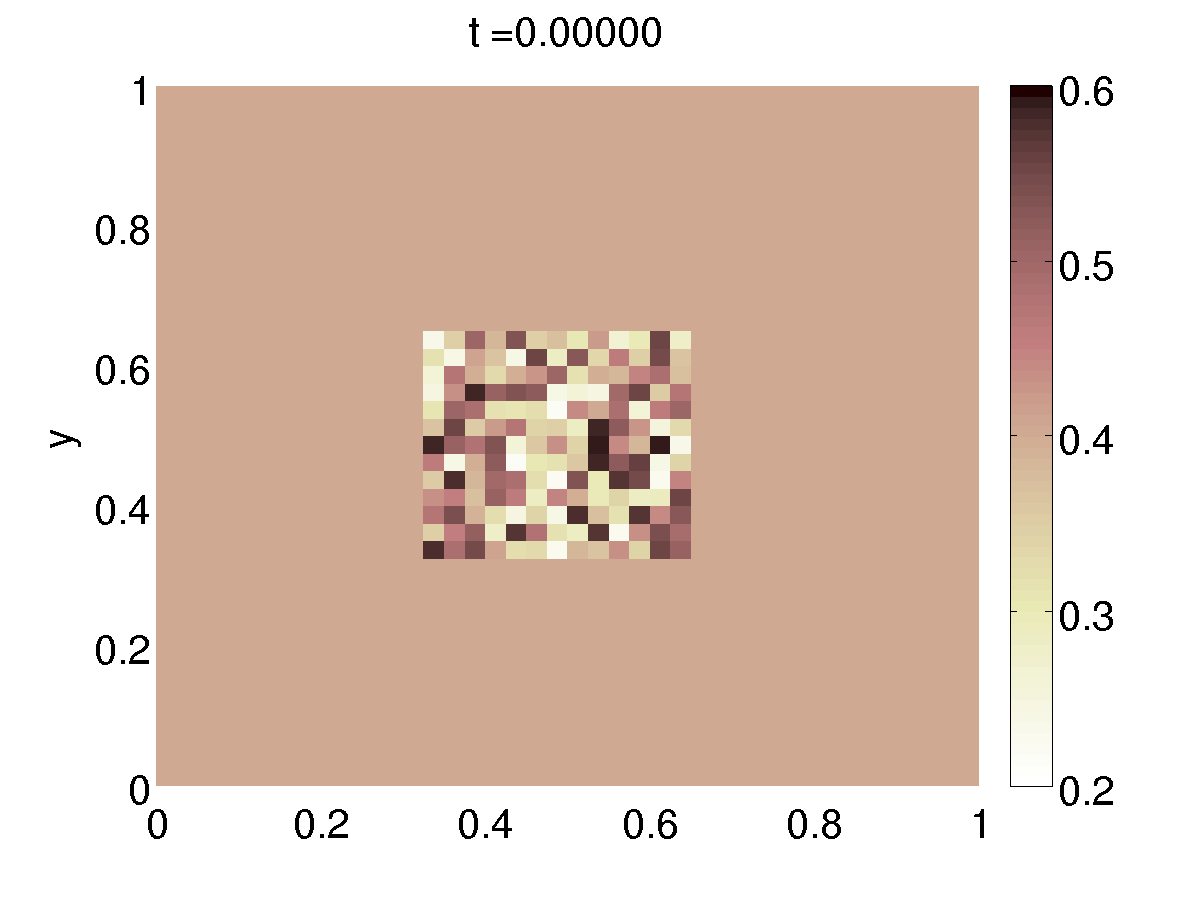}}
\caption{Crowd model: Initial densities $\rho_+$ (Fig. \ref{fig:20:a}), $\rho_-$ (Fig. \ref{fig:20:b}) as
  functions of $x$ and $y$. The densities are color-coded, with color map indicated to the right of the figures.}
 \label{fig:20}
\end{figure}

The following test problem shows how two flows with opposite desired directions interact. The initial data correspond to a flow at rest with a random perturbation on the density.  The steady flow at rest is defined as follows:
\begin{gather*}
  \rho_{s+}(x)=0.4,\quad 
  \boldsymbol{q}_{s+}(x)=0, \quad 
  \boldsymbol{w}_{s+}(x)=\boldsymbol{e}_x,\\
\rho_{s-}(x)=0.4,\quad
  \boldsymbol{q}_{s-}(x)=0, \quad 
    \boldsymbol{w}_{s-}(x)=-\boldsymbol{e}_x,
\end{gather*}
where pedestrians heading to the right (resp. left) have desired velocity along the $x$ axis pointing in the positive (resp. negative) direction. We perturb the density in the square $[1/3,2/3]\times [1/3,2/3]$ by a random noise to model some inhomogeneities in the crowd. The perturbation $\rho_{r+}$ is sampled out of a uniform distribution in the interval $[-0.19,0.19]$ and is constant on squares grouping 25 cells of the simulation mesh (i.e. squares made of 5 spatial steps on each side). 
Finally, the initial conditions are set up such that 
\begin{eqnarray*}
& &  \rho_+(x,0) = \rho_{s+}(x) + \rho_{r+}(x), \quad \rho(x,0) = 0.8, \quad   \rho_-(x,0) = \rho(x,0) - \rho_+(x,0),  \\
& &   \boldsymbol{q}_{\pm}(x,0) = \boldsymbol{q}_{s\pm}(x), \quad \boldsymbol{w}_{\pm}(x,0) = \boldsymbol{w}_{s\pm}(x).
\end{eqnarray*}
The non-uniform initial densities allow for some non-uniform motion to develop. The low density regions give room to pedestrians to pass through while, due to the density constraint, the high density regions act as bottlenecks preventing pedestrians to go through. To illustrate the process more clearly, we choose a relatively slow relaxation process $\beta=0.5$. The mesh size is chosen to be such that $\Delta x=\Delta y = 0.005$ and the time step is  $\Delta t=0.0005$.  Fig. \ref{fig:20} provides a representation of the initial densities of the right-going pedestrians $\rho_+$ (fig. \ref{fig:20:a}) and of the left-going ones $\rho_-$  (fig. \ref{fig:20:b}) in the two-dimensional domain, using a color code. The perturbation of the uniform initial densities inside the inner rectangle takes the form of a checkerboard with cells of random color ranging from light beige color (small densities) to dark brown ones (large densities).

To solve this system, the splitting method is used in the same way as described in section \ref{sec_numerical_method}. During the conservative step, the equations for $(\rho,\boldsymbol{q})$ and $\boldsymbol{w}$ are decoupled and are written as follows:  
\begin{align}
&\rho_{+,t} + \nabla_{\boldsymbol{x}}\cdot \boldsymbol{q}_+ = 0, \label{eq:cons_peds_dens_+}\\
&\boldsymbol{q}_{+,t} +
\nabla_{\boldsymbol{x}}\cdot\left(\frac{\boldsymbol{q}_+\otimes
    \boldsymbol{q}_+ }{\rho_+}\right) +  \nabla_{\boldsymbol{x}}
(p^\varepsilon_0(\rho)) +\nabla_{\boldsymbol{x}} (
p^\varepsilon_1(\rho)) =0, \label{eq:cons_peds_vel_+}\\
&(\rho_+ \boldsymbol{ w}_+)_t+\nabla_{\boldsymbol{x}}\cdot\left({\boldsymbol{w}_+\otimes
    \boldsymbol{q}_+ }\right)=0, \label{eq:cons_peds_desvel_+}\\
&\rho_{-,t} + \nabla_{\boldsymbol{x}}\cdot \boldsymbol{q}_- = 0, \label{eq:cons_peds_dens_-}\\
    &\boldsymbol{q}_{-,t} +
\nabla_{\boldsymbol{x}}\cdot\left(\frac{\boldsymbol{q}_-\otimes
    \boldsymbol{q}_- }{\rho_-}\right) +  \nabla_{\boldsymbol{x}}
(p^\varepsilon_0(\rho)) +\nabla_{\boldsymbol{x}} (
p^\varepsilon_1(\rho)) = 0, \label{eq:cons_peds_vel_-}\\
&(\rho_- \boldsymbol{ w}_-)_t+\nabla_{\boldsymbol{x}}\cdot\left({\boldsymbol{w}_-\otimes
    \boldsymbol{q}_- }\right)=0, \label{eq:cons_peds_desvel_-}
\end{align}
where the pressure is decomposed into $p^\varepsilon(\rho) = p_0^\varepsilon(\rho) + p_1^\varepsilon(\rho) $ as in (\ref{eq_p_decomp}), (\ref{eq_p_decomp_1}).
We can add (\ref{eq:cons_peds_dens_+}) to (\ref{eq:cons_peds_dens_-}) and (\ref{eq:cons_peds_vel_+}) to (\ref{eq:cons_peds_vel_-}) and get the following conservation eqs. for the total density $\rho$ and total momentum $\boldsymbol{q}=\boldsymbol{q}_++\boldsymbol{q}_-$ as follows: 
\begin{align}
&\rho_{t} + \nabla_{\boldsymbol{x}}\cdot \boldsymbol{q} = 0, \nonumber\\
&\boldsymbol{q}_{t} +
\nabla_{\boldsymbol{x}}\cdot\left(\frac{\boldsymbol{q}_+\otimes
    \boldsymbol{q}_+ }{\rho_+}\right) +
\nabla_{\boldsymbol{x}}\cdot\left(\frac{\boldsymbol{q}_-\otimes
    \boldsymbol{q}_- }{\rho_-}\right) +  2\nabla_{\boldsymbol{x}}
(p^\varepsilon_0(\rho)) +2\nabla_{\boldsymbol{x}} (
p^\varepsilon_1(\rho)) =0. \nonumber
\end{align}
In a similar way as in the previous sections, we can implement the AP schemes for this system. The semi-discretization can be written as follows:
\begin{eqnarray}
& & \hspace{-1cm} \frac{\rho^{n+1}-\rho^{n}}{\Delta t} + \nabla_{\boldsymbol{x}}\cdot \boldsymbol{q}^{n+1} = 0, \label{eq:cons_totaldens} \\
& & \hspace{-1cm} \frac{\boldsymbol{q}^{n+1}-\boldsymbol{q}^{n}}{\Delta t} +
\nabla_{\boldsymbol{x}}\cdot\left(\frac{\boldsymbol{q}^{n}_+\otimes
    \boldsymbol{q}^{n}_+ }{\rho_+^{n}}\right) +
\nabla_{\boldsymbol{x}}\cdot\left(\frac{\boldsymbol{q}^{n}_-\otimes
    \boldsymbol{q}^{n}_- }{\rho^{n}_-}\right) +  2\nabla_{\boldsymbol{x}}
(p^\varepsilon_0(\rho^{n})) \nonumber\\
&& \hspace{8.5cm} +2\nabla_{\boldsymbol{x}} (
p^\varepsilon_1(\rho^{n+1})) =0. \label{eq:cons_totalmom}
\end{eqnarray}
The strategy for solving this system is the same as before. The elimination of  $\boldsymbol{q}^{n+1}$ between (\ref{eq:cons_totaldens}) and (\ref{eq:cons_totalmom}) leads to an elliptic equation for the pressure $p_1^{\varepsilon, n+1}$. From $p_1^{\varepsilon, n+1}$, the total density $\rho^{n+1}$ can be obtained by inverting the relation $p_1^\varepsilon(\rho^{n+1}) = p_1^{\varepsilon, n+1}$. Then, the momenta $\boldsymbol{q}_\pm$ can be updated thanks to the discrete, semi-implicit versions of (\ref{eq:cons_peds_vel_+}) and (\ref{eq:cons_peds_vel_-}). Finally, the densities $\rho_\pm$ can be updated thanks to the discrete, implicit version of (\ref{eq:cons_peds_dens_+}) and (\ref{eq:cons_peds_dens_-}). The equations for $\boldsymbol{w}_\pm$ (\ref{eq:cons_peds_desvel_+}) and (\ref{eq:cons_peds_desvel_-}) are decoupled and can be solved easily by standard shock capturing methods. We leave the details to the reader. 

The second step of the splitting method is the relaxation step. It is written as follows: 
\begin{gather*}
  \rho_{\pm,t}=0, \quad w_{\pm,t} = 0, \quad 
  \boldsymbol{q}_{\pm,t}=
\frac{1}{\beta}(\rho_{\pm} \boldsymbol{
  w}_{\pm}-\boldsymbol{q}_{\pm}),
\end{gather*}
and can be solved easily.

\begin{figure}[htbp]
  \centering
\subfigure[density $\rho_+$ at
$t=0.025$]{\label{fig:12:a}\includegraphics[scale=0.3]{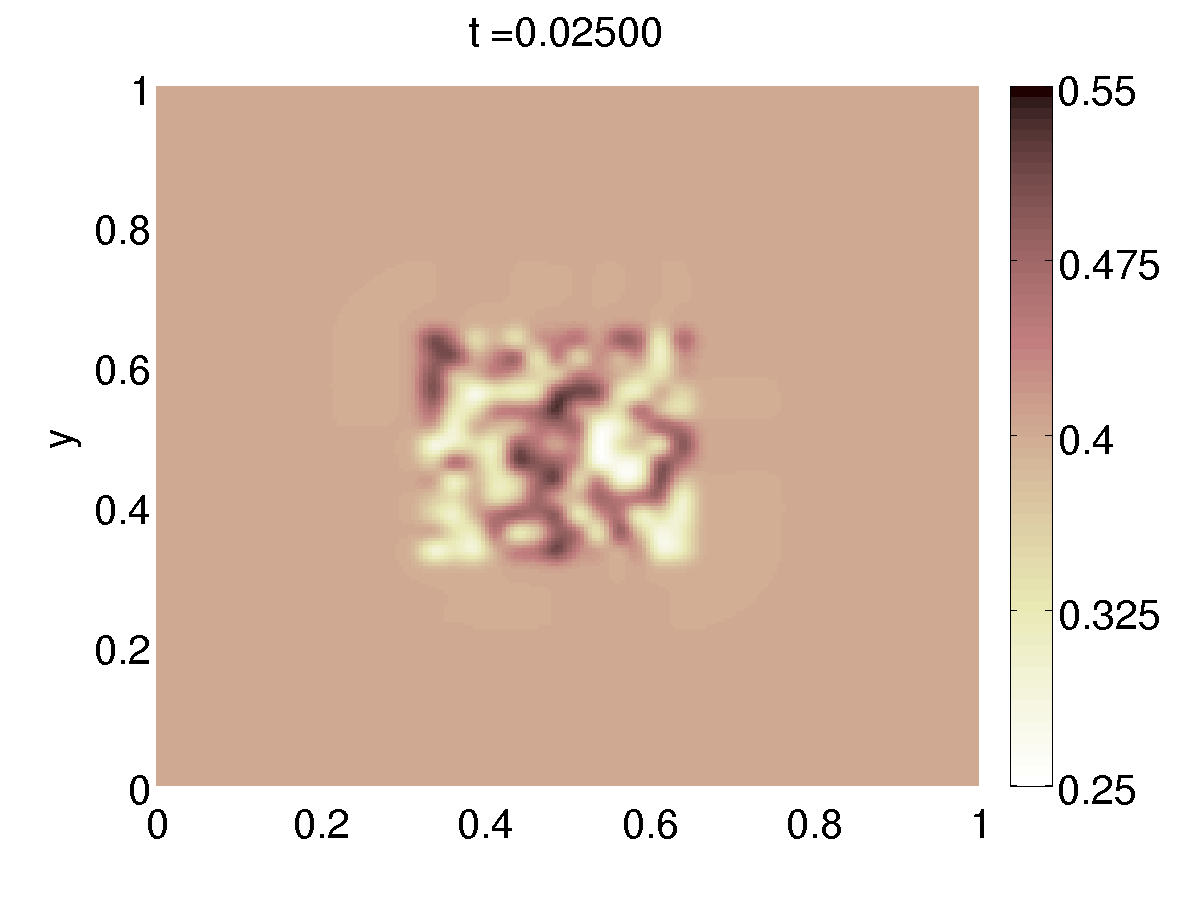}}
\subfigure[density $\rho_+$ at
$t=0.05$]{\label{fig:12:b}\includegraphics[scale=0.3]{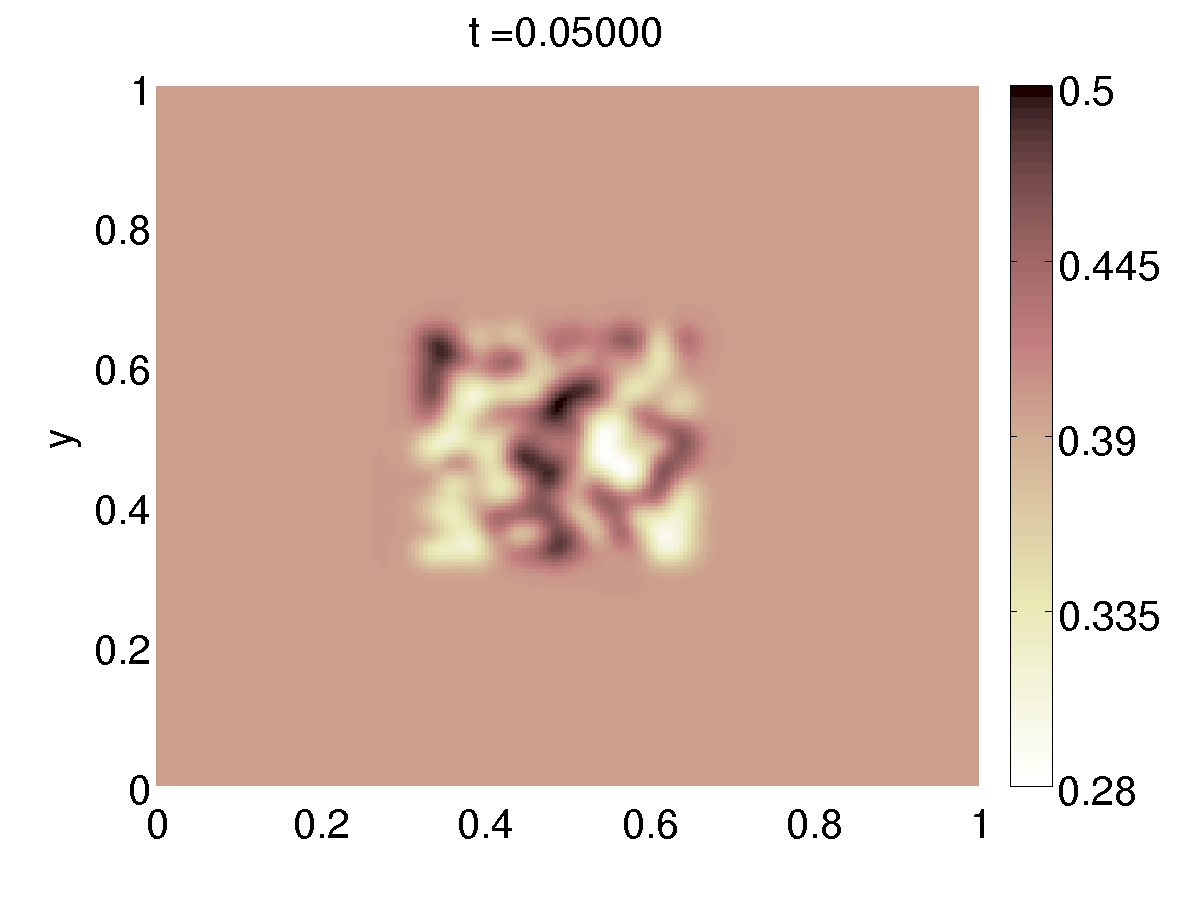}}
\caption{Crowd model: Density $\rho_+$  as
a  function of $x$ and $y$ at times $t=0.025$ (Fig. \ref{fig:12:a}) and $t=0.05$ (Fig. \ref{fig:12:b}). The densities are color-coded, with color map indicated to the right of the figures.}
  \label{fig:12}
\end{figure}

The right-going pedestrian density $\rho_+$ is shown in Fig. \ref{fig:12} at two different times (fig. \ref{fig:12:a}: $t=0.025$ ; fig. \ref{fig:12:b}: $t=0.05$) in the two-dimensional domain, using a color code. 
Due to the relaxation to the desired velocities, the flow is set into motion. The sharp edges between the checkerboard squares in the initial density $\rho_+$ (see Fig. \ref{fig:20:a}) progressively fade away in Figs. \ref{fig:12:a} and \ref{fig:12:b} as the two species of pedestrians mix together. 

In order to highlight the flow structure, we compute the following two quantities which measure which of the left or right going flows is dominant: 
\begin{gather*}
  D\rho=\rho_+-\rho_-, \quad Dq_1=\boldsymbol{q}_{+,1}+\boldsymbol{q}_{-,1},
\end{gather*}
where $\boldsymbol{q}_{\pm,1}$ is the first component of $\boldsymbol{q}_{\pm}$. The quantity $D\rho$ measures which pedestrian species is dominant ($D \rho >0$ if there are more right-going than left-going pedestrians at this point, and vice-versa if $D \rho <0$). The quantity $D q_1$ measures which flow is dominant ($D q_1 >0$ if the right-going flow is dominant and vice-versa if $D q_1 <0$). We expect that $D\rho$ and $Dq_1$ fluctuate about zero since initially $\rho_+ \approx \rho_-$ up to random fluctuations and the initial desired velocities are opposite. But the morphology of the fluctuations results from the dynamics. In particular, we expect that alternating regions where right- or left-going pedestrians dominate will form. In these regions, the dominant flow will be right-going or left-going respectively. 

\begin{figure}[htbp]
  \centering
\subfigure[$Dq1$ and $D\rho$ at
$t=0$]{\label{fig:21:a}\centering\includegraphics[viewport=150 0 450
  430,clip,scale=0.7]{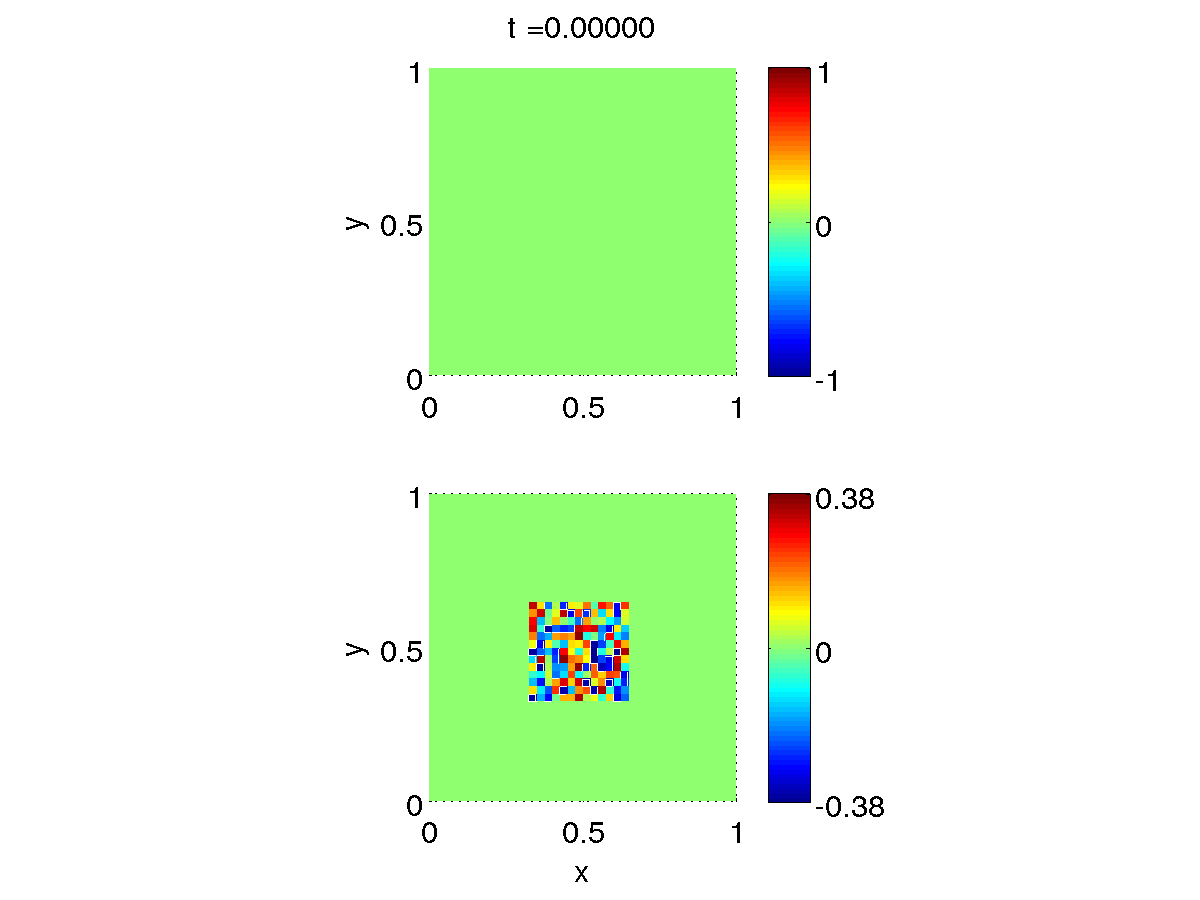}}%
\subfigure[$Dq1$ and $D\rho$ at
$t=0.025$]{\label{fig:21:b}\centering\includegraphics[viewport=150 0 450
  430,clip,scale=0.7]{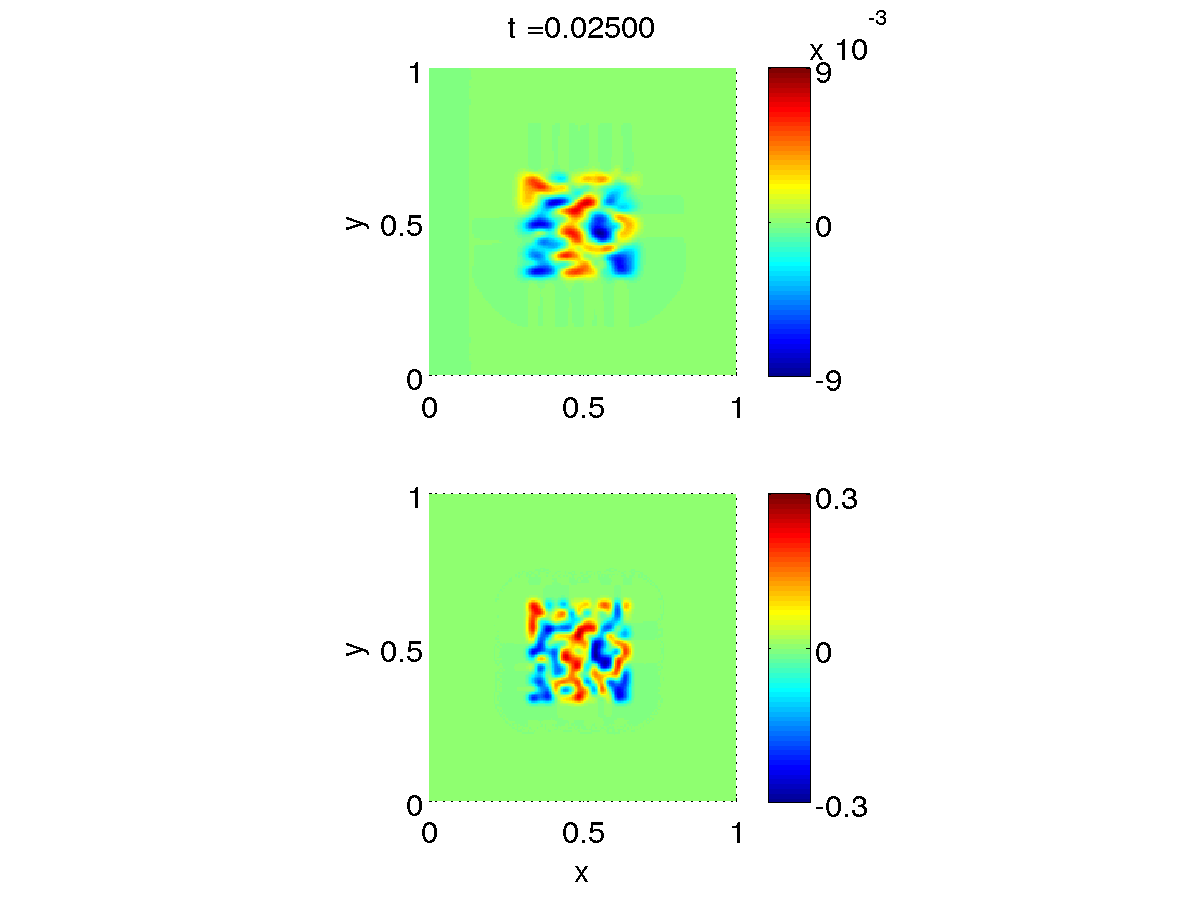}}
\caption{Crowd model: (a)-(b) $x$-component of the momentum difference
  $Dq_1$ (top) and density difference $D \rho$ (bottom) as functions
  of $x$ and $y$ at times $t=0$ (Fig. \ref{fig:21:a}) and $t=0.025$ (Fig. \ref{fig:21:b}). The numerical values are color-coded, with color map indicated to the right of the figures.}
\label{fig:21}
\end{figure}

\begin{figure}[htbp]
  \centering
\subfigure[$Dq1$ and $D\rho$ at
$t=0.05$]{\label{fig:21:c}\includegraphics[viewport=150 0 450
  430,clip,scale=0.7]{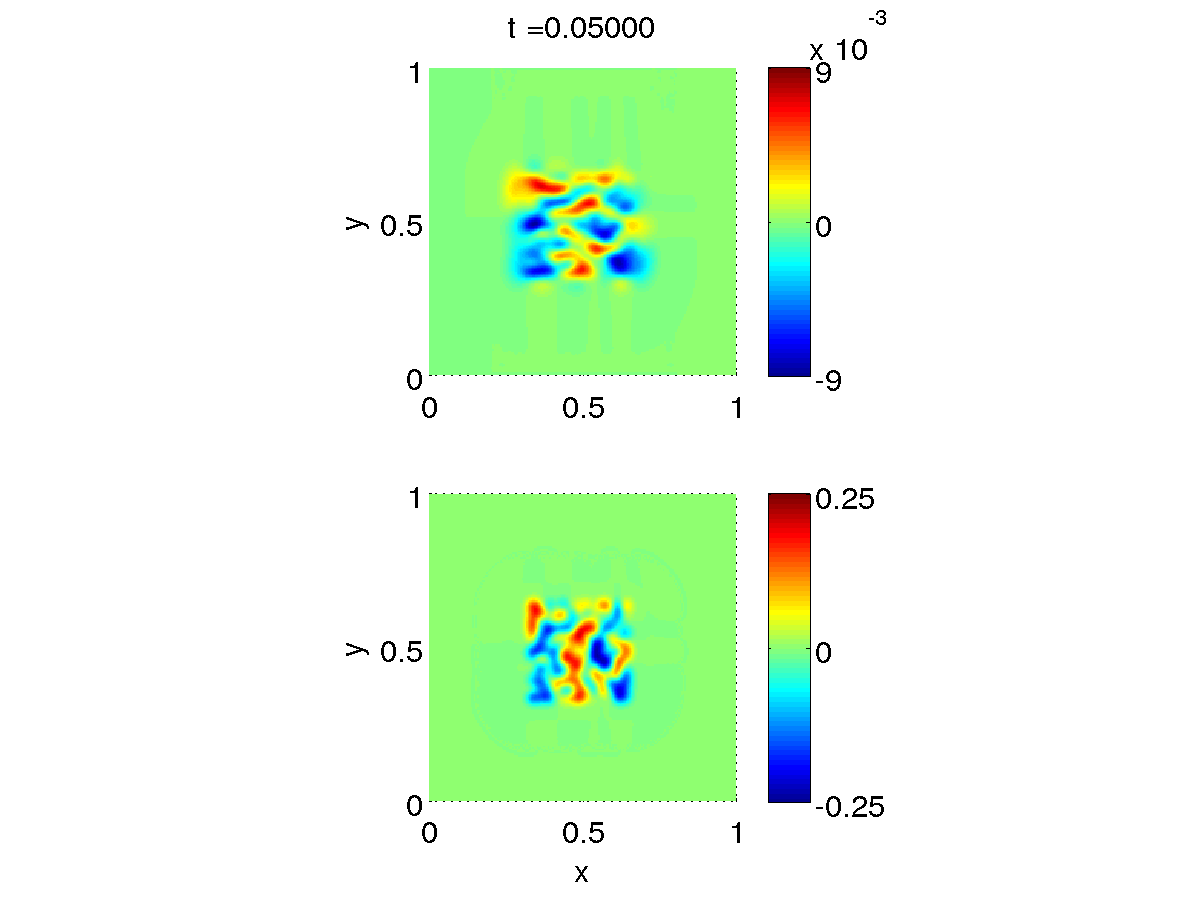}}%
\subfigure[$Dq1$ and $D\rho$ at $t=0.075$]{\label{fig:21:d}\includegraphics[viewport=150 0 450
  430,clip,scale=0.7]{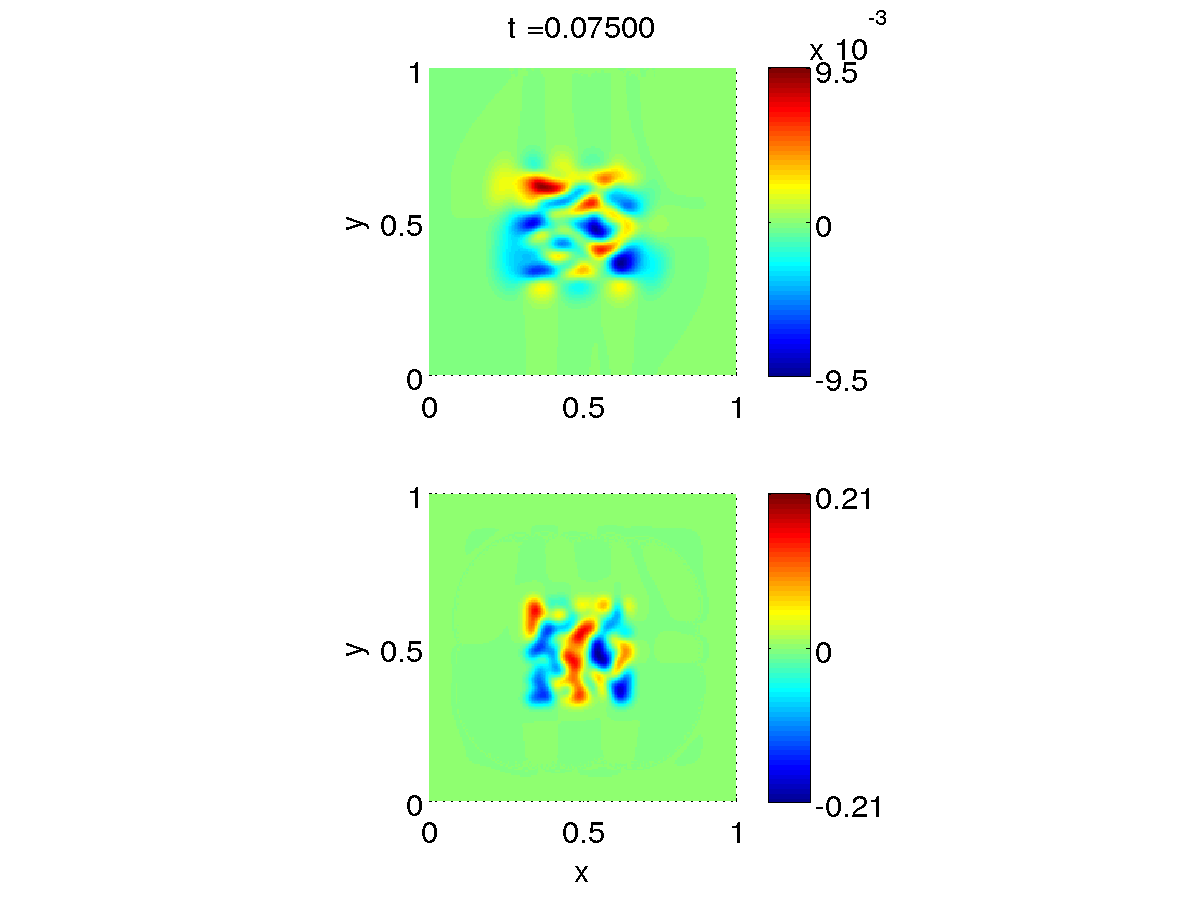}}
\caption{Crowd model: (a)-(b) $x$-component of the momentum difference
  $Dq_1$ (top) and density difference $D \rho$ (bottom) as functions
  of $x$ and $y$ at times $t=0.05$ (Fig. \ref{fig:21:c}) and $t=0.075$ (Fig. \ref{fig:21:d}). The numerical values are color-coded, with color map indicated to the right of the figures.}
\label{fig:21_bis}
\end{figure}

In Fig. \ref{fig:21} and \ref{fig:21_bis}, $D\rho$ and
$Dq_1$  are plotted at different times as functions of $x$ and $y$ using a color code (Fig. \ref{fig:21:a}:  $t=0$ ; Fig. \ref{fig:21:b}:  $t=0.025$ ; Fig. \ref{fig:21:c}:  $t=0.05$~; Fig. \ref{fig:21:d}:  $t=0.075$). In each of these figures, the top picture shows $Dq_1$ and the bottom picture, $D\rho$. The color code is indicated by the color bar next to the figures (the zero value is green, red colors show positive values, while blue colors, negative ones). This figures show the emergence of a phenomenon of path formation. Paths appear as elongated patches of the same color in $Dq_1$ (top pictures). Blue (resp. red) regions  materialize paths of pedestrians moving from right to left (resp. from left to right). Path formation indeed results from the balanced influence of the relaxation towards the target velocity which forces pedestrians to move in one given direction and congestion avoidance which drives pedestrians towards the regions of lower density. Alternating regions of high densities of right-going and left-going pedestrians are materialized in the bottom figures by patches of red and blue colors. These patches reproduce the structure of the patches of $Dq_1$ but not exactly, meaning that some pedestrians are forced to adopt a motion in the transverse direction ($y$-direction) because their desired motion is impeded by large density clusters of opposite pedestrians. 
These simulations show the ability of the model to generate paths from initial random density fluctuations. 

\setcounter{equation}{0}
\section{Conclusion}
\label{sec_conclu}

In this paper, we have investigated the Self-Organized Hydrodynamic (SOH) model derived in \cite{Degond_Motsch_M3AS08}. Short-range repulsion is modeled by a singular pressure which becomes infinite at the congestion density. The singular limit of an infinite pressure stiffness leads to phase transitions from compressible to incompressible dynamics. We have proposed an Asymptotic-Preserving scheme which allows to treat the singular pressure with a very small but finite value of the stiffness parameter. We have performed numerical simulations which illustrate the efficiency of the scheme to treat the occurrence of congestions. A two-fluid variant has been proposed to model path formation in crowds.

Future work will be devoted to the physical enrichment of the model. It will make it quantitatively more accurate for the modeling of herds of gregarious animals like ovine or bovine. It will also allow a more precise description of pedestrian flow and crowd behavior. To this aim, more detailed information from ethology and cognitive sciences will be included in the model. Another direction of improvement concerns the ability of the numerical method to treat vacuum regions and to provide sharp transitions from uncongested to congested areas free of any spurious oscillations.

\setcounter{equation}{0}
\section*%
{Appendix 1: The two dimensional full time and space
  discretization}
\label{sec:2d-case}

We consider the 2D case with domain $\Omega= [0,1]\times
[0,1]$. We denote by $(x_i,y_j)=(i\Delta x,j \Delta y)$, 
$i=0,\cdots,M_1$, $j=0,\cdots,M_2$, where $M_1=1/\Delta x$, $M_2=1/\Delta
y$. Let $U=(\rho,{\boldsymbol{q}})^T$, ${\boldsymbol{q}}=(q_1,q_2)^T$ and $U_{i,j}=U(x_j,y_j)$. To
simplify the exposition, we define
\begin{gather*}
  \boldsymbol{F}(U)=
  \begin{pmatrix}
   c\frac{q_1^2}{\rho}+\lambda p^\varepsilon_0(\rho)\\
c\frac{q_1q_2}{\rho}
  \end{pmatrix}, \quad
\boldsymbol{G}(U)=
  \begin{pmatrix}
   c\frac{q_1q_2}{\rho}\\
c\frac{q_2^2}{\rho}+\lambda p^\varepsilon_0(\rho)
  \end{pmatrix}.
\end{gather*}
Then the left-hand side of \eqref{Eq:Euler_rho_eps} and \eqref{Eq:Euler_q_eps} can be written as
\begin{align*}
&\partial_t \rho+ \partial_x q_1+\partial_y q_2=0,\\
&\partial_t{\boldsymbol{q}} +\partial_x \boldsymbol{F}(U)+\partial_y \boldsymbol{G}(U)+\lambda\nabla_{\boldsymbol{x}}(p^\varepsilon_1(\rho))=0.
\end{align*}
We denote by
\begin{align*}
   &\boldsymbol{F}^{n}=
  \begin{pmatrix}
c\frac{(q_1^n)^2}{\rho^n}+\lambda p^\varepsilon_0(\rho^n)\\
c\frac{q_1^nq_2^n}{\rho^n}
  \end{pmatrix},\quad 
\boldsymbol{G}^{n}=
  \begin{pmatrix}
c\frac{q_1^nq_2^n}{\rho^n}\\
c\frac{(q_2^n)^2}{\rho^n}+\lambda p^\varepsilon_0(\rho^n)
  \end{pmatrix},\\
&\nabla_{i,j}(p^\varepsilon_1)^{n+1}=
\begin{pmatrix}
D_{i,j}^x( p_1^\varepsilon (\rho))^{n+1}\\
D_{i,j}^y(p^\varepsilon_1(\rho))^{n+1}
\end{pmatrix}=
\begin{pmatrix}
D_{i,j}^x( p_1^\varepsilon (\rho^{n+1}))\\
D_{i,j}^y(p^\varepsilon_1(\rho^{n+1}))
\end{pmatrix},
\end{align*}
where $D_{i,j}^xu$, $D_{i,j}^yu$ are the centered difference operators, which, for
any scalar function $u$, are defined as follows:
\begin{gather*}
  D_{i,j}^xu=\frac{u_{i+1,j}-u_{i-1,j}}{2\Delta x},\quad D_{i,j}^yu=\frac{u_{i,j+1}-u_{i,j-1}}{2\Delta y}. 
\end{gather*}
We also define the eigenvalues of the Jacobian matrix for two
one-dimensional hyperbolic system as follows:
\begin{gather*}
  \lambda^{(1)}=c\frac{q_1}{\rho},c\frac{q_1}{\rho}\pm \sqrt{(c^2-c)\frac{q_{1}^2}{\rho^2}+
   \lambda (p^\varepsilon_0)'(\rho)}, \quad \lambda^{(2)}=c\frac{q_2}{\rho},c\frac{q_2}{\rho}\pm \sqrt{(c^2-c)\frac{q_{2}^2}{\rho^2}+\lambda(p^\varepsilon_0)'(\rho)}.
\end{gather*}
With these notations, the full discretization of the scheme takes the following form in 2D:
\begin{align*}
&\frac{\rho_{i,j}^{n+1}-\rho_{i,j}^{n}}{\Delta t}+ \frac{1}{\Delta x}\bigg(Q^{n+\frac{1}{2}}_{i+\frac{1}{2},j}-Q^{n+\frac{1}{2}}_{i-\frac{1}{2},j}\bigg)+ \frac{1}{\Delta y}\bigg(\tilde{Q}^{n+\frac{1}{2}}_{i,j+\frac{1}{2}}-\tilde{Q}^{n+\frac{1}{2}}_{i,j-\frac{1}{2}}\bigg)=0,\\
   &\frac{{\boldsymbol{q}}_{i,j}^{n+1}-{\boldsymbol{q}}_{i,j}^{n}}{\Delta t}+ \frac{1}{\Delta x}\bigg(\boldsymbol{F}^{n}_{i+\frac{1}{2},j}-\boldsymbol{F}^{n}_{i-\frac{1}{2},j}\bigg)+ \frac{1}{\Delta y}\bigg(\boldsymbol{G}^{n}_{i,j+\frac{1}{2}}-\boldsymbol{G}^{n}_{i,j-\frac{1}{2}}\bigg)+\nabla_{i,j}(p^\varepsilon_1)^{n+1}=0,
 \end{align*}
where the fluxes are 
\begin{align*}
&Q^{n+\frac{1}{2}}_{i+\frac{1}{2},j}=\frac{1}{2}\left\{(q_1)^{n+1}_{i+1,j}+(q_1)^{n+1}_{i,j}\right\}-\frac{1}{2}C_{i+\frac{1}{2},j}(\rho^n_{i+1,j}-\rho^n_{i,j}),\\
&\tilde{Q}^{n+\frac{1}{2}}_{i,j+\frac{1}{2}}=\frac{1}{2}\left\{(q_2)^{n+1}_{i,j+1}+(q_2)^{n+1}_{i,j}\right\}-\frac{1}{2}C_{i,j+\frac{1}{2}}(\rho^n_{i,j+1}-\rho^n_{i,j}),\\
&\boldsymbol{F}^{n}_{i+\frac{1}{2},j}=\frac{1}{2}\left\{\boldsymbol{F}^{n}_{i+1,j}+\boldsymbol{F}^{n}_{i,j}\right\}-\frac{1}{2}C_{i+\frac{1}{2},j}({\boldsymbol{q}}^n_{i+1,j}-{\boldsymbol{q}}^n_{i,j}),\\
&\boldsymbol{G}^{n}_{i,j+\frac{1}{2}}=\frac{1}{2}\left\{\boldsymbol{G}^{n}_{i,j+1}+\boldsymbol{G}^{n}_{i,j}\right\}-\frac{1}{2}C_{i,j+\frac{1}{2}}({\boldsymbol{q}}^n_{i,j+1}-{\boldsymbol{q}}^n_{i,j}),
\end{align*}
and
\begin{gather*}
  C_{i+\frac{1}{2},j}=\max
  \{|\lambda^{(1)}_{i,j}|,|\lambda^{(1)}_{i+1,j}|,|\lambda^{(2)}_{i,j}|,|\lambda^{(2)}_{i+1,j}|\},\\
 C_{i,j+\frac{1}{2}}=\max
  \{|\lambda^{(1)}_{i,j}|,|\lambda^{(1)}_{i,j+1}|,|\lambda^{(2)}_{i,j}|,|\lambda^{(2)}_{i,j+1}|\}.
\end{gather*}

Similarly to the one-dimensional case, by inserting the momentum equation into the density equation, we can get
the following discrete elliptic equation:
\begin{align*}
  \begin{split}
    &\rho^{n+1}_{i,j}  - \frac{\Delta t^2}{4}\bigg\{\frac{1}{\Delta
      x^{2}}\left[p^\varepsilon_1(\rho_{i+2,j}^{n+1}) -2 \varepsilon
      p_1(\rho_{i,j}^{n+1}) + \varepsilon
      p_1(\rho_{i-2,j}^{n+1})\right]\\
&\qquad +\frac{1}{\Delta y^{2}}\left[p^\varepsilon_1(\rho_{i,j+2}^{n+1}) -2 p^\varepsilon_1(\rho_{i,j}^{n+1}) + p^\varepsilon_1(\rho_{i,j-2}^{n+1})\right]\bigg\}\\
   = &\ \rho^{n}_{i,j}-\Delta t(D^x_{i,j}q_1^{n} +
   D^y_{i,j}q_2^{n})\\
&+\frac{\Delta t^2}{2} \bigg\{ \frac{1}{\Delta x^2}\left[(\boldsymbol{F}_{i+3/2,j}^{n})^{(1)} - (\boldsymbol{F}_{i+1/2,j}^{n})^{(1)}
      - (\boldsymbol{F}_{i-1/2,j}^{n})^{(1)} + (\boldsymbol{F}_{i-3/2,j}^{n})^{(1)}\right]\\
&\qquad +\frac{1}{\Delta x\Delta y}\left[(\boldsymbol{G}_{i+1,j+1/2}^{n})^{(1)} - (\boldsymbol{G}_{i+1,j-1/2}^{n})^{(1)}
      - (\boldsymbol{G}_{i-1,j+1/2}^{n})^{(1)} + (\boldsymbol{G}_{i-1,j-1/2}^{n})^{(1)}\right]\\
&\qquad +\frac{1}{\Delta x\Delta y}\left[(\boldsymbol{F}_{i+1/2,j+1}^{n})^{(2)} - (\boldsymbol{F}_{i-1/2,j+1}^{n})^{(2)}
      - (\boldsymbol{F}_{i+1/2,j-1}^{n})^{(2)} + (\boldsymbol{F}_{i-1/2,j-1}^{n})^{(2)}\right]\\
&\qquad +\frac{1}{\Delta y^2}\left[(\boldsymbol{G}_{i,j+3/2}^{n})^{(2)} - (\boldsymbol{G}_{i,j+1/2}^{n})^{(2)}
      - (\boldsymbol{G}_{i,j-1/2}^{n})^{(2)} +
      (\boldsymbol{G}_{i,j-3/2}^{n})^{(2)}\right]\bigg\}\\
&+\frac{\Delta t}{2\Delta x}\bigg
[C_{i+\frac{1}{2},j}(\rho^n_{i+1,j}-\rho^n_{i,j})-C_{i-\frac{1}{2},j}(\rho^n_{i,j}-\rho^n_{i-1,j})\bigg]\\
&+\frac{\Delta t}{2\Delta y}\bigg [C_{i,j+\frac{1}{2}}(\rho^n_{i,j+1}-\rho^n_{i,j})-C_{i,j-\frac{1}{2}}(\rho^n_{i,j}-\rho^n_{i,j-1})\bigg].
  \end{split}
\end{align*}
Here $(\boldsymbol{F}_{i+1/2,j}^{n})^{(1)}$ is the first component of the vector
$\boldsymbol{F}_{i+1/2,j}^{n}$. Also, like in the one-dimensional case, we solve
the above elliptic equation to get first $p_1^{n+1}$ then $\rho^{n+1}$.
Once $\rho^{n+1}$ is known, we can get ${q}^{n+1}$ explicitly by solving
 \begin{gather*}
 \boldsymbol{q}_{i,j}^{n+1}= \boldsymbol{q}_{i,j}^{n}- \frac{\Delta
   t}{\Delta
   x}\bigg(\boldsymbol{F}^{n}_{i+\frac{1}{2},j}-\boldsymbol{F}^{n}_{i-\frac{1}{2},j}\bigg)-
 \frac{\Delta t}{\Delta
   y}\bigg(\boldsymbol{G}^{n}_{i,j+\frac{1}{2}}-\boldsymbol{G}^{n}_{i,j-\frac{1}{2}}\bigg)-\Delta
 t \nabla_{i,j}(p^\varepsilon_1)^{n+1}.
 \end{gather*}

\end{document}